\documentclass[11pt,a4paper]{article}
\pdfoutput=1
\usepackage{jheppub}
\usepackage{rotating}
\usepackage{array}
\usepackage{amsmath}
\usepackage[normalem]{ulem}
\usepackage{slashed}
\usepackage{booktabs}
\usepackage[pdftex,table]{xcolor}
\usepackage{units}

\usepackage{multirow}

\definecolor{lightgrey}{gray}{0.95}

\newcommand{\TMDM}{2MDM}

\preprint{DESY-16-113}

\title{How to save the WIMP: global analysis of a dark matter model with two \texorpdfstring{$\boldsymbol{s}$}{s}-channel mediators}

\author[a]{Michael Duerr,}
\author[a]{Felix Kahlhoefer,}
\author[a]{Kai Schmidt-Hoberg,}
\author[b]{Thomas Schwetz,}
\author[b]{Stefan~Vogl}

\affiliation[a]{DESY, Notkestra\ss e 85, D-22607 Hamburg, Germany}
\affiliation[b]{Institut f\"ur Kernphysik, Karlsruher Institut f\"ur Technologie (KIT), D-76021 Karlsruhe, Germany}

\emailAdd{michael.duerr@desy.de}
\emailAdd{felix.kahlhoefer@desy.de}
\emailAdd{kai.schmidt-hoberg@desy.de}
\emailAdd{schwetz@kit.edu}
\emailAdd{stefan.vogl@kit.edu}
\abstract{A reliable comparison of different dark matter~(DM) searches requires models that satisfy certain consistency requirements like gauge invariance and perturbative unitarity. As a well-motivated example, we study two-mediator DM (2MDM). The model is based on a spontaneously broken $U(1)'$ gauge symmetry and contains a Majorana DM particle as well as two $s$-channel mediators, one vector (the $Z'$) and one scalar (the dark Higgs). We perform a global scan over the parameters of the model assuming that the DM relic density is obtained by thermal freeze-out in the early Universe and imposing a large set of constraints: direct and indirect DM searches, monojet, dijet and dilepton searches at colliders, Higgs observables, electroweak precision tests and perturbative unitarity. We conclude that thermal DM is only allowed either close to an $s$-channel resonance or if at least one mediator is lighter than the DM particle. In these cases a thermal DM abundance can be obtained although DM couplings to the Standard Model are tiny. Interestingly, we find that vector-mediated DM--nucleon scattering leads to relevant constraints despite the velocity-suppressed cross section, and that indirect detection can be important if DM annihilations into both mediators are kinematically allowed.}


\begin{document}
\maketitle

\flushbottom

\section{Introduction}

The idea that dark matter~(DM) communicates with the Standard Model~(SM)
via the exchange of additional new particles (so-called dark mediators)
has recently received large amounts of interest~\cite{Frandsen:2012rk,Alves:2013tqa,
Arcadi:2013qia, Garny:2014waa, Chala:2015ama, Alves:2015mua,Ghorbani:2015baa,Fairbairn:2016iuf,
Jacques:2016dqz, Bell:2016fqf}. This framework allows for a consistent
interpretation of simplified DM models~\cite{Buchmueller:2013dya,
Harris:2014hga, Buckley:2014fba,
Abdallah:2015ter,Abercrombie:2015wmb,Kahlhoefer:2015bea,
Englert:2016joy,Boveia:2016mrp} and thus for a reliable comparison of DM
searches at the LHC~\cite{Alves:2015dya, Jacques:2015zha,
Harris:2015kda, Bell:2015rdw, Haisch:2016usn, Brennan:2016xjh} with
other experimental constraints on the interactions of
DM~\cite{Buchmueller:2014yoa, Fairbairn:2014aqa,Alves:2015pea, Choudhury:2015lha,
Blennow:2015gta, Heisig:2015ira}, as well as with the interaction
strength required to obtain the correct relic density from thermal
freeze-out~\cite{Busoni:2014gta}. For a model with a single $s$-channel
mediator the typical conclusion is that the couplings of the mediator to
SM particles have to be rather small in order to be in agreement with
experimental bounds and as a result DM overproduction can only be
avoided in rather special corners of parameter
space~\cite{Chala:2015ama,Jacques:2016dqz}.

It has been pointed out, however, that for a meaningful comparison of different constraints the dark mediator model needs to be consistent with gauge invariance and perturbative unitarity~\cite{Kahlhoefer:2015bea}. As a straight-forward way to satisfy these requirements we consider a DM model containing a new $U(1)'$ gauge group and a dark Higgs that breaks the $U(1)'$ and generates the mass of the fermionic DM particle as well as the mass of the $Z'$ gauge boson. Both the $Z'$ and the dark Higgs can also couple to SM particles and thereby mediate the interactions of DM. For the $Z'$ such a coupling can arise if some or all of the SM fermions are charged under the $U(1)'$ gauge group, while the dark Higgs can couple to SM states via mixing with the SM Higgs.

The model we consider can therefore be thought of as a combination of two simplified models, one with a spin-1 $s$-channel mediator and one with a spin-0 $s$-channel mediator. Since we require all coupling structures and masses to result from a perturbative and gauge-invariant UV completion, however, the model has to satisfy certain relations between the different masses and couplings not usually imposed on simplified DM models. In particular, these relations typically imply that it is not possible to completely decouple one of the two mediators while keeping the remaining masses and couplings fixed. We will therefore refer to this framework as two-mediator DM~(\TMDM). The aim of the present paper is to classify how the simultaneous presence of two mediators changes the conclusions derived for a single mediator.

In the specific example model for \TMDM{} that we study in detail in this paper, we assume vector-like and flavour-diagonal couplings of the $Z'$ to SM quarks. Hence the $U(1)'$ can be identified with gauged baryon number~\cite{Pais:1973mi}. Recently, realistic gauge theories for baryon number that are in agreement with all experimental constraints have been discussed in~\cite{Duerr:2013dza,Perez:2014qfa}. These models contain a fermionic DM candidate~\cite{Duerr:2013lka,Duerr:2014wra,Ohmer:2015lxa,Duerr:2015vna}, and in the limit of neglecting the additional fermions relevant for anomaly cancellation they reduce to the example that we consider.

The generic prediction is that both mediators contribute to the DM annihilation cross section that sets the DM relic abundance. On the other hand, certain experimental bounds may constrain only one of the two mediators. For example, the observed properties of the SM Higgs boson constrain the couplings of the spin-0 mediator, whereas LHC searches for monojets and dijets are more sensitive to the properties of the spin-1 mediator. This compelling interplay provides the model with additional freedom to evade experimental constraints and therefore reduces the tension with the requirement to reproduce the observed relic abundance.

A particularly interesting situation occurs if only one of the two mediators has sizeable interactions with SM states, while the other one is very weakly coupled to the SM. In this case, the weakly-coupled mediator can be much lighter than the DM particle and therefore provide a new final state for DM annihilation. This configuration opens up new parameter space where the DM relic abundance can be reproduced. Such a configuration has been called ``secluded WIMP''~\cite{Pospelov:2007mp} or ``indirect Higgs portal''~\cite{LopezHonorez:2012kv} in the context of Higgs portal DM. We refer to such a weakly-coupled light mediator as a \emph{dark terminator}, because it terminates rather than mediates the interactions of DM.

Models containing only a DM particle and a dark terminator are notoriously difficult to constrain due to the smallness of the interactions between DM and SM particles. The situation is different in our context, because the second mediator can still have sizeable interactions, which can be probed by a range of different DM searches. Hence, the dark terminator can be thought of as a tool to relax the relic density constraints on the interactions of the second mediator in a controlled way and to study the phenomenology of the second mediator in the extended viable parameter space.

The presence of a dark terminator can also lead to novel experimental signatures. Two particularly intriguing examples are indirect detection signals from cascade annihilations, which can avoid the $p$-wave suppression typically present for Majorana DM~\cite{Martin:2014sxa,Bell:2016fqf}, and production of dark terminators at the LHC from final state radiation of other dark sector states~\cite{Autran:2015mfa,Buschmann:2015awa, Bai:2015nfa, Gupta:2015lfa, Buschmann:2016hkc}.

This paper is structured as follows. In section~\ref{sec:model} we introduce our model and discuss the relations between the various couplings. In section~\ref{sec:spin-1} we then consider the case where the interactions between DM and SM particles are dominantly mediated by the $Z'$ and the dark Higgs can be a terminator. Section~\ref{sec:spin-0} focuses on the converse case, i.e.\ a dark Higgs mediator and a $Z'$ terminator. Then, in section~\ref{sec:two-mediators} we study additional effects that may be important if both mediators have sizeable couplings to SM states, or if both act as dark terminators. In section~\ref{sec:2med-scan} we present the results of a global scan of the parameters of the model, with the goal to identify the regions of parameter space which are consistent with a thermal relic abundance, all experimental constraints and perturbative unitarity. We summarise in section~\ref{sec:conclusions}. Supplementary material is provided in four appendices.

\section{Two-mediator DM (\TMDM)}
\label{sec:model}

\subsection{Model set-up}

The aim of this paper is to study the interaction of a Majorana DM
particle~$\chi$ with two different mediators, namely a massive vector
boson $Z'$ and a real scalar $s$. We consider a set-up where the SM
gauge group is extended by an additional $U(1)'$ that is broken by the
vacuum expectation value~(vev) $w$ of a complex scalar field. In such a
set-up the real scalar $s$ corresponds to the Higgs boson from the
spontaneous breaking of the $U(1)'$ and the $Z'$ is the additional
gauge boson that acquires a mass proportional to $w$. This
set-up offers a natural framework for the simultaneous presence of both
mediators and is motivated by basic requirements such as gauge
invariance and perturbative unitarity~\cite{Kahlhoefer:2015bea}.

We consider the following interactions of the two mediators with DM and SM quarks:
\begin{align}
\mathcal{L}_\chi & \supset - \frac{g_\chi}{2}  \bar{\chi} \gamma^\mu \gamma^5 \chi Z'_\mu - \frac{y_\chi}{2 \sqrt{2}} \bar{\chi} \chi s \; , \label{eq:LDM}\\
\mathcal{L}_q & \supset - \sum_q \left(g_q \bar{q} \gamma^\mu q Z'_\mu + \sin \theta \frac{m_q}{v} \bar{q} q s \right)\; , \label{eq:Lq}
\end{align}
where $v = \unit[246]{GeV}$ is the electroweak vev and $\theta$ is the mixing angle between the dark and the SM Higgs. Here we only describe the most important features of our example model for \TMDM, more details are given in appendix~\ref{app:Model}. Note that we restrict our discussion to vector couplings of the $Z'$ to quarks. In principle it is possible that the $Z'$ has additional couplings to the SM, such as axial-vector couplings to quarks, vector and/or axial-vector couplings to leptons, or kinetic mixing with the SM hypercharge gauge bosons. These alternative possibilities are discussed in some detail in appendix~\ref{app:mixing}. In general they are strongly constrained by electroweak precision tests~(EWPT) and dilepton searches at the LHC~\cite{Kahlhoefer:2015bea}. We show in appendix~\ref{app:mixing} that these constraints typically imply that neither axial couplings nor kinetic mixing can give large enough DM annihilation cross sections to avoid DM overproduction. To avoid constraints from flavour-changing processes, we assume the vector couplings to quarks to be flavour-independent.

The flavour-universal vector couplings to quarks that we consider correspond to the case of gauged baryon number, and a detailed discussion of anomaly cancellation and full viable solutions can be found in~\cite{Liu:2011dh,Duerr:2013dza,Perez:2014qfa,Ekstedt:2016wyi}. Note that a Majorana DM can easily be realised in these models by an appropriate choice of quantum numbers, with different possibilities for the additional fermion content~\cite{Duerr:2015vna,Ohmer:2015lxa}. Let us emphasise that our model is still simplified in the sense that we do not specify
the additional particles needed for anomaly cancellation, hence our example model corresponds to a subsector of these models.
The coupling structure imposed by the underlying symmetries of the model ensure that there is no colour anomaly~\cite{Kahlhoefer:2015bea}. As a
result, no new coloured states are required in order to make the
$U(1)'$ anomaly-free and the additional states are expected not to
affect the phenomenology of the model significantly.\footnote{Note, however, that loop-mediated processes such as annihilation into gamma-ray lines may depend upon the detailed particle content of the model~\cite{Duerr:2015vna}.}

At first sight, the interactions in eqs.~\eqref{eq:LDM} and \eqref{eq:Lq} resemble the ones usually considered for $s$-channel simplified models. However, in the present context, there are a number of important differences. Most importantly, the two couplings of DM to the $Z'$ and the dark Higgs are not independent, because the DM Yukawa coupling can be re-expressed in terms of the DM mass and the vev of the dark Higgs, $y_\chi = \sqrt{2} \, m_\chi / w$, and $w$ can be re-expressed in terms of the $Z'$ mass and the $Z'$--DM coupling, $w = m_{Z'} / (2 g_\chi)$. We therefore obtain the important equality
\begin{equation}\label{eq:dark_couplings}
 \frac{y_\chi}{m_\chi} = 2 \sqrt{2} \, \frac{g_\chi}{m_{Z'}} \; .
\end{equation}
This relation is essential to ensure that processes like $\chi \chi \rightarrow Z'_L Z'_L$, where $Z'_L$ denotes a $Z'$ with longitudinal polarisation, do not violate unitarity at large energies~\cite{Kahlhoefer:2015bea}. From the practical point of view, having only one independent coupling in the dark sector significantly simplifies the analysis.

The interactions between DM and SM states are therefore characterised by 6 independent parameters: the three masses $m_\chi$, $m_{Z'}$ and $m_s$, the two couplings between the mediators and the SM, $g_q$ and $\sin \theta$, and the single dark sector coupling, either $g_\chi$ or $y_\chi$. This number should be compared to the four parameters conventionally required in simplified models with a single $s$-channel mediator. The additional two parameters provide the freedom to interpolate between regimes where one of the mediators dominates the phenomenology and to describe the case that both mediators give a relevant contribution. The parameters are summarised in table~\ref{tab:parameters}.

\begin{table}
\begin{center}
\begin{tabular}{ll p{0.5cm} lc}
\toprule
\multicolumn{2}{c}{particle masses} & &
\multicolumn{2}{c}{coupling constants} \\ 
\midrule
DM mass & $m_\chi$ & & dark-sector coupling & $g_\chi$ or $y_\chi$ \\
$Z'$ mass & $m_{Z'}$ && quark--$Z'$ coupling & $g_q$ \\
dark Higgs mass & $m_s$ && Higgs mixing angle & $\theta$ \\
\bottomrule
\end{tabular}
\end{center}
\caption{Summary of the 6 independent parameters of our model. For given masses, there is only one independent dark-sector coupling, since $g_\chi$ and $y_\chi$ are related via eq.~\eqref{eq:dark_couplings}.}
\label{tab:parameters}
\end{table}

In principle, it would be conceivable to simply implement the interactions given above, without specifying the interactions of the new particles with any other SM state. For the purpose of this paper, however, we prefer a different approach, where we include the additional contributions predicted by the simple UV completion in terms of a spontaneously broken $U(1)'$, see appendix~\ref{app:Model} for details. The additional effects compared to the simplified model introduced in eqs.~\eqref{eq:LDM} and \eqref{eq:Lq} can be understood as follows:
\begin{enumerate}
 \item An essential feature of our model is that the two mediators are linked by their common origin from the broken $U(1)'$. This implies in particular that the two mediators can interact with each other, leading to processes like for example $\chi \chi \rightarrow Z'^\ast \rightarrow Z' s$ or $\chi \chi \rightarrow s^{(\ast)} \rightarrow Z' Z'$.
 \item The interactions between the dark Higgs $s$ and SM quarks arise from mixing between the two Higgs bosons. Such a mixing necessarily introduces couplings of all SM particles to the dark Higgs proportional to their masses, leading in particular to the possibility that DM particles can annihilate into SM gauge bosons.\footnote{Note that our approach provides a gauge-invariant realisation of a simplified model with a spin-0 $s$-channel mediator.} At the same time, this mixing changes the couplings of the SM Higgs to other SM particles and furthermore couples the SM Higgs to both the $Z'$ and the DM particle. This leads to additional constraints on $\sin \theta$ from the observed properties of the SM Higgs boson.
 \item In order to ensure that the model remains perturbative, one needs to consider not only $g_q$ and $g_\chi$, but also $y_\chi$ and $\sin \theta$. In particular, requiring that all interactions between the two Higgs bosons remain perturbative leads to non-trivial bounds on the coefficients in the scalar potential, which can be translated into bounds on the mixing angle $\theta$ in terms of the various masses, see section~\ref{sec:perturb} below.
 \item In this model the stability of the DM particle is a consequence of the $U(1)'$ gauge symmetry. Even after symmetry breaking a $Z_2$ symmetry remains unbroken, which protects the DM particle from decay. 
 \item Finally, a general expectation for such a model is that the new $U(1)'$ can mix with the $U(1)_Y$ hypercharge gauge group of the SM. In principle, the kinetic mixing parameter $\epsilon$ could be taken as an additional free parameter of our model. However, the magnitude of this mixing is very tightly constrained, so that it typically cannot have a large effect on the DM phenomenology of the model. We therefore assume in the main text that kinetic mixing is absent at some high scale and only introduced radiatively by quark loops. Tree-level kinetic mixing is discussed in appendix~\ref{app:mixing}. 
\end{enumerate}

We structure the discussion by considering the couplings between the two mediators and SM states. For $g_q \gg \sin\theta$, we expect the $Z'$ to play the dominant role in the phenomenology of the model (discussed in section~\ref{sec:spin-1} below). Conversely, for $\sin\theta \gg g_q$, the dark Higgs will be responsible for mediating the interactions of DM (section~\ref{sec:spin-0}). Finally, if both couplings are comparable, we can expect an interesting interplay between the two mediators, depending on the ratio of the different masses (section~\ref{sec:two-mediators}). This set-up is illustrated in table~\ref{tab:illustration}.

\begin{table}
\begin{center}
\begin{tabular}{l p{0.2cm} c c c}
\toprule
& & $g_q \gg \sin \theta$ & $g_q \sim \sin \theta$ & $\sin \theta \gg g_q$ \\
\midrule
& & & & \\
\multirow{2}{*}{$m_s \gg m_{Z'}$} & & \cellcolor{lightgrey} Spin-1 mediator & & \cellcolor{lightgrey} Spin-0 mediator with \\
& & \cellcolor{lightgrey} simplified model & & \cellcolor{lightgrey} spin-1 terminator \\
& & & & \\
$m_{Z'} \sim m_s$ & & & Two-mediator model & \\
& & & & \\
\multirow{2}{*}{$m_{Z'} \gg m_s$} & & \cellcolor{lightgrey} Spin-1 mediator with & & \cellcolor{lightgrey} Spin-0 mediator \\
& & \cellcolor{lightgrey} spin-0 terminator & & \cellcolor{lightgrey} simplified model\\
& & & & \\
\bottomrule
\end{tabular}
\end{center}
\caption{Simple illustration of the different regimes of the two-mediator model considered in this paper. Note that this table ignores the mass of the DM particle, which ultimately determines whether one of the two mediators can act as a dark terminator.}
\label{tab:illustration}
\end{table}

\subsection{The relic density constraint}

Throughout this work we will adopt the hypothesis that the total
observed abundance of DM is produced by thermal freeze-out (``WIMP
hypothesis''). This implies that for each point in the parameter space
all available annihilation channels are completely determined by the
structure of the model and included self-consistently. We provide
analytic approximations for the most relevant annihilation
cross sections in appendix~\ref{app:annihilation}. For the numerical
study we have implemented the model in \texttt{MicrOMEGAs\_v4.2.5}~\cite{Belanger:2014vza} in
order to obtain an accurate calculation of the predicted relic
density.\footnote{\texttt{MicrOMEGAs} does not take non-perturbative effects due to the multiple exchange of light mediators into account. A detailed study of these effects is beyond the scope of this paper. We note that while such effects might be relevant in some parts of the parameter space, we do not expect these effects to significantly change our conclusions.} Then we vary the parameters of the model to match the observed
relic density $\Omega_\text{DM} h^2 = 0.1188\pm0.0010$~\cite{Ade:2015xua}.

\subsection{Perturbative unitarity}
\label{sec:perturb}

Partial wave perturbative unitarity can be used to derive a number of conditions that the couplings and masses in a given model have to satisfy in order to obtain a consistent description~\cite{Kahlhoefer:2015bea}. Considering the process $\chi \chi \rightarrow \chi \chi$ in the limit of large centre-of-mass energy $\sqrt{s} \gg m_\chi$ and neglecting terms proportional to $\log(s/m_{Z'}^2)$ one gets the following unitarity bounds:
\begin{equation}\label{eq:unit_coupling}
 g_\chi < \sqrt{4\pi} \; ,\qquad y_\chi < \sqrt{8\pi} \; .
\end{equation}
Since $g_\chi$ and $y_\chi$ are related via
eq.~\eqref{eq:dark_couplings}, those inequalities imply also a
constraint on the masses. For example the second
inequality can be rewritten as $ g_\chi \, m_\chi / m_{Z'} < \sqrt{\pi}$.

Perturbative unitarity also constrains the couplings $\lambda_s$, $\lambda_h$ and $\lambda_{hs}$ appearing in the scalar potential, which is fully determined once the dark Higgs mass $m_s$, the mixing parameter $\sin \theta$ and the dark vev $w$ have been specified (see appendix~\ref{app:Model}). By considering the scattering processes $s s \rightarrow s s$ and $h h \rightarrow h h$, it is possible to derive a combined constraint on the three couplings appearing in the scalar potential:\footnote{Our final result differs from the one in~\cite{Hedri:2014mua} by a factor of 2, because we find a factor 1/2 from phase space for two identical final-state particles (see also~\cite{Lee:1977eg}). Note also that stronger bounds could be obtained by also considering the scattering of $W^+W^-$, $ZZ$ and $Z'Z'$~\cite{Kang:2013zba}.}
\begin{equation}\label{eq:unit_HiggsPot}
3 (\lambda_h + \lambda_s) \pm \sqrt{9 (\lambda_h - \lambda_s)^2 + \lambda_{hs}^2} < 16 \pi \; .
\end{equation}
Expressions for the scalar couplings in terms of the free parameters considered in this section are given in eqs.~(\ref{eq:lambdah})--(\ref{eq:lambdahs}). In the absence of Higgs mixing ($\lambda_{hs} = 0$), the inequality \eqref{eq:unit_HiggsPot} simply gives an upper bound on the mass of the dark Higgs: $m_s < \sqrt{16\pi/3} \, w = \sqrt{4\pi/3} \, m_{Z'} / g_\chi$. We note that this expression can also be rewritten as $y_\chi < \sqrt{32\pi/3} \, m_\chi / m_s$, thus giving an upper bound on $y_\chi$ for fixed ratio $m_\chi / m_s$. For $m_s / m_\chi > \sqrt{4\pi/3} \approx 2.05$ this bound is stronger than the bound on the Yukawa coupling given in eq.~(\ref{eq:unit_coupling}).

For the numerical analysis below we are going to impose the conditions specified in eqs.~\eqref{eq:unit_coupling} and \eqref{eq:unit_HiggsPot} to determine the region of perturbative unitarity. Note that we do not consider the running of the dark gauge coupling,
which may become relevant for $g_\chi$ close to the perturbative
boundary. For such large values of $g_\chi$ it is expected that the dark
gauge coupling develops a Landau pole at some high energy. Requiring
that $g_\chi$ remains perturbative up to high scales would therefore
lead to slightly stronger perturbativity bounds. For the largest part of
the parameter space of the model, however, the running is sufficiently
slow so that these effects are completely negligible.

\section{Spin-1 mediation}
\label{sec:spin-1}

We begin by focusing on spin-1 mediation, i.e.\ we consider the case that the Higgs mixing angle $\theta$ is sufficiently small to be negligible for the relic density calculation. If the dark Higgs is significantly heavier than the DM particle, it plays a negligible role in the thermal freeze-out. This scenario has been considered to some extent in~\cite{Kahlhoefer:2015bea}. Here we extend this framework by considering also the situation of $m_s \ll m_\chi$. We start by summarising the most important constraints on the model in sections~\ref{sec:vector-DD} and \ref{sec:vector-collider}, and discuss our results in section~\ref{sec:vector-results}.

\subsection{Direct detection}
\label{sec:vector-DD}

For our assumptions of Majorana DM and vector couplings between the $Z'$ and quarks we obtain an effective four-fermion interaction relevant for DM--nucleus scattering of the axial--vector type:\footnote{We note that for the vector quark current the renormalisation group effects discussed in~\cite{D'Eramo:2016atc} are irrelevant and therefore the high-energy Lagrangian can be directly matched at the hadronic scale.}
\begin{equation}
 \mathcal{L}_\text{eff} \supset  \sum_q \frac{g_\chi \, g_q}{2 \, m_{Z'}^2} \bar{\chi} \gamma^\mu \gamma^5 \chi \, \bar{q} \gamma_\mu q \; , \label{eq:DDoperator}
\end{equation}
where we have integrated out the $Z'$. For this operator, the DM--nucleon scattering cross section is suppressed in the non-relativistic limit. In most studies the resulting constraints from direct detection experiments have therefore simply been neglected. Here we show, however, that these constraints can in fact be relevant.

The leading non-relativistic DM--nucleus operators have been classified systematically~\cite{Fitzpatrick:2012ix}. It was shown that in the non-relativistic limit the operator~(\ref{eq:DDoperator}) can be decomposed into a piece proportional to the DM velocity $\vec{v}$ and a piece proportional to the momentum transfer $\vec{q}$. In the notation of~\cite{Anand:2013yka} one finds 
 \begin{equation}
   \bar{\chi} \gamma^\mu \gamma^5 \chi \, \bar{q} \gamma_\mu q
   \quad\rightarrow\quad
   2 \vec{v}^{\bot} \cdot \vec{S}_{\chi} + 2 i \vec{S}_{\chi} \cdot \left( \vec{S}_{N} \times \frac{\vec{q}}{m_{N}} \right) \;,
\end{equation}
where $\vec{v}^{\bot}= \vec{v} + \frac{ \vec{q}}{2 \mu_{\chi N}}$ with $\mu_{\chi N}$ being
the DM--nucleon reduced mass, and $\vec{S}_{N}$ and
$\vec{S}_{\chi}$ denote the spin of the nucleus and the DM
particle, respectively. This corresponds to the operators
$\mathcal{O}_8$ and $\mathcal{O}_9$ of~\cite{Anand:2013yka}. Crucially, the first term is independent of the nucleus spin and therefore receives a coherent enhancement proportional to the mass of the target nucleus squared. This enhancement factor, which can be as large as $2 \times 10^4$ for xenon-based experiments, partially compensates for the velocity suppression and allows direct detection experiments to retain some sensitivity to the interaction in eq.~(\ref{eq:DDoperator}).

Since these operators lead to a recoil spectrum that differs substantially from the standard spin-(in)dependent interactions, the limits
reported by direct detection experiments do not apply
directly. Currently LUX~\cite{Akerib:2015rjg} has the best sensitivity,
and we translate their results into an upper bound on the interactions
considered here.

As we do not have access to detailed information on the observed events or the background expectation, we cannot repeat the full likelihood analysis performed by the collaboration. In our simplified analysis we  exploit the fact that most of the events observed by LUX fall above  the median in the $S1$--$S2$ plane (see Fig.~2 of~\cite{Akerib:2015rjg})
and limit the region of interest to the lower half of the LUX search region. We extract 
the detection efficiency as a function of the nuclear recoil energy from Fig.~1 of~\cite{Akerib:2015rjg}.
Given that the efficiency  is close to unity for a substantial range of nuclear recoils, the smaller signal region can be accounted for by reducing the total efficiency by a factor $\frac{1}{2}$. We use the implementation of the nuclear response functions provided by~\cite{Anand:2013yka} to calculate the differential recoil rate in a xenon detector and, after convolving the differential rate with the  efficiency of LUX, we obtain the number of expected events in the signal region. Taking one observed event below the median in the $S1$--$S2$ plane and assuming no background (which gives the weakest upper limit on the signal) we find a 90\%~CL upper limit on the expected number of events of $3.89$. In order to validate our results we repeat our analysis with the LUX 2013 data~\cite{Akerib:2013tjd} and recover the bound on the effective operator in eq.~\eqref{eq:DDoperator} 
reported in~\cite{DelNobile:2013sia}.

\subsection{Collider constraints}
\label{sec:vector-collider}

Important collider constraints on the model arise from a number of  searches at the LHC and at LEP. Some of the signatures are  directly related to the production of DM, such as monojet searches, while other observables, for example searches for dijet and dilepton resonances, constrain new physics interactions between SM particles. Finally, indirect precision measurements of SM relations, in particular EWPT, can be relevant.

\bigskip

{\it Monojets.} Searches for the production of DM in association with jets have been conducted by ATLAS and CMS both at 8 and \unit[13]{TeV}~\cite{Khachatryan:2014rra,Aad:2015zva,CMS:2016tns,Aaboud:2016tnv}. Since the \unit[13]{TeV} results are not yet competitive with the  final limits from the \unit[8]{TeV} run of the LHC we focus on the \unit[8]{TeV} CMS results~\cite{Khachatryan:2014rra} in our analysis.
We simulate monojet events with
\texttt{CalcHEP~v3}~\cite{Belyaev:2012qa} and pass them to  \texttt{Pythia~v8}~\cite{Sjostrand:2007gs} for showering and hadronisation. Detector effects are taken into account with the help of \texttt{DELPHES~v3}~\cite{deFavereau:2013fsa}.

\bigskip

{\it Dijets.} A combination of dijet searches from ATLAS and CMS at \unit[8]{TeV} and \unit[13]{TeV}~\cite{Khachatryan:2015sja,Aad:2014aqa,Khachatryan:2015dcf,ATLAS:2015nsi,Khachatryan:2016ecr} was recently performed in~\cite{Fairbairn:2016iuf}. The resulting model-independent bounds on the $Z'$ coupling as a function of its mass and width can be directly applied to our case. These bounds are valid as long as $\Gamma_{Z'} / m_{Z'} \leq 0.3$, which is sufficient for all coupling combinations considered in this paper (with the exception of the global scans discussed in section~\ref{sec:2med-scan}). While larger widths can in principle be probed using dijet angular correlations (see~\cite{ATLAS:2015nsi}), we do not consider these constraints here. We also note that there are presently no LHC bounds on dijet resonances with an invariant mass below $\unit[500]{GeV}$. In principle, this mass range can be constrained using dijet searches from previous hadron colliders and LHC searches for dijet resonances produced in association with SM gauge bosons~\cite{Chala:2015ama}, but we find that stronger constraints are obtained from the searches for dilepton resonances discussed next.

\bigskip

{\it Bounds due to gauge boson kinetic mixing.} It has been known for a long time that  
fermions charged under hypercharge and a $U(1)^\prime$ induce kinetic mixing between the corresponding gauge bosons~\cite{Holdom:1985ag} which can be parametrized by
\begin{equation}\label{eq:kin-mix}
\mathcal{L} = -\frac{1}{2} \sin \epsilon \, F'^{\mu\nu} B_{\mu\nu} \; ,
\end{equation} 
where $B_{\mu\nu}$ and $F'^{\mu\nu}$ are the field strength tensors of the SM hypercharge $U(1)_Y$ and the $U(1)^\prime$, respectively. This interaction respects the full gauge symmetry of the theory and could therefore  be present at tree level so that $\epsilon$ would be a free parameter. This option is discussed in appendix~\ref{app:mixing}, where we show that it is strongly constrained. Here we adopt the natural assumption that kinetic mixing is absent at some high scale (see e.g.~\cite{Babu:1996vt}) and consider only the contribution expected from loops. Under the assumption that $\epsilon$ vanishes at a scale $\Lambda$ the induced kinetic mixing at a scale $\mu$ is given by~\cite{Carone:1995pu}
\begin{equation}
\epsilon(\mu) = \frac{e \, g_q}{2 \pi^2 \, \cos \theta_\text{W}} \log \frac{\Lambda}{\mu} \simeq   0.02 \, g_q \, \log \frac{\Lambda}{\mu} \; .
\end{equation}
Hence, in this approach there is a natural size for $\epsilon$ that depends only logarithmically on $\Lambda$. To estimate the magnitude of this effect, we set $\Lambda = \unit[10]{TeV}$ and identify $\mu$ with $m_{Z^\prime}$ in the following. Effectively, $\epsilon$ is then no longer a free parameter.

As detailed in appendix~\ref{app:U(1)mixing}, the mixing between the
$Z$ and the $Z^\prime$ generates a coupling between leptons and the
$Z^\prime$ which can be searched for using dilepton resonances.  In
our numerical analysis we include the \unit[8]{TeV} ATLAS search for dilepton
resonances~\cite{Aad:2014cka} and a CDF search~\cite{Aaltonen:2008ah},
which is more sensitive in the low-mass regime.  

Furthermore, kinetic
mixing also modifies the properties of the $Z$, see appendix~\ref{app:U(1)mixing} for details. In the gauge boson
sector any deviation from the SM expectation is tightly constrained by
precision measurements at LEP.  The impact of heavy new physics
is conveniently parametrized by the $S$ and $T$ parameters. For $m_{Z^\prime}^2 > 2 m_Z^2$ we confront the model predictions with
the limits from the combined fit of $S$ and $T$ given
in~\cite{Agashe:2014kda}.  For lighter $m_{Z^\prime}$ this procedure
becomes unreliable and we use the limit on the $\rho$ parameter $\rho \equiv m_W^2 / (m_Z^2 c_\mathrm{W}^2) = 1.0004 \pm
0.00024$~\cite{Agashe:2014kda} instead. Our bound on $\epsilon$ is in
good agreement with the limit obtained from a general fit to
precision observables presented in~\cite{Hook:2010tw}.
We emphasise that the bounds from dilepton searches as well as EWPT could be modified due to the UV sensitivity of the kinetic mixing. The resulting constraints are discussed in more detail in appendix~\ref{app:mixing}.

\subsection{Results}
\label{sec:vector-results}

\begin{figure}[t]
\centering
\includegraphics[height=0.28\textheight]{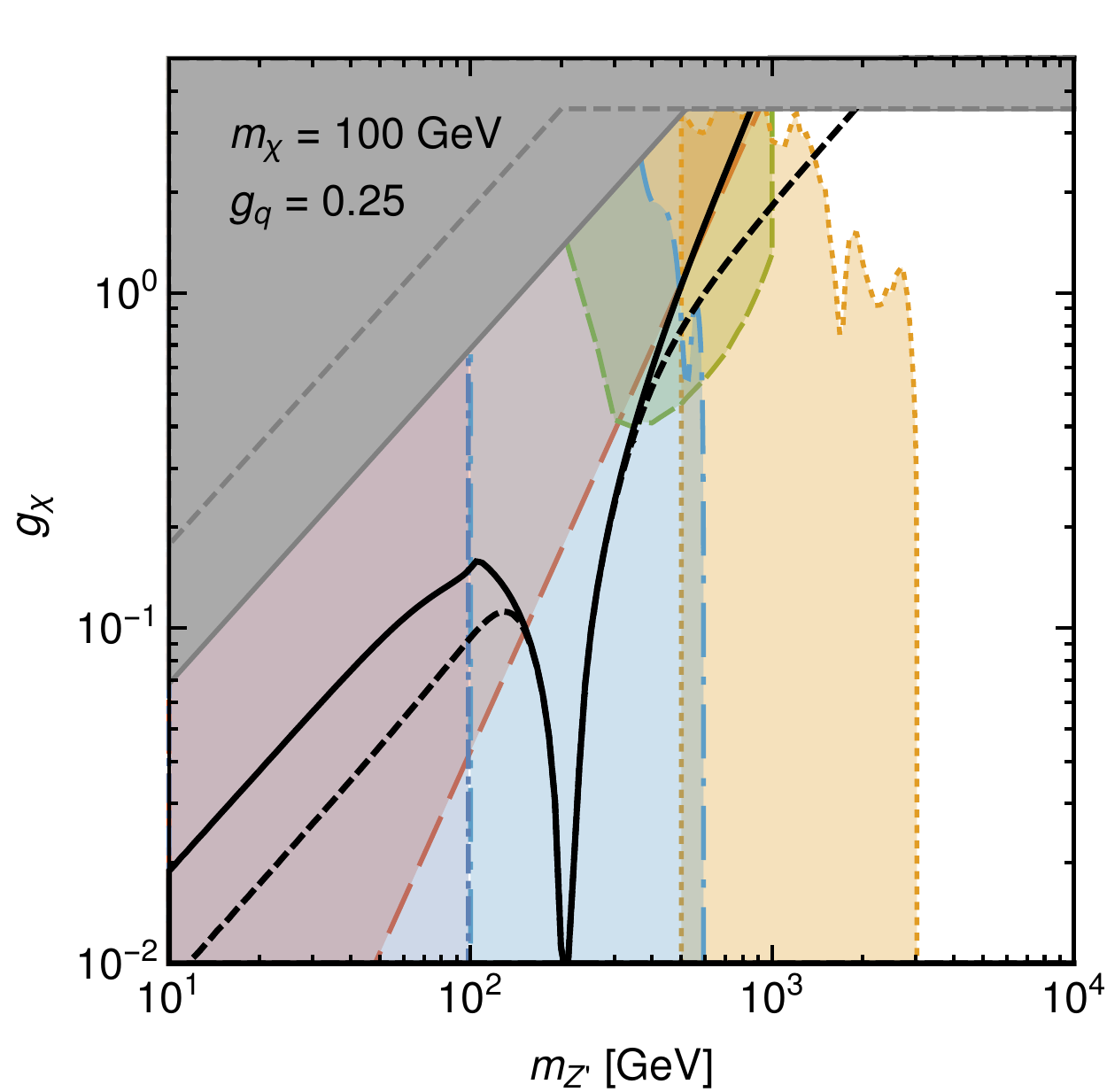}\qquad\includegraphics[height=0.28\textheight]{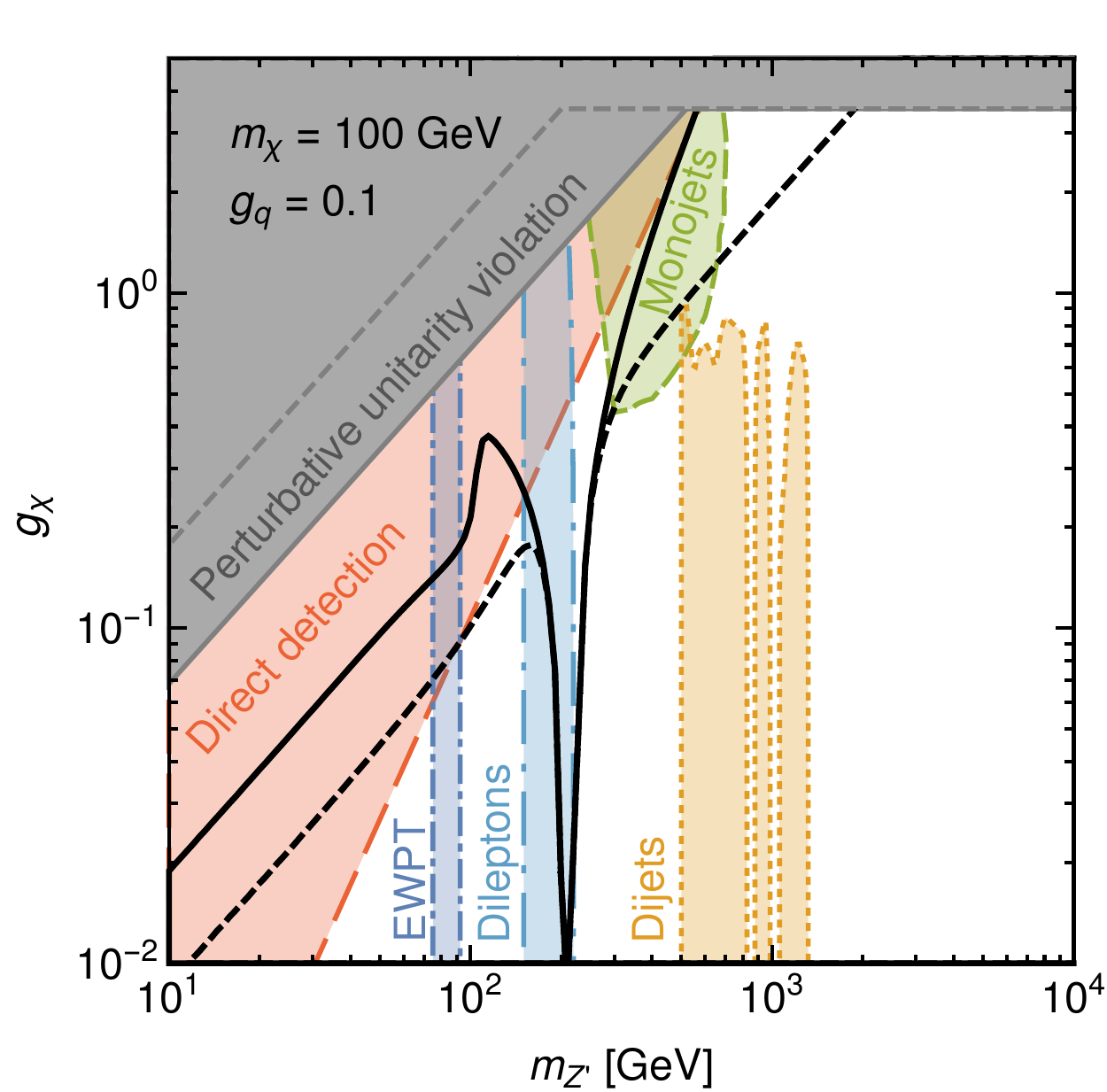}
\includegraphics[height=0.28\textheight]{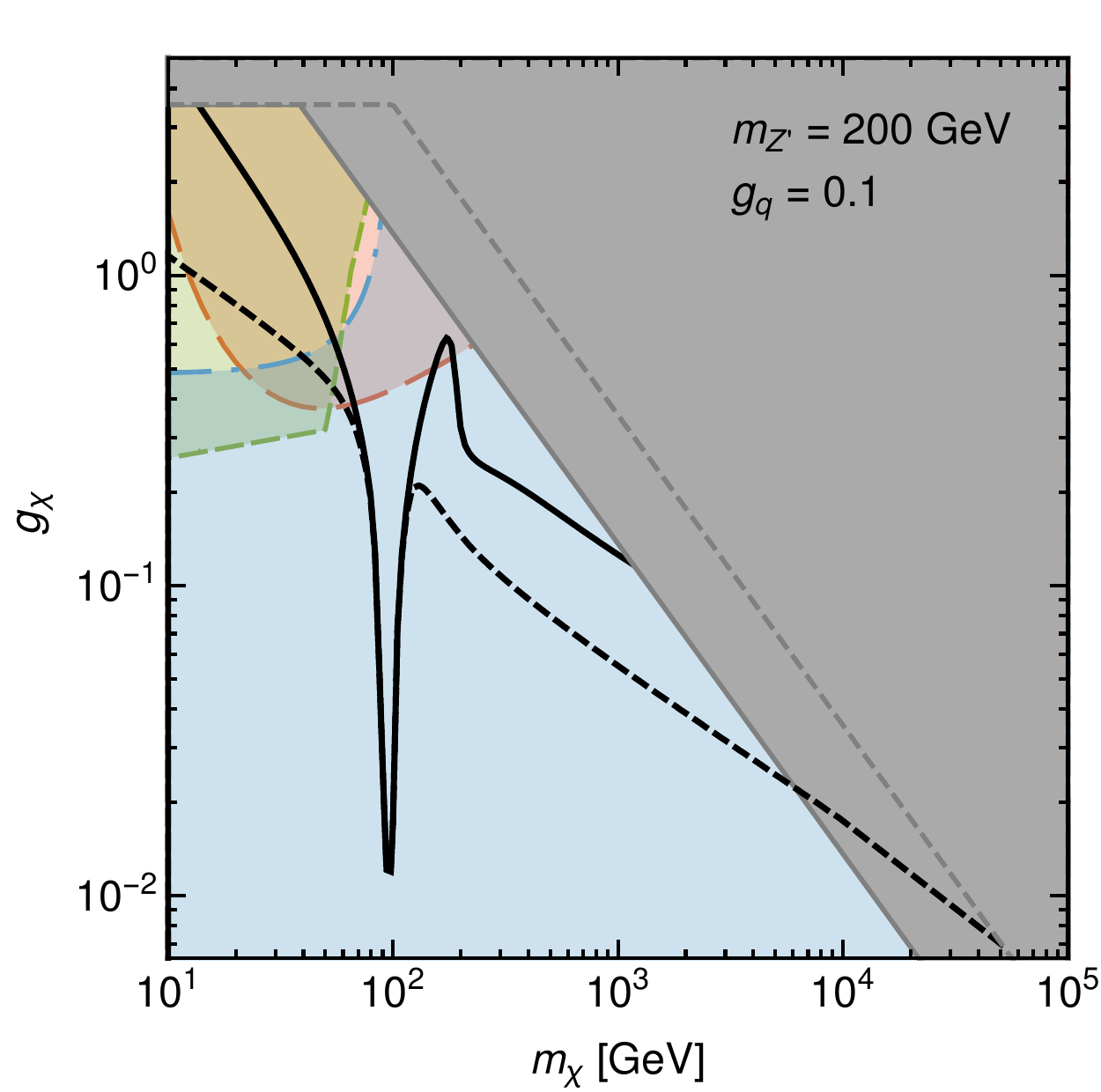}\qquad\includegraphics[height=0.28\textheight]{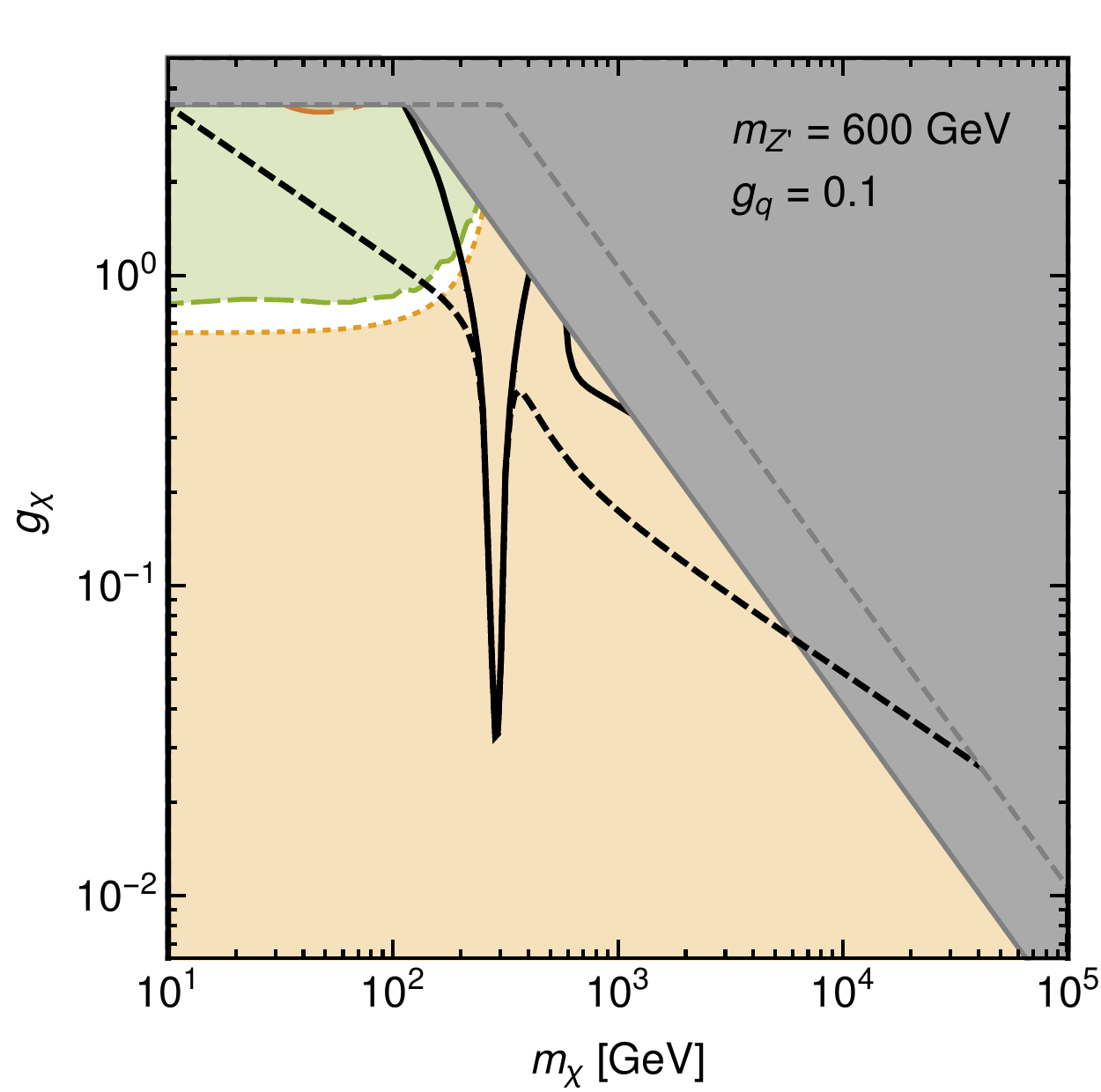}
\caption{Constraints for the case of a $Z'$ mediator in the $m_{Z'}$--$g_\chi$ plane (top) and $m_{\chi}$--$g_\chi$ plane (bottom).
The black line indicates the parameter values for which the correct relic abundance is achieved for $m_s=3 \, m_\chi$ (solid line) and $m_s=0.1 \, m_\chi$ (dashed line). 
In the grey shaded regions at least one of the couplings leads to violation of perturbative unitarity (with again $m_s=3 \, m_\chi$ (solid) and $m_s=0.1 \, m_\chi$ (dashed)). The red shaded regions (long dashed) are excluded by LUX, the green shaded regions (short dashed) are excluded by monojets, the orange shaded regions (dotted) are excluded by dijets. Dark blue regions (short dash-dotted) are excluded by EWPT and light blue regions (long dash-dotted) are excluded by dilepton searches, both assuming loop-induced kinetic mixing.}
\label{fig:gDM}
\end{figure}

All DM annihilation processes involve the single dark sector coupling
$g_\chi$. As a result, it is always possible to fix this coupling (for
given masses $m_{Z'}$, $m_\chi$ and $m_s$ and fixed coupling $g_q$) by
the requirement that the observed relic abundance is reproduced (up to
unitarity constraints). This reduces the dimensionality of the
parameter space and thus we can get a better qualitative
understanding of the relevance of the various constraints. 

Figure~\ref{fig:gDM} shows the value of $g_\chi$ obtained from the relic density requirement as a function of $m_{Z'}$ (top row) and of $m_\chi$ (bottom row) for different values of the remaining masses and couplings. For the dark Higgs boson mass we consider two representative cases: the solid black line corresponds to the case that the dark Higgs is significantly heavier than the DM particle ($m_s=3 \, m_\chi$), whereas the dashed black line corresponds to the case that it is much lighter ($m_s=0.1 \, m_\chi$). It can be seen that away from the resonance corresponding to $m_\chi \approx m_{Z'} / 2$ smaller values of $g_\chi$ are required in case of a light dark Higgs. This is because DM can very effectively annihilate into final states with dark terminators, e.g.\ $\chi \chi \rightarrow s s$, as will be discussed in more detail below.

We also show in figure~\ref{fig:gDM} the constraints discussed above both in the $m_{Z'}$--$g_\chi$ and in the $m_{\chi}$--$g_\chi$ parameter plane. Here and in the following, all constraints are shown at 95\% confidence level, except for direct and indirect detection constraints, which are shown at 90\% confidence level as customary. As we consider negligible mixing of the dark Higgs with the SM Higgs, these constraints are independent of the value of $m_s$, with the only exception of the unitarity constraint, which is indicated by a solid grey line for $m_s=3 \, m_\chi$ and by a dashed grey line for $m_s=0.1 \, m_\chi$. For $m_s = 3 \, m_\chi$ and $\theta \approx 0$ the unitarity bound from the scalar potential given in eq.~(\ref{eq:unit_HiggsPot}) is always stronger than the direct bound on the Yukawa coupling in eq.~(\ref{eq:unit_coupling}) and leads to the strongest constraint for $m_{Z'} < 500\:\text{GeV}$ (top row) and $m_\chi > 40\:\text{GeV}$ (bottom row). For larger values of $m_{Z'}$ or smaller values of $m_\chi$ this bound becomes weaker, so that the unitarity bound results simply from the requirement $g_\chi < \sqrt{4\pi}$. For $m_s = 0.1 \, m_\chi$ the requirement of perturbative unitarity of the scalar potential is less constraining and the relevant bound for small $m_{Z'}$ and large $m_\chi$ comes from the condition $y_\chi < \sqrt{8\pi}$.

In the lower panels of figure~\ref{fig:gDM} the required thermal coupling $g_\chi$, as well as the unitarity bound on $g_\chi$, are seen to decrease with increasing $m_\chi$ in a seemingly counter-intuitive manner. This behaviour can be understood once we consider the dominant annihilation channels. For $m_{Z^\prime} \ll m_\chi$ the relic abundance is either set by $\chi \chi \rightarrow Z^\prime_L Z^\prime_L$ or by $\chi \chi \rightarrow s Z^\prime_L $ and both these processes are controlled  by the Yukawa coupling  $y_\chi \propto g_\chi \frac{m_\chi}{m_{Z^\prime}}$ which grows with $m_\chi$ just as expected. We note that for $m_s \ll m_\chi$ the DM mass can be as large as $m_\chi \sim \unit[40\text{--}50]{TeV}$ before the requirement of perturbative unitarity becomes incompatible with thermal freeze-out~\cite{Griest:1989wd}. 

\begin{figure}[t]
\centering
\includegraphics[height=0.28\textheight]{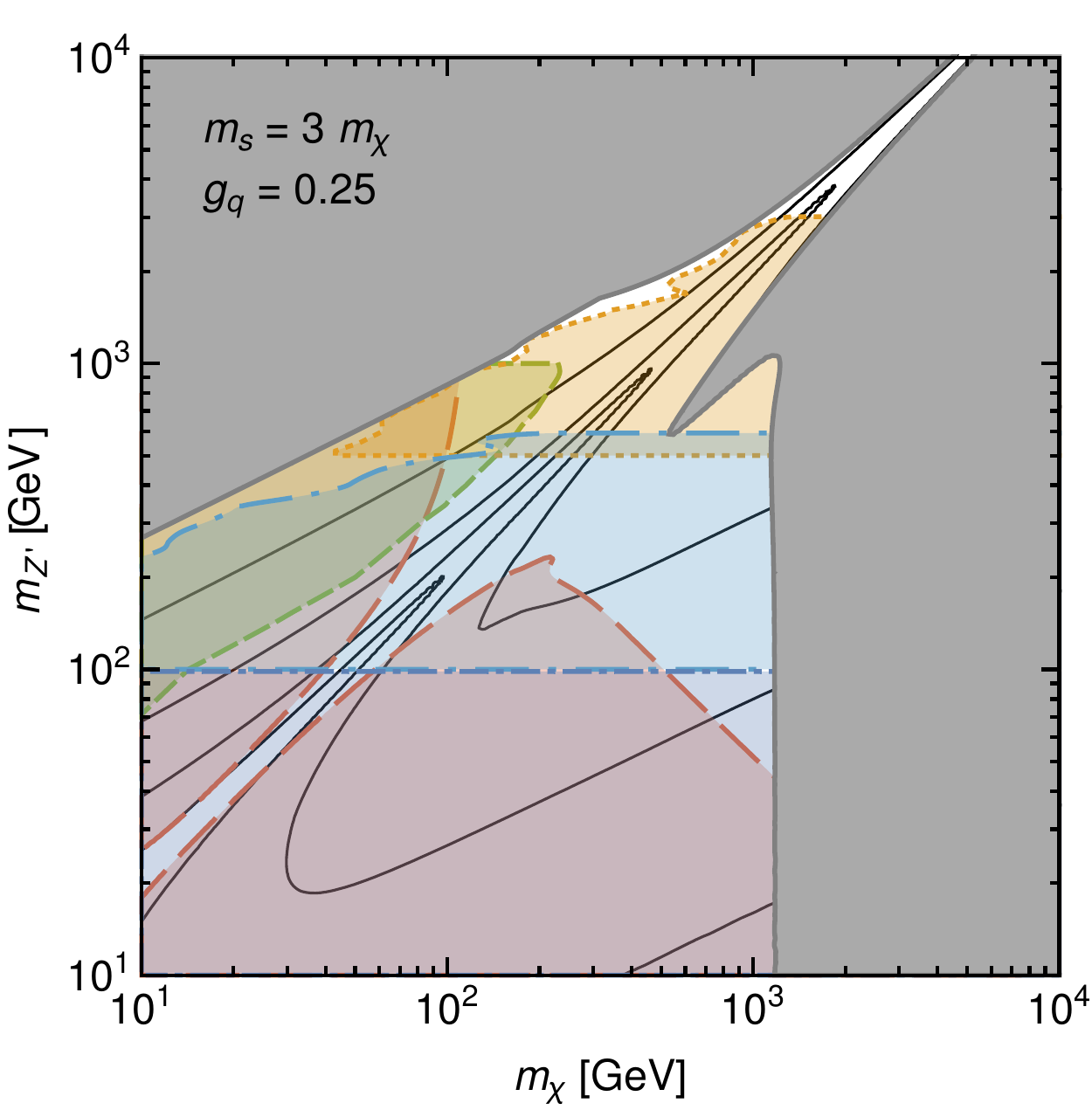}\qquad\includegraphics[height=0.28\textheight]{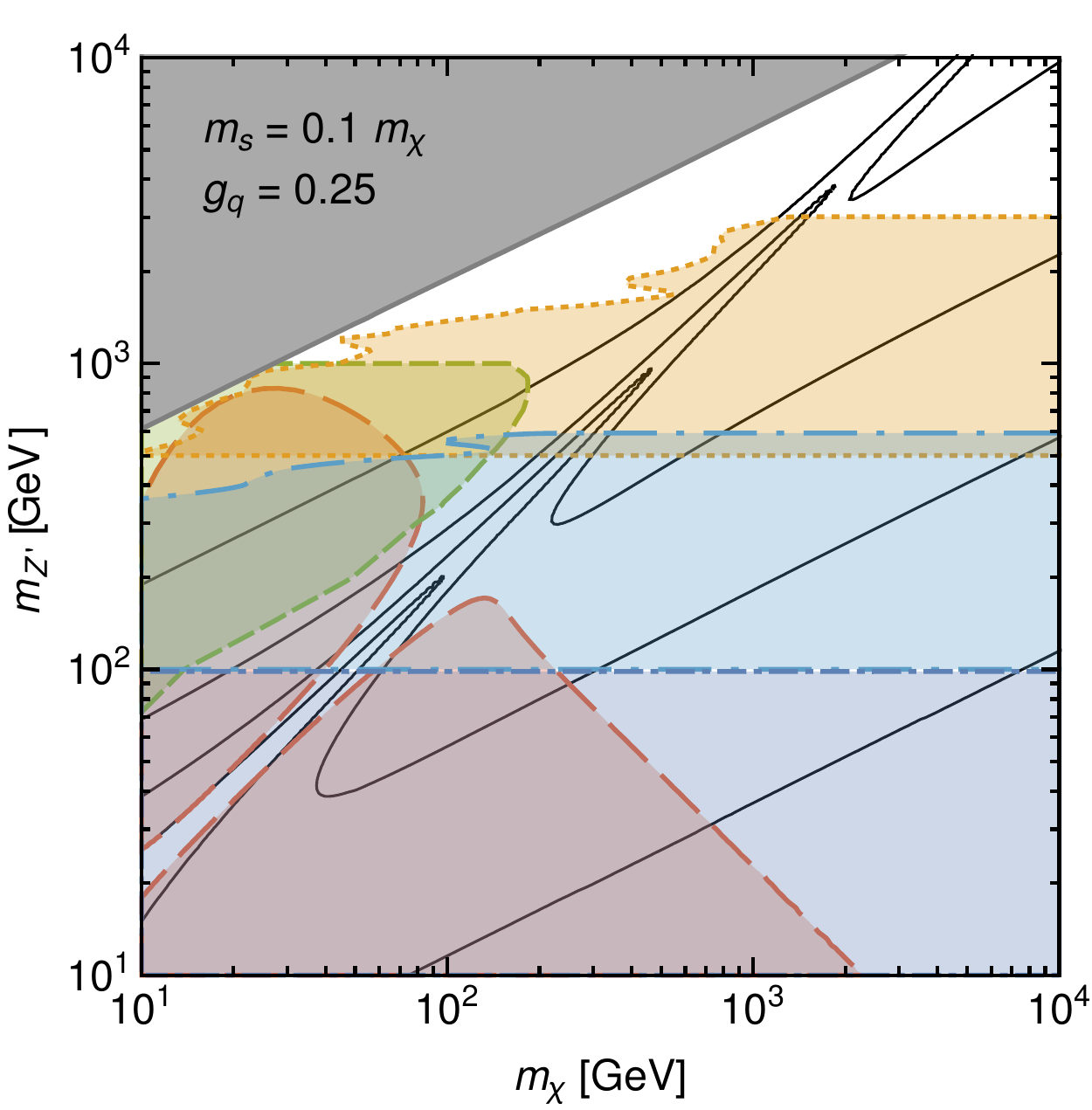}
\includegraphics[height=0.28\textheight]{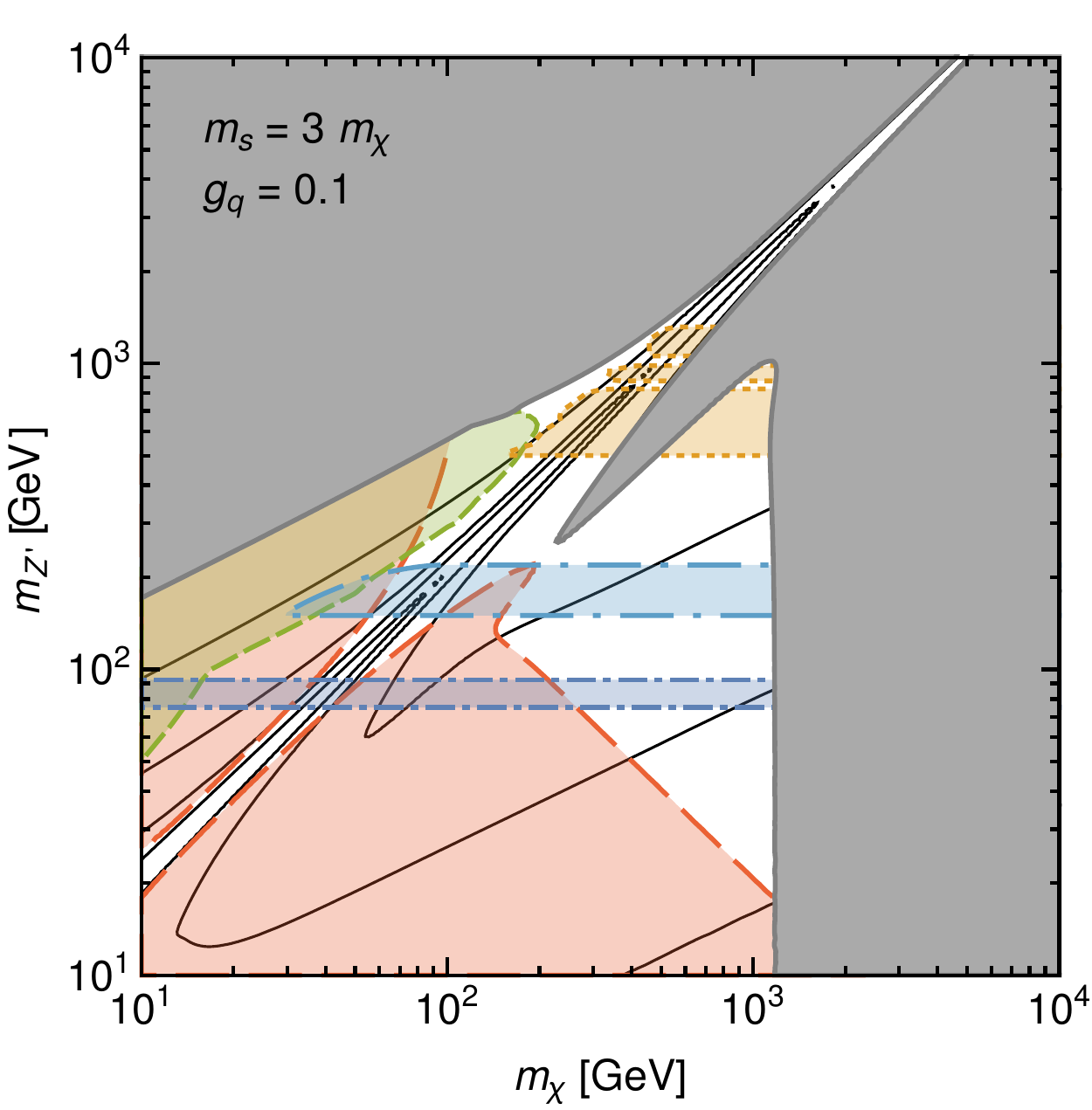}\qquad\includegraphics[height=0.28\textheight]{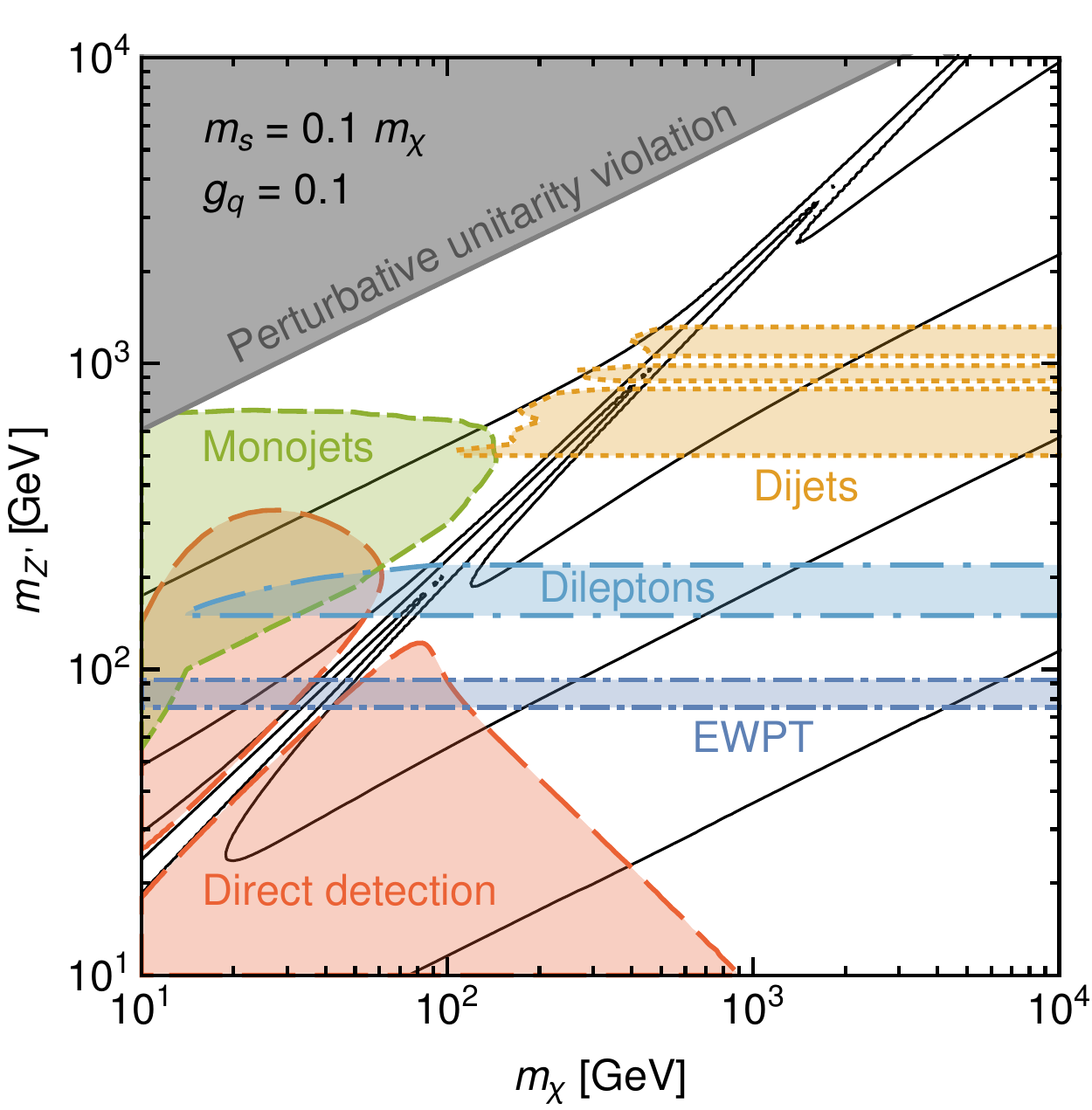}
\caption{Constraints for the case of a $Z'$ mediator, with the DM--$Z'$ coupling determined by the requirement to reproduce the observed relic abundance. 
Lines (black solid) represent constant $g_\chi$, showing the values $g_\chi = 0.01,\,0.05,\,0.2,\,1$ (bottom to top). 
In the grey shaded regions (solid line) at least one of the couplings leads to violation of perturbative unitarity. The red shaded regions (long dashed) are excluded by LUX, the green shaded regions (short dashed) are excluded by monojets, the orange shaded regions (dotted) are excluded by dijets. Dark blue regions (short dash-dotted) are excluded by EWPT and light blue regions (long dash-dotted) are excluded by dilepton searches, both assuming loop-induced kinetic mixing.}
\label{fig:spin1-relic}
\end{figure}

We now fix the dark sector coupling via the relic density requirement as shown in figure~\ref{fig:gDM}. We can then study the phenomenology of the model in the $m_\chi$--$m_{Z'}$ parameter plane. The resulting constraints are shown in figure~\ref{fig:spin1-relic}. The two rows correspond to different values of $g_q$, the two columns to different ratios of $m_s / m_\chi$. The constraints can be understood qualitatively by referring to figure~\ref{fig:gDM}: as expected, the weakest constraints are found for the resonance region $m_{Z'} \sim 2 \, m_\chi$, where $g_\chi$ can be very small. For increasing $m_{Z'}$ larger values of $g_\chi$ are required to reproduce the relic abundance up to the point where $g_\chi$ is in conflict with the requirement of perturbative unitarity. For small $Z'$ masses, on the other hand, constraints from direct detection and EWPT are very strong. Dijet and dilepton constraints are most relevant for $\unit[100]{GeV} \leq m_{Z'} \leq \unit[3]{TeV}$. Finally, monojet constraints dominate for $m_\chi < m_{Z'} / 2$, where DM production at the LHC receives a resonant enhancement.

In the left column, we have taken $m_s = 3 \, m_\chi$. For such a large mass (and the assumed small mixing), the dark Higgs remains completely irrelevant for the phenomenology of the model. Its presence becomes nevertheless visible in the fact that large DM masses are excluded by the requirement of perturbative unitarity of the scalar potential. In other words, if we insist on $m_s \gg m_\chi$, the dark Higgs self-coupling necessarily becomes non-perturbative for DM masses at the TeV scale. This constraint can be significantly relaxed if we instead consider the case $m_s = 0.1 \, m_\chi$, as shown in the right column of figure~\ref{fig:spin1-relic}. In this case, the bound on $y_\chi$ from perturbative unitarity only requires $m_\chi \lesssim \unit[40\text{--}50]{TeV}$ (see figure~\ref{fig:gDM}).

Taking the Higgs to be light compared to the DM particle leads to additional changes, which are visible even for DM masses well below the unitarity bound. The reason is that, once the dark Higgs mass becomes comparable to or lighter than the DM mass, DM particles can annihilate into it. These processes can happen even for very small Higgs mixing, as long as the dark Higgs ultimately decays into SM particles. The two processes of interest are $\chi \chi \rightarrow s s$ and $\chi \chi \rightarrow s Z'$. The corresponding annihilation cross sections are given in appendix~\ref{app:annihilation}. Prospects for observing these processes with indirect detection experiments are discussed in section~\ref{sec:indirect}.

It is important to note that the dominant contribution to the annihilation into two dark Higgs bosons results from the diagram with a $t$-channel DM particle, so that the resulting annihilation cross section is proportional to $y_\chi^4 / m_\chi^2 \propto g_\chi^4 \, m_\chi^2 / m_{Z'}^4$. Consequently, the requirement that $g_\chi$ remains perturbative still places an upper bound on $m_{Z'}$ even if the dark Higgs is the dominant annihilation channel.

\section{Spin-0 mediation}
\label{sec:spin-0}

Let us now consider the case $g_q \ll 1$, such that the dark Higgs is the only state that can mediate sizeable interactions between DM and SM particles. If the $Z'$ couples only very weakly to SM states, it also makes sense to assume that kinetic mixing is negligible. The independent parameters of importance for this case are therefore the three masses $m_{Z'}$, $m_\chi$ and $m_s$, the coupling between DM and the dark Higgs $y_\chi$, and the Higgs mixing parameter $\sin \theta$. Note that in contrast to the previous section, we now take $y_\chi$ as the independent dark sector coupling, which is related to $g_\chi$ via eq.~\eqref{eq:dark_couplings}. We now review the relevant constraints for this case.

\subsection{Higgs mixing and signal strength}

The magnitude of the mixing angle $\theta$ between the SM Higgs and the dark Higgs is constrained by the observed Higgs signal strength $\mu$, which is required to satisfy $\mu > 0.89$ at 95\% confidence level~\cite{AtlasCMSHiggsSignalStrength}. In our model, there are three separate effects that potentially lead to a reduction of the signal strength:
\begin{itemize}
 \item Mixing between the two Higgs bosons reduces the production cross section for the SM Higgs.
 \item For $m_\chi < m_h / 2$, the SM Higgs can have a sizeable invisible branching fraction. In principle, such invisible decays can be tested directly by looking for invisibly decaying Higgs bosons produced either in vector boson fusion or in association with a $Z$ boson. The resulting constraints are however generally weaker than the ones obtained from the Higgs signal strength.
 \item For $m_s < m_h / 2$ or $m_{Z'} < m_h / 2$, the SM Higgs can decay into two singlets or two $Z'$ bosons, which depletes the cross section in the remaining channels. Such decays would give rise to decays of the form $h \rightarrow 4f$ (with $f$ being SM quarks or leptons), which should in principle be observable. A detailed study of such decay processes is beyond the scope of this work, so we only consider the impact on the Higgs signal strength here.
\end{itemize}

The partial decay width for invisible decays is given by (assuming $m_\chi < m_h / 2$)
\begin{equation}
\Gamma_\text{inv} = \frac{y_\chi^2 \, m_h \, \sin^2 \theta}{32 \pi} \left(1 - \frac{4 \, m_\chi^2}{m_h^2}\right)^{3/2} \; ,
\end{equation}
while the partial decay width for decays into a pair of dark Higgs bosons with $m_s < m_h / 2$ is
\begin{equation}
\Gamma_{ss} = \frac{(m_h^2 + 2 \, m_s^2)^2 \, \sin^2 2 \theta}{128 \pi \, m_h}  \left(1 - \frac{4 \, m_s^2}{m_h^2}\right)^{1/2} \left(\frac{1}{w} \cos \theta + \frac{1}{v} \sin \theta \right)^2 \; .
\end{equation}
Note that $\Gamma_{ss}$ is sensitive to the sign of $\theta$ due to an interference between contributions proportional to $1/w$ and $1/v$. A negative sign of $\theta$ leads to a partial cancellation between these contributions and thus suppresses the resulting bounds. The partial decay width for decays into a pair of $Z'$ bosons with $m_{Z'} < m_h/2$ is given by
\begin{equation}
\Gamma_{Z'Z'} = \frac{g_\chi^2 \, \sin^2 \theta}{8\pi \, m_h \, m_{Z'}^2 }  \left(m_h^4 - 4 \, m_h^2 \, m_{Z'}^2 + 12 \, m_{Z'}^4\right) \sqrt{1 - \frac{4\,m_{Z'}^2}{m_h^2}} \; .
\end{equation}
Finally, if $\Gamma_0$ denotes the total width of the SM Higgs for $\cos \theta = 1$, the partial width into all SM final states will be given by
\begin{equation}
 \Gamma_\text{SM} = \cos^2 \theta \, \Gamma_0 \; .
\end{equation}
Furthermore, for $\cos \theta < 1$, the Higgs production cross section will be reduced proportional to $\cos^2 \theta$.

Putting all of these contributions together, we find that the predicted Higgs signal strength in our model is given by
\begin{equation}
 \mu = \cos^2 \theta \times \text{BR}(h \rightarrow \text{SM}) = \frac{\cos^2 \theta \, \Gamma_\text{SM}}{\Gamma_\text{SM} + \Gamma_{ss} + \Gamma_{Z'Z'} + \Gamma_\text{inv}} \; .
\end{equation}
Even for $\Gamma_{ss} = \Gamma_{Z'Z'} = \Gamma_\text{inv} = 0$, the observed lower bound on $\mu$~\cite{AtlasCMSHiggsSignalStrength} implies $\theta < 0.34$. If additional decay channels are kinematically allowed, even stronger bounds will be obtained.

The mixing angle $\theta$ is also constrained by resonance searches
for additional Higgs bosons. These bounds depend on the mass of the
dark Higgs and can be somewhat more stringent than the mass-independent bound of $\theta < 0.34$ for $\unit[200]{GeV} \lesssim m_s \lesssim \unit[500]{GeV}$~\cite{Pelliccioni:2015hva,Aad:2015kna}.  However, for the
values of $\theta$ considered in the remainder of this paper these
bounds play no role. A similar comment applies to bounds on $\theta$
from EWPT, see e.g.~\cite{Falkowski:2015iwa}.

\subsection{Direct detection}

The scalar mediator induces unsuppressed spin-independent DM--nucleus interactions, which are strongly constrained by direct detection experiments. Hence these experiments provide a stringent bound on the Higgs mixing angle $\theta$. The spin-independent DM--nucleon scattering cross section is given by
\begin{equation}
 \sigma^\text{SI} = \frac{\mu_{\chi N}^2 \, m_N^2 \, f_N^2 \, y_\chi^2 \, \sin^2 \theta \, \cos^2 \theta}{2 \pi \, v^2}  \left(\frac{1}{m_h^2} - \frac{1}{m_s^2}\right)^2~,
\end{equation}
where $\mu_{\chi N} = m_N \, m_\chi / (m_N + m_\chi)$ is the reduced mass of the DM--nucleon system and $f_N = 0.3$ is the effective coupling of DM to nucleons~\cite{Boveia:2016mrp}. Recent results from the LUX experiment~\cite{Akerib:2015rjg} place strong bounds on $\sigma^\text{SI}$, which lead to constraints on $\theta$ that can be comparable or even more constraining than the ones from the Higgs signal strength. We implement this constraint by comparing our predicted value of $\sigma^\text{SI}$ to the 90\%~CL upper bound from LUX~\cite{Akerib:2015rjg}.

\subsection{Results}

Let us first consider the case that the $Z'$ is significantly heavier than the DM particle. Since we consider very small quark couplings, the $Z'$ then plays no role for the phenomenology of the model. In the presence of Higgs mixing, however, both the dark Higgs and the SM Higgs can mediate DM annihilation processes (see appendix~\ref{app:annihilation}). Moreover, for $m_\chi > m_s$, the DM particles can also annihilate into pairs of dark Higgs bosons, which subsequently decay into other SM particles (see section~\ref{sec:spin-1}). Crucially, all these annihilation channels depend on $y_\chi$, so it is always possible (for given masses and mixing angle) to fix the DM Yukawa coupling in such a way that the observed DM relic abundance is reproduced (up to perturbativity). The required value of $y_\chi$ can then be used to compare to the various constraints discussed above.

\begin{figure}
\centering
\includegraphics[height=0.26\textheight]{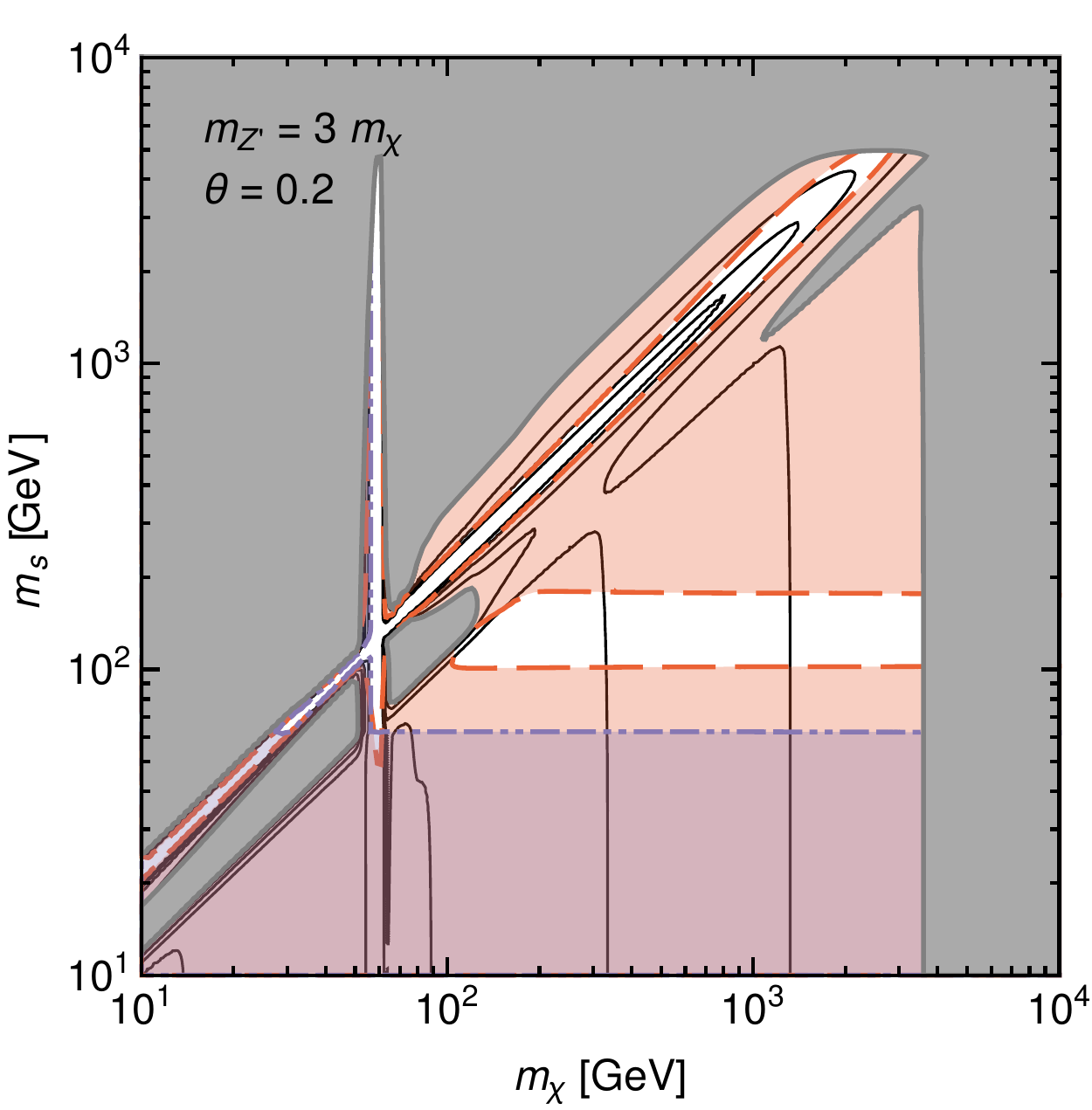}\qquad\includegraphics[height=0.26\textheight]{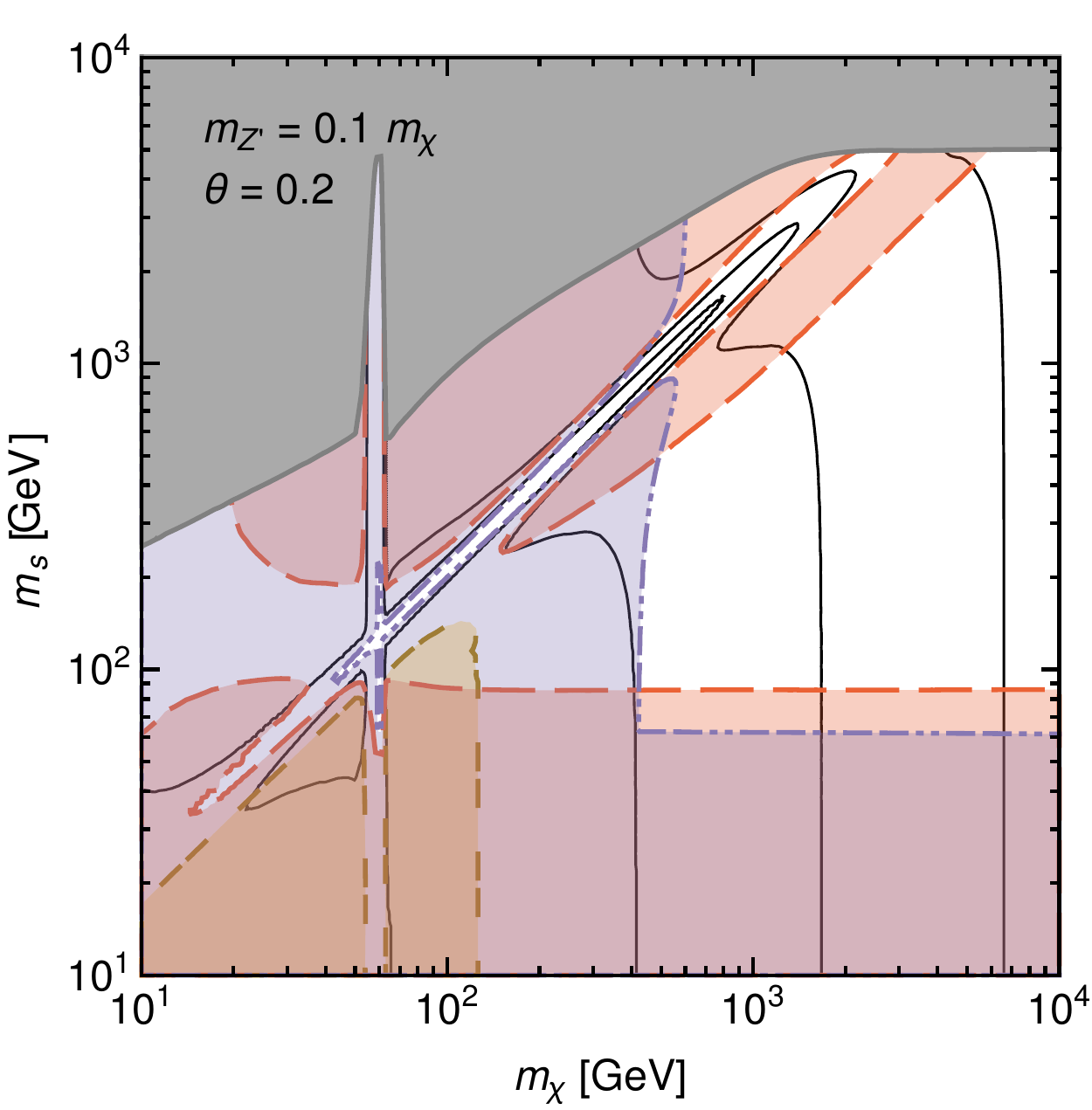}
\includegraphics[height=0.26\textheight]{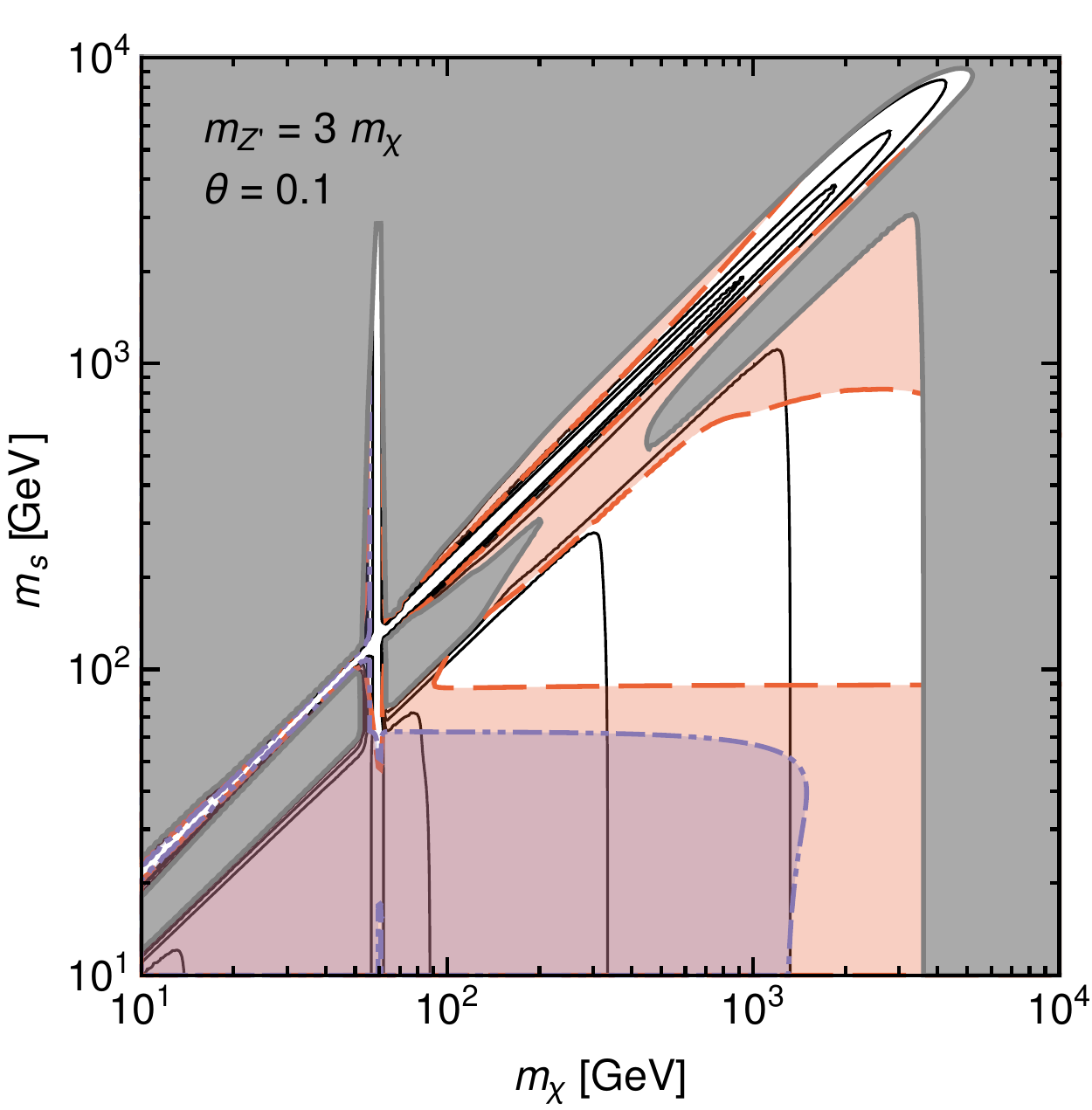}\qquad\includegraphics[height=0.26\textheight]{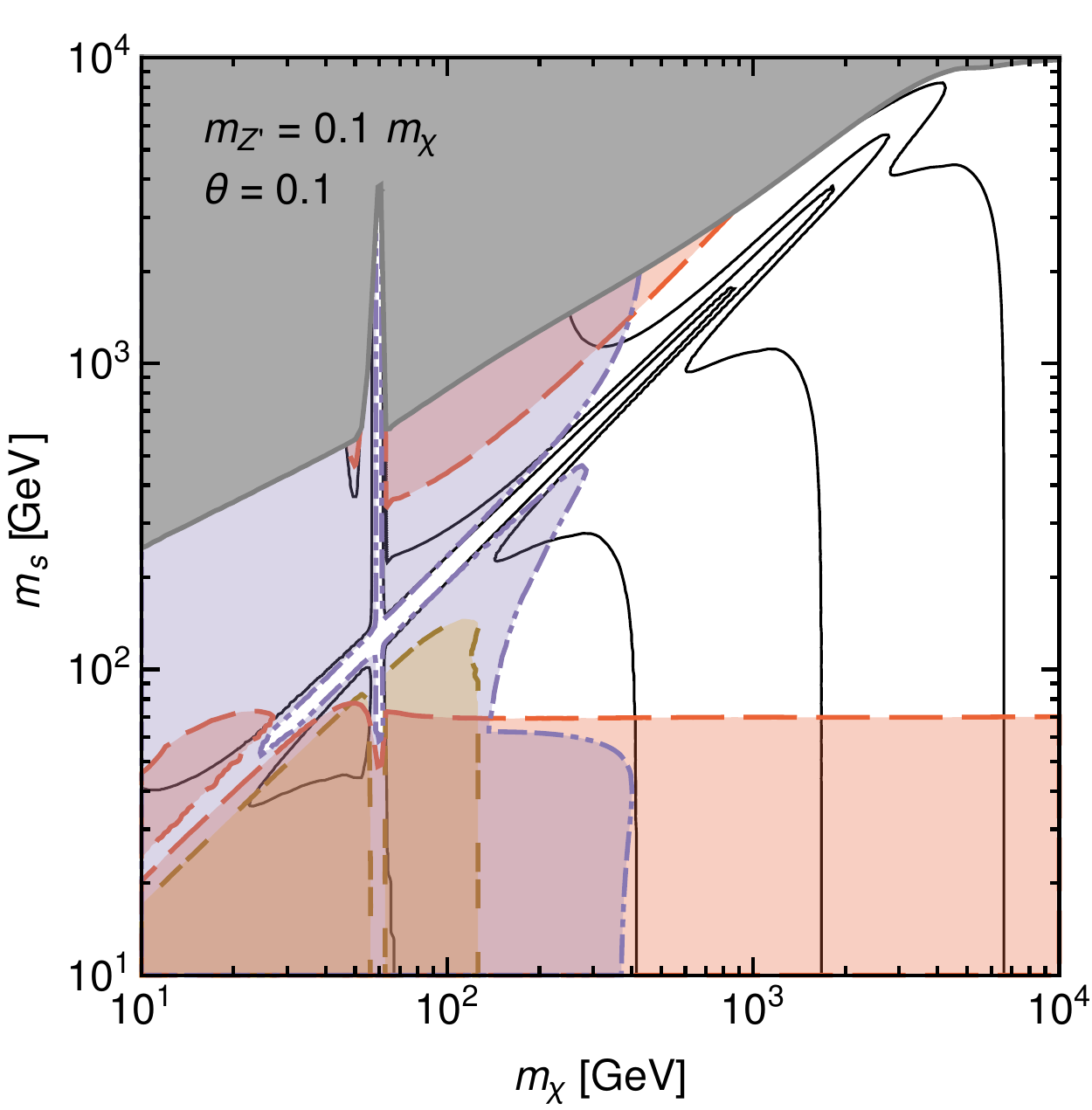}
\includegraphics[height=0.26\textheight]{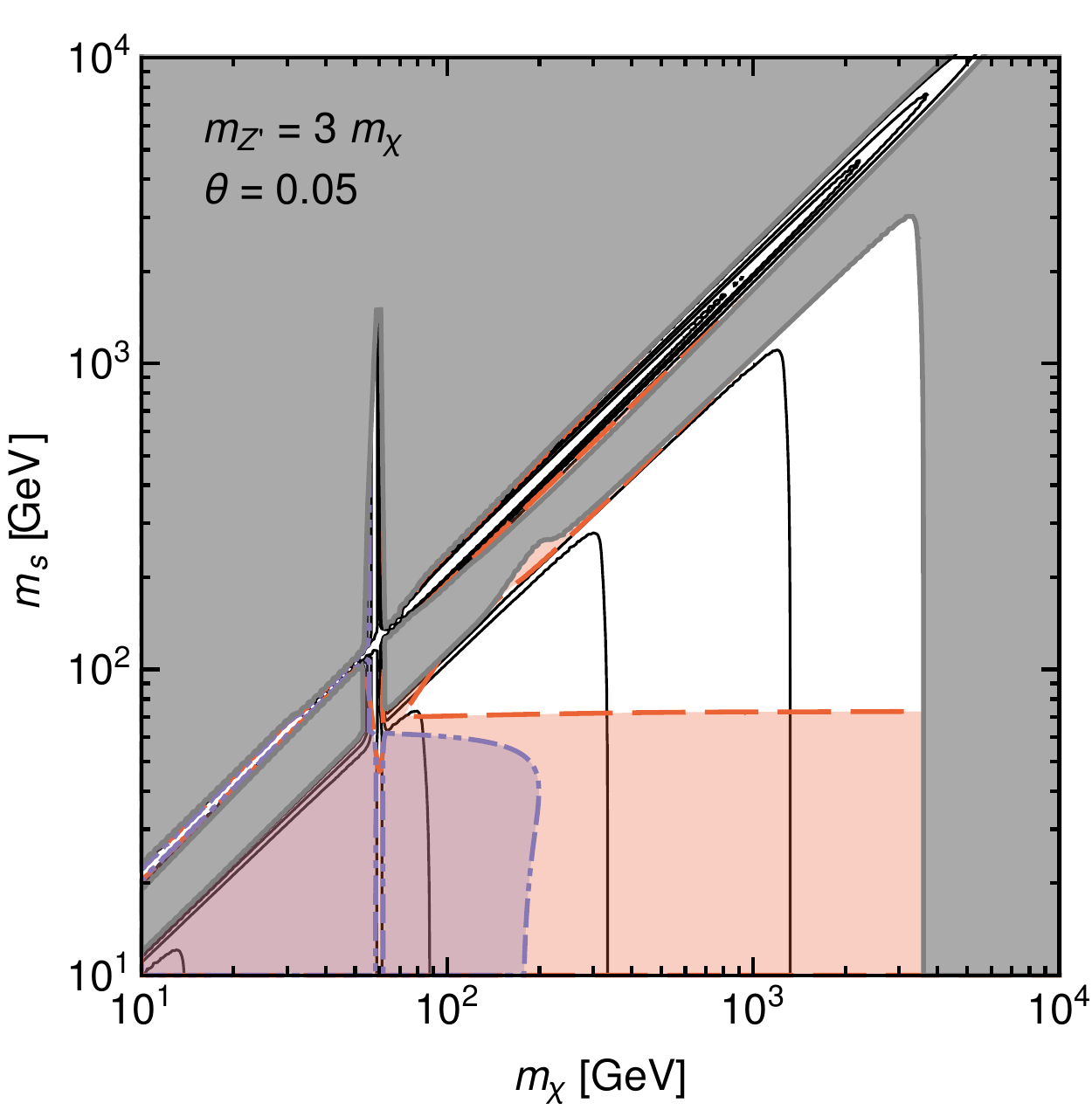}\qquad\includegraphics[height=0.26\textheight]{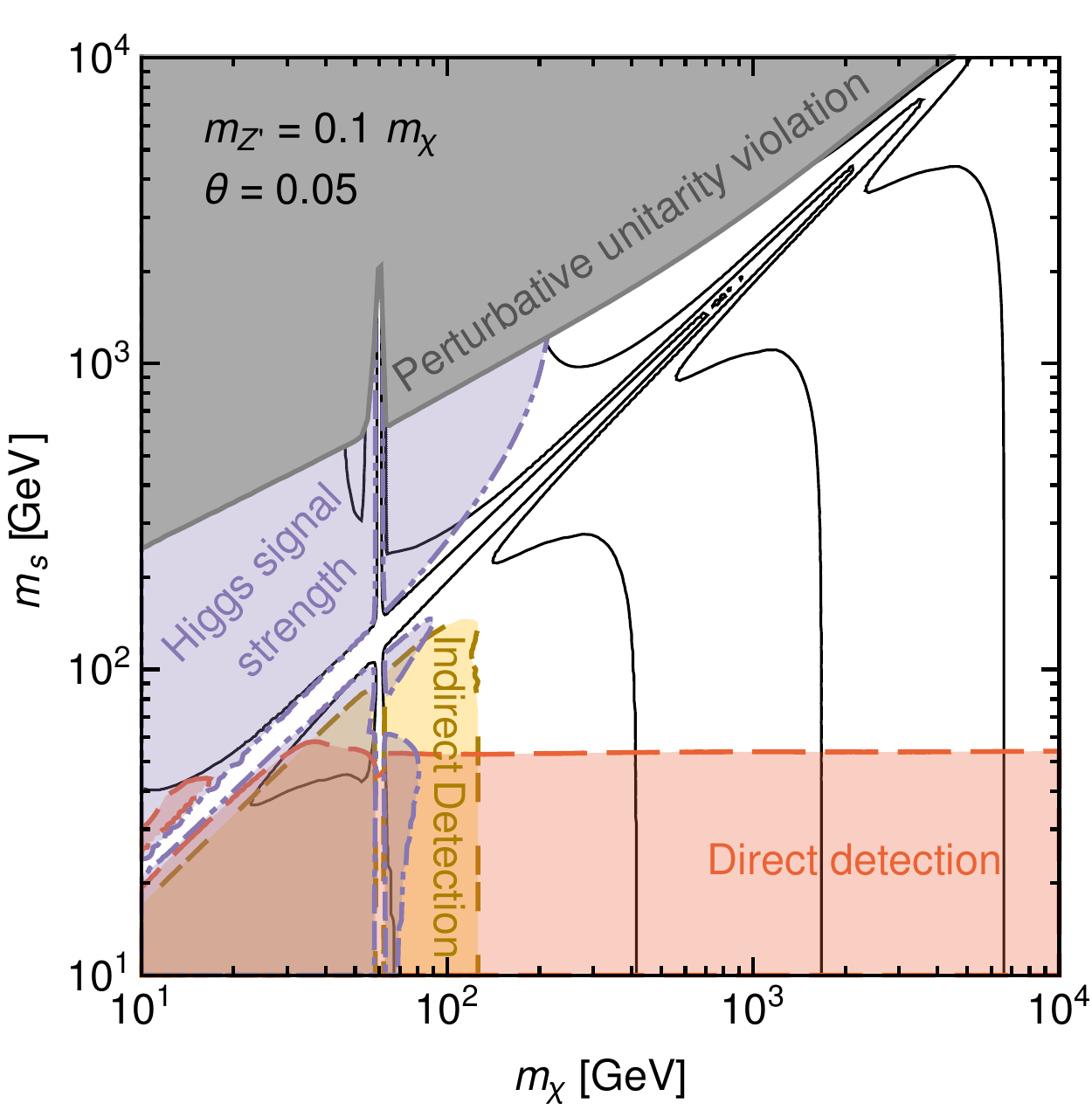}
\caption{Constraints for the case of a dark Higgs mediator, with the DM Yukawa coupling determined by the requirement to reproduce the observed relic abundance. 
Lines (black solid) represent constant $y_\chi$, showing the values $y_\chi = 0.2,\,0.5,\,1,\,2$ (bottom to top). 
In the grey shaded regions (solid line) at least one of the couplings leads to violation of perturbative unitarity. The red shaded regions (long dashed) are excluded by LUX, the purple shaded regions (double dash-dotted) by the observed Higgs signal strength, and the yellow regions (double dashed) by indirect detection.}
\label{fig:spin0-relic}
\end{figure}

The results of such an analysis are shown in the left column of figure~\ref{fig:spin0-relic} as a function of $m_\chi$ and $m_s$ for different values of $\theta$ (and $m_{Z'}$ large enough that the precise value is irrelevant). The following features can be identified: For $m_\chi \sim m_h / 2$, annihilation via the SM Higgs receives a resonant enhancement, whereas for $m_\chi \sim m_s / 2$ annihilation via the dark Higgs is resonantly enhanced. For $m_\chi > m_s$ direct annihilation into two dark Higgs bosons dominates. In this parameter region, the relic density is independent of $\sin \theta$ and therefore the constraints from direct detection and the Higgs signal strength can be weakened arbitrarily by making $\sin \theta$ very small. Nevertheless, for $m_{Z'} = 3 m_\chi$ the required value of $g_\chi$ may violate perturbative unitarity, limiting the allowed parameter region to $m_\chi < 3.5\:\text{TeV}$. For small values of $m_\chi$, on the other hand, large regions of parameter space violate the perturbativity condition on $y_\chi$. The requirement of perturbative unitarity in the scalar potential only becomes relevant if both $m_s$ and $m_\chi$ are large.

Let us now turn to the case where the $Z'$ is significantly lighter than the DM particle, so that DM can directly annihilate into a pair of $Z'$ bosons. Since we assume that the quark coupling of the $Z'$ is very small, these $Z'$ bosons can have a very long lifetime, but it is easily possible to ensure that they decay before Big Bang nucleosynthesis. As shown in appendix~\ref{app:annihilation}, the relic density becomes independent of $m_{Z'}$ in the limit $m_{Z'} \rightarrow 0$, so we simply fix $m_{Z'} = 0.1 \, m_\chi$ for convenience. In this case, the bound $g_\chi < \sqrt{4\pi}$ can easily be satisfied up to very large DM masses, so only the bound on $y_\chi$ and the requirement of a perturbative scalar potential remain visible.

In the case that $2 \, m_\chi > m_{Z'} + m_s$, one also needs to consider to the process $\chi \chi \rightarrow s Z'$. In contrast to all other processes considered so far, this annihilation process can proceed via $s$-wave and is enhanced for $m_\chi \gg m_{Z'}$  due to the DM Yukawa coupling becoming large. In analogy, also the process $\chi \chi \rightarrow h Z'$ can become large in the presence of Higgs mixing. It is then not even necessary for the dark Higgs to be light. 

As shown in the right column of figure~\ref{fig:spin0-relic}, having a light spin-1 terminator significantly relaxes the constraints. In particular, since smaller values of $y_\chi$ are sufficient to reproduce the observed relic abundance, the requirement of perturbative unitarity is less constraining. Nevertheless, it is still not possible to raise $m_s$ arbitrarily above $m_\chi$. If the channel $\chi\chi\to s Z'$ is open, indirect detection experiments give relevant constraints for $m_\chi \lesssim \unit[100]{GeV}$, which are independent of both $\theta$ and $g_q$ (see section~\ref{sec:indirect}). The excluded parameter region is shown in yellow in figure~\ref{fig:spin0-relic}.

\begin{figure}
\centering
\includegraphics[height=0.26\textheight]{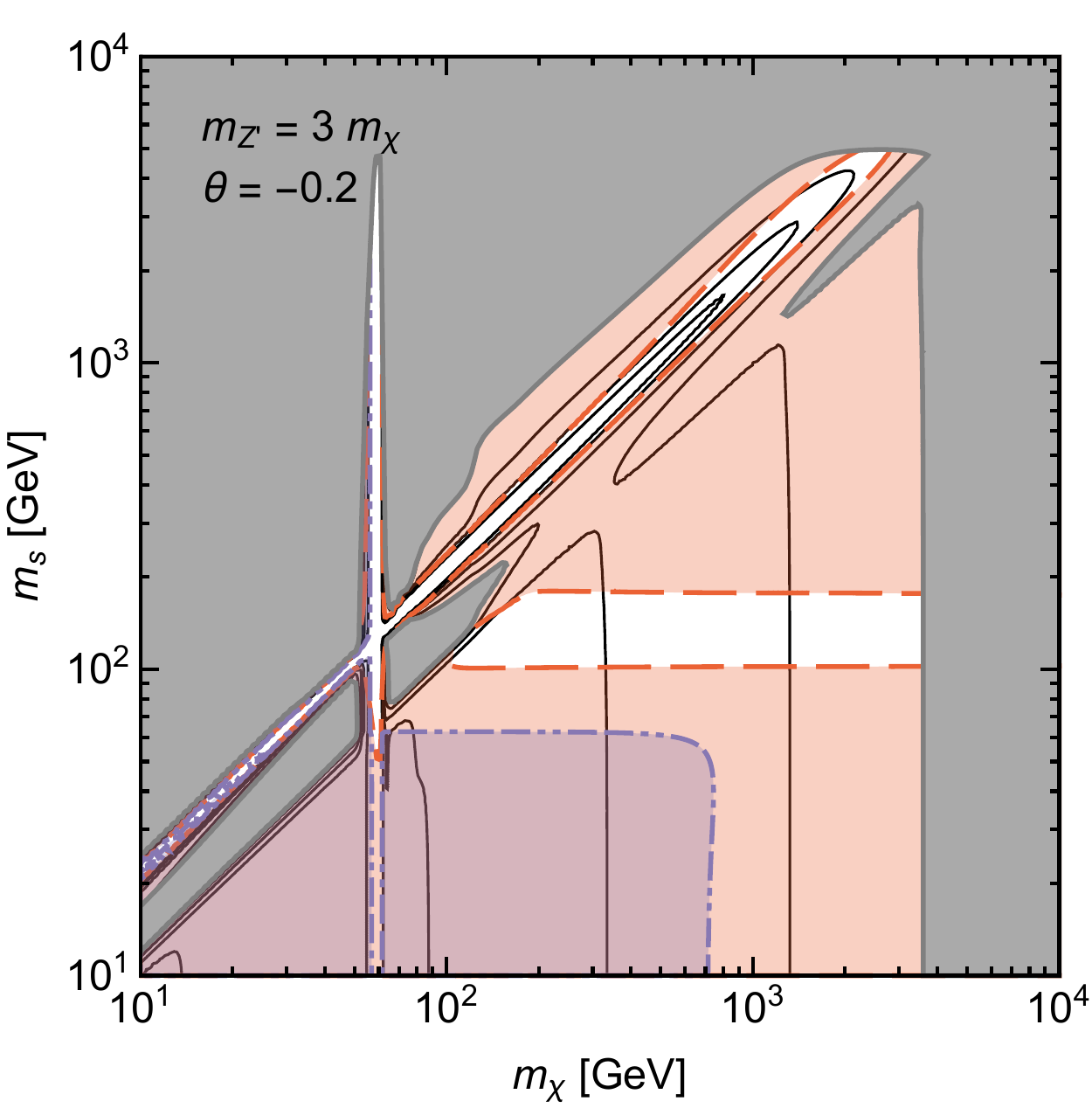}\qquad\includegraphics[height=0.26\textheight]{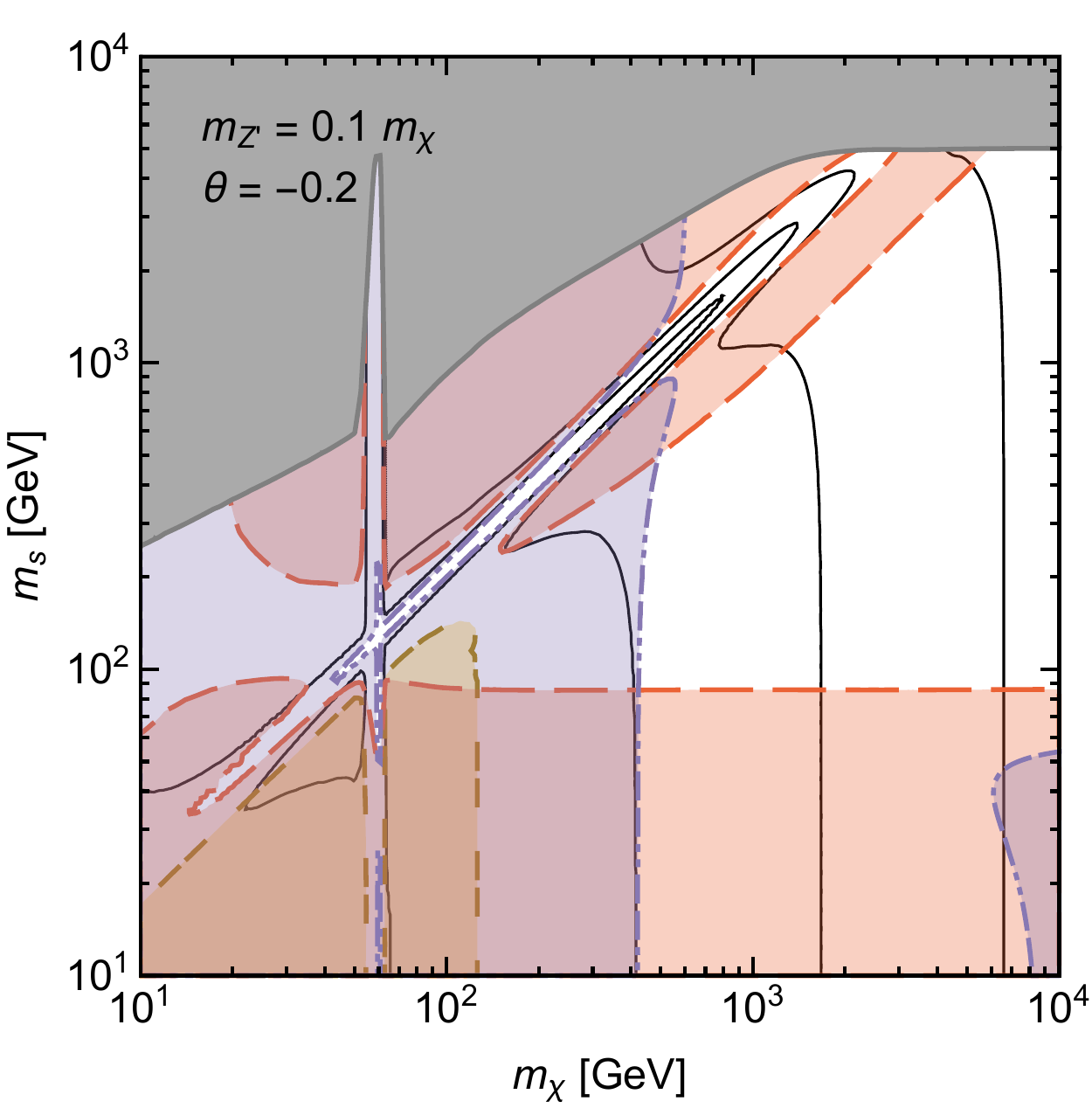}
\caption{Same as the first row of figure~\ref{fig:spin0-relic} but with opposite sign of the mixing angle $\theta$. The only sizeable difference is the strength of the bound from $h \rightarrow s s$, which is relaxed for the negative sign.}
\label{fig:spin0-relic2}
\end{figure}

To conclude this section, we return to the choice of sign for $\theta$. As discussed above, this sign is relevant for the trilinear vertices coupling one SM Higgs to two dark Higgs and vice versa. Considering $\theta < 0$ thus modifies the prediction for the decay $h \rightarrow s s$ as well as for the annihilation process $\chi \chi \rightarrow h h$ via an $s$-channel dark Higgs. The latter process, however, is typically subdominant compared to $\chi \chi \rightarrow W^+ W^-$, so that the relic density calculation is not significantly affected by the sign of $\theta$ (this is also true for the process $\chi \chi \rightarrow h s$). To make the impact of the sign choice for $\theta$ explicit, we show in figure~\ref{fig:spin0-relic2} the case $\theta = -0.2$ for $m_{Z'} = 3\,m_\chi$ and $m_{Z'} = 0.1\,m_\chi$, which can be directly compared to the first row of figure~\ref{fig:spin0-relic}. We observe that the only visible change is that the bound from the Higgs signal strength for $m_s < m_h / 2$ is significantly relaxed for the negative sign. Nevertheless, this parameter region is independently excluded by direct detection experiments (which are not sensitive to the sign of $\theta$). For smaller values of $|\theta|$ the differences between positive and negative sign are even smaller. Since considering $\theta < 0$ does not open up any additional parameter space, we will focus on the case $\theta > 0$ for the remainder of this work.

\section{Two mediators}
\label{sec:two-mediators}

Having discussed the case of a $Z'$ mediator with $g_q \gg \sin \theta$ in section~\ref{sec:spin-1} and the case of a dark Higgs mediator with $\sin \theta \gg g_q$ in section~\ref{sec:spin-0}, we now turn to the case where the couplings of the two mediators are comparable. To visualise this situation we consider the $m_s$--$m_{Z'}$ plane, i.e.\ we vary the masses of both mediators, while keeping the DM mass fixed. 

We will be particularly interested in the case that both $g_q$ and $\sin \theta$ are very small (typically of order 0.01 or smaller), so that both mediators are secluded from the SM. In this case, most experimental constraints can be suppressed. However, as recently pointed out in~\cite{Bell:2016fqf}, it may still be possible to probe this set-up in indirect detection experiments, which we now discuss in more detail.

\subsection{Indirect detection}
\label{sec:indirect}

For almost all annihilation channels discussed above the relic density
is dominantly set via $p$-wave processes. An exception is the $s$-wave
contribution from $\chi \chi \rightarrow Z' Z'$ which is comparable to
the $p$-wave one for $0.5 \, m_\chi < m_{Z^\prime} < 0.9 \, m_\chi $.
For smaller $m_{Z^\prime}$ the $p$-wave contribution is enhanced by
$m_{\chi}^4 / m_{Z^\prime}^4$ and dominates the cross section during
freeze-out.  As a result, the annihilation rate in the present
Universe is suppressed compared to the
thermal cross section.\footnote{This observation differs
  from~\cite{Martin:2014sxa}, where it was found that the process
  $\bar{\chi} \chi \rightarrow Z' Z'$ proceeds dominantly via
  $s$-wave. The origin of this difference is that we consider Majorana
  DM with axial couplings to the $Z'$, whereas~\cite{Martin:2014sxa}
  considers Dirac DM with vector couplings, so that the Goldstone
  modes do not contribute to the $p$-wave annihilation cross
  section. Note also that we neglect the possibility that the $Z'$ is
  sufficiently light compared to the DM mass that the annihilation
  cross section in the present Universe receives a Sommerfeld
  enhancement.}

However, one interesting possibility is the case that both the $Z'$ and the dark Higgs are sufficiently light that the annihilation process $\chi \chi \rightarrow Z' s$ is allowed. This process proceeds dominantly via $s$-wave, with an annihilation cross section given by (for $\theta \rightarrow 0$ and $m_{Z'}, m_s \ll m_\chi$)
\begin{equation}\label{eq:cc-to-sZp}
\sigma v_\chi (\chi \chi \rightarrow sZ') \simeq \frac{y_\chi^4}{64 \pi \, m_\chi^2}\; .
\end{equation}
When kinematically allowed, this process is typically the dominant one for thermal freeze-out. In this case, observable indirect detection signals may be obtained from cascade annihilations~\cite{Martin:2014sxa}, i.e.\ the subsequent decays of the $Z'$ and the dark Higgs into SM particles. 

Fermi Large Area Telescope (LAT) observations of Milky Way dwarf spheroidals~\cite{Ackermann:2015zua} place tight constraints on the $\gamma$-ray flux from DM annihilations. To study these constraints, we calculate this $\gamma$-ray spectrum as a function of $g_\chi$ and the three masses $m_\chi$, $m_{Z'}$ and $m_s$ using \texttt{MicrOMEGAs\_v4.2.5}~\cite{Belanger:2014vza}. For each set of parameters we then calculate the likelihood using the publicly available likelihood functions from Fermi-LAT~\cite{Ackermann:2015zua}. Finally, we combine the likelihoods from 15 different dwarf spheroidals using the $J$-factors from~\cite{Ackermann:2015zua}.

We find that for $m_{Z'},m_s < m_\chi \lesssim \unit[100]{GeV}$, the resulting constraints are typically sensitive to the thermal cross section, so that the possibility that the DM relic abundance is set by the process $\chi \chi \rightarrow Z' s$ in this mass region can be excluded.

\subsection{Results for fixed couplings}
\label{sec:2med-fixed}

For the remainder of this section, we proceed in the same way as above by fixing the couplings $\theta$ and $g_q$ and then calculating $g_\chi$ as a function of the three masses by imposing the relic density requirement. As expected, the various constraints are found to depend sensitively on the chosen values of the couplings. In section~\ref{sec:2med-scan} we therefore perform a global scan, where for each combination of masses we vary all three couplings simultaneously in order to determine whether there is a combination that reproduces the observed relic density while evading all experimental constraints.

We first consider the case where both $g_q$ and $\sin \theta$ are sizeable, i.e.\ of order 0.1. As discussed above, significantly larger values of $g_q$ would be in strong tension with searches for dijet resonances. Significantly larger values of $\theta$, on the other hand, would be in conflict with the observed Higgs signal strength and electroweak precision observables. For $g_q$ and $\sin\theta$ being sizeable, both the dark Higgs and the $Z'$ can mediate the interactions of DM with SM particles and all the constraints discussed in the two previous sections apply.

\begin{figure}[tb]
\centering
\includegraphics[height=0.28\textheight]{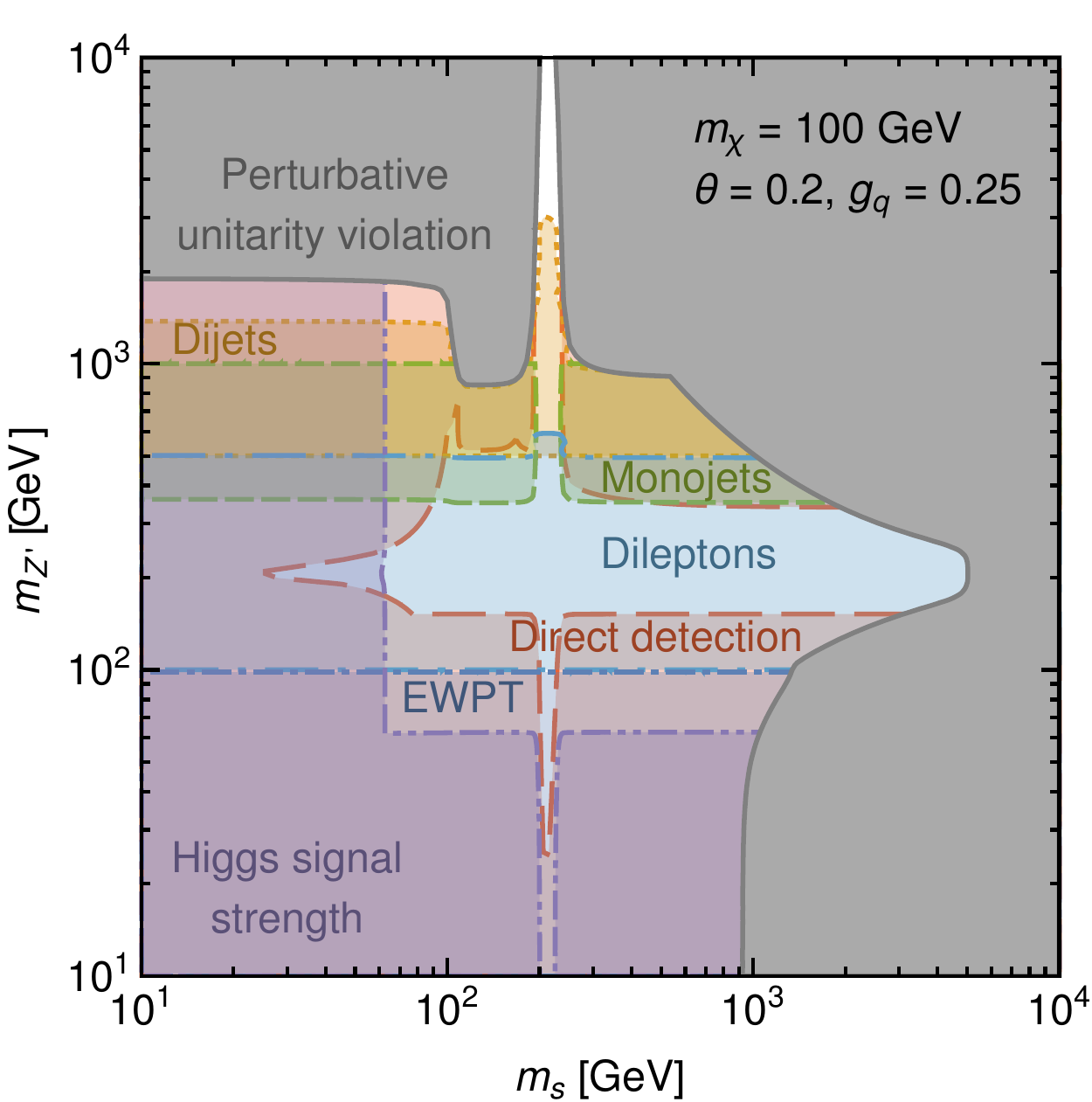}\qquad\includegraphics[height=0.28\textheight]{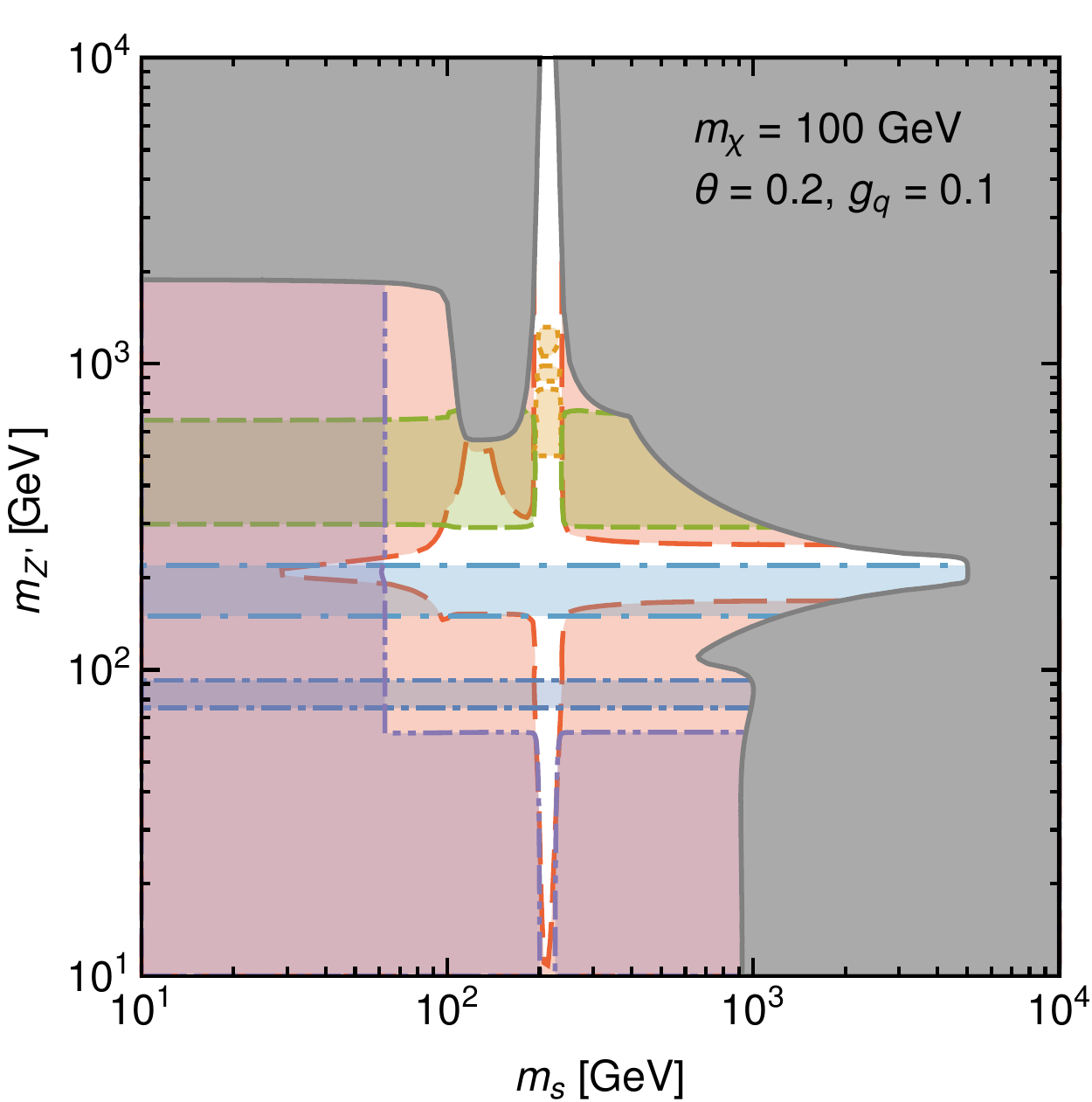}
\includegraphics[height=0.28\textheight]{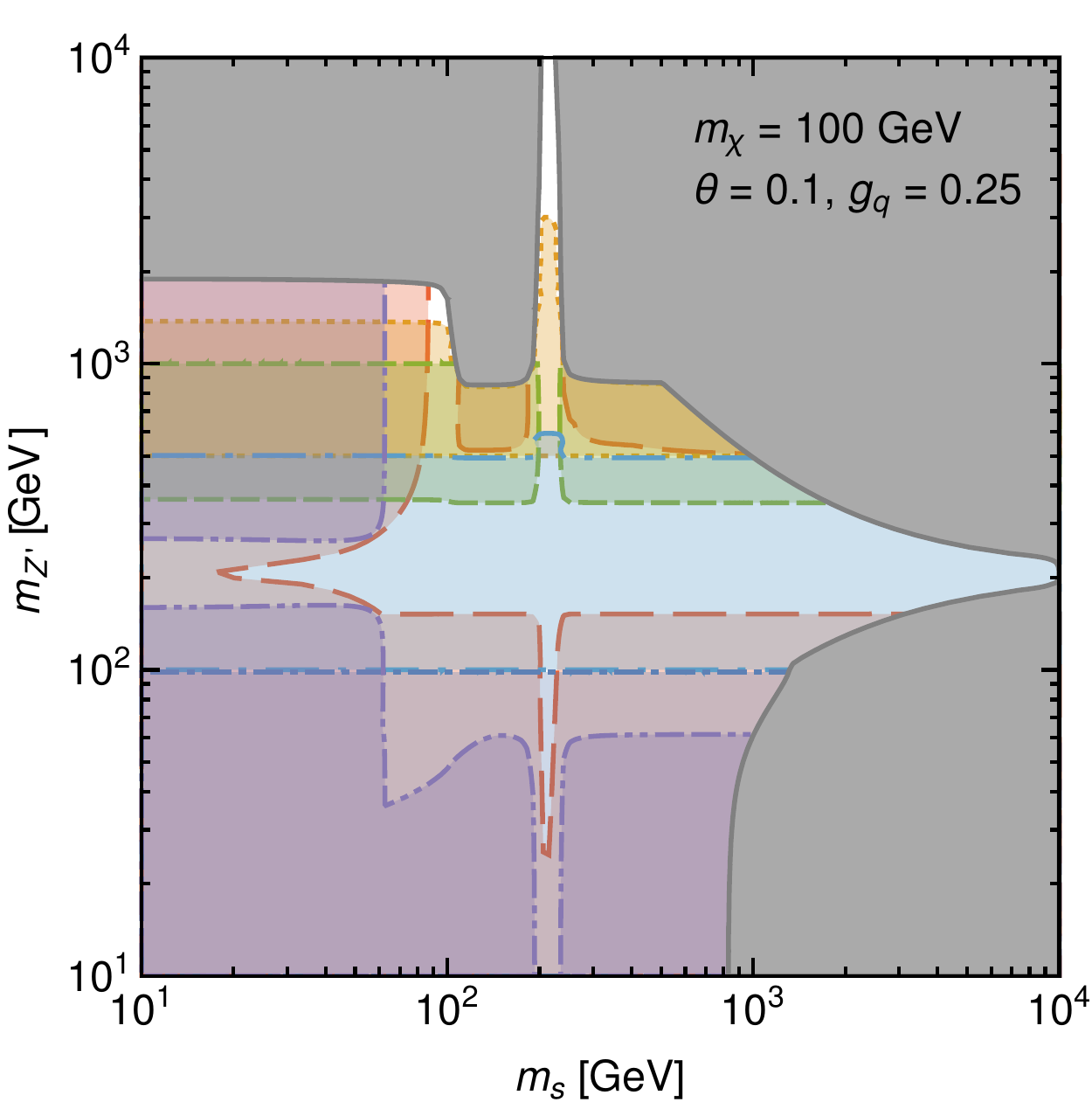}\qquad\includegraphics[height=0.28\textheight]{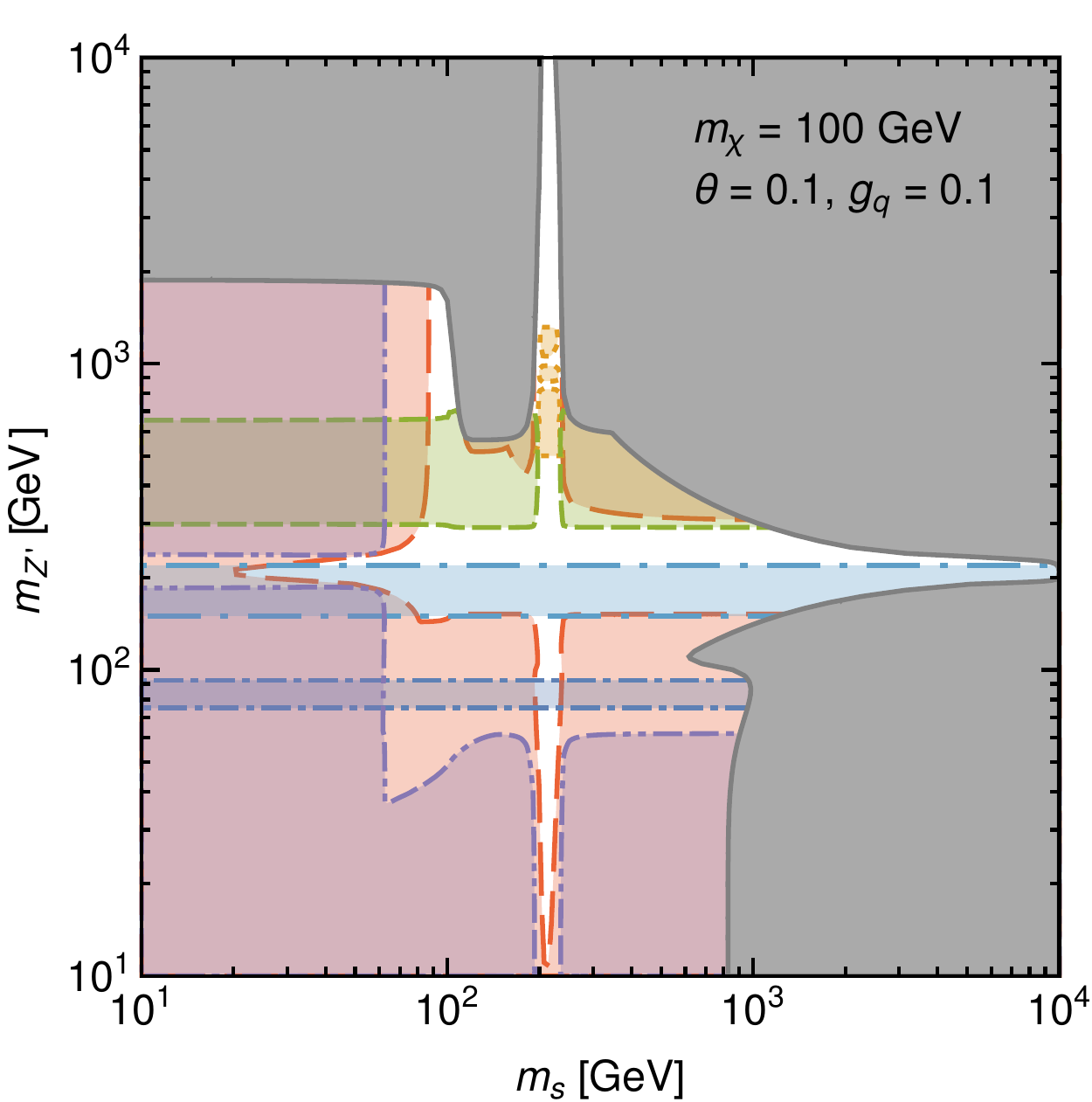}
\caption{Constraints for the case of two mediators, with the (common) DM--mediator coupling determined by the requirement to reproduce the observed relic abundance. Plots in the same row correspond to constant $\theta$, plots in the same column correspond to constant $g_q$. In all panels, we have fixed $m_\chi = \unit[100]{GeV}$. 
In the grey shaded regions (solid lines) at least one of the couplings leads to violation of perturbative unitarity. The yellow shaded regions (dotted) are excluded by dijet searches, the green shaded regions (short dashed) by monojet searches, the red shaded regions (long dashed) by direct detection, the purple shaded regions (double dash-dotted) by the observed Higgs signal strength and the bound on invisible Higgs decays. The dark blue regions (short dash-dotted) are excluded by EWPT and the light blue regions (long dash-dotted) by dilepton searches, both for loop-induced kinetic mixing.}
\label{fig:mediators-relic}
\end{figure}

\begin{figure}[tb]
\centering
\includegraphics[height=0.28\textheight]{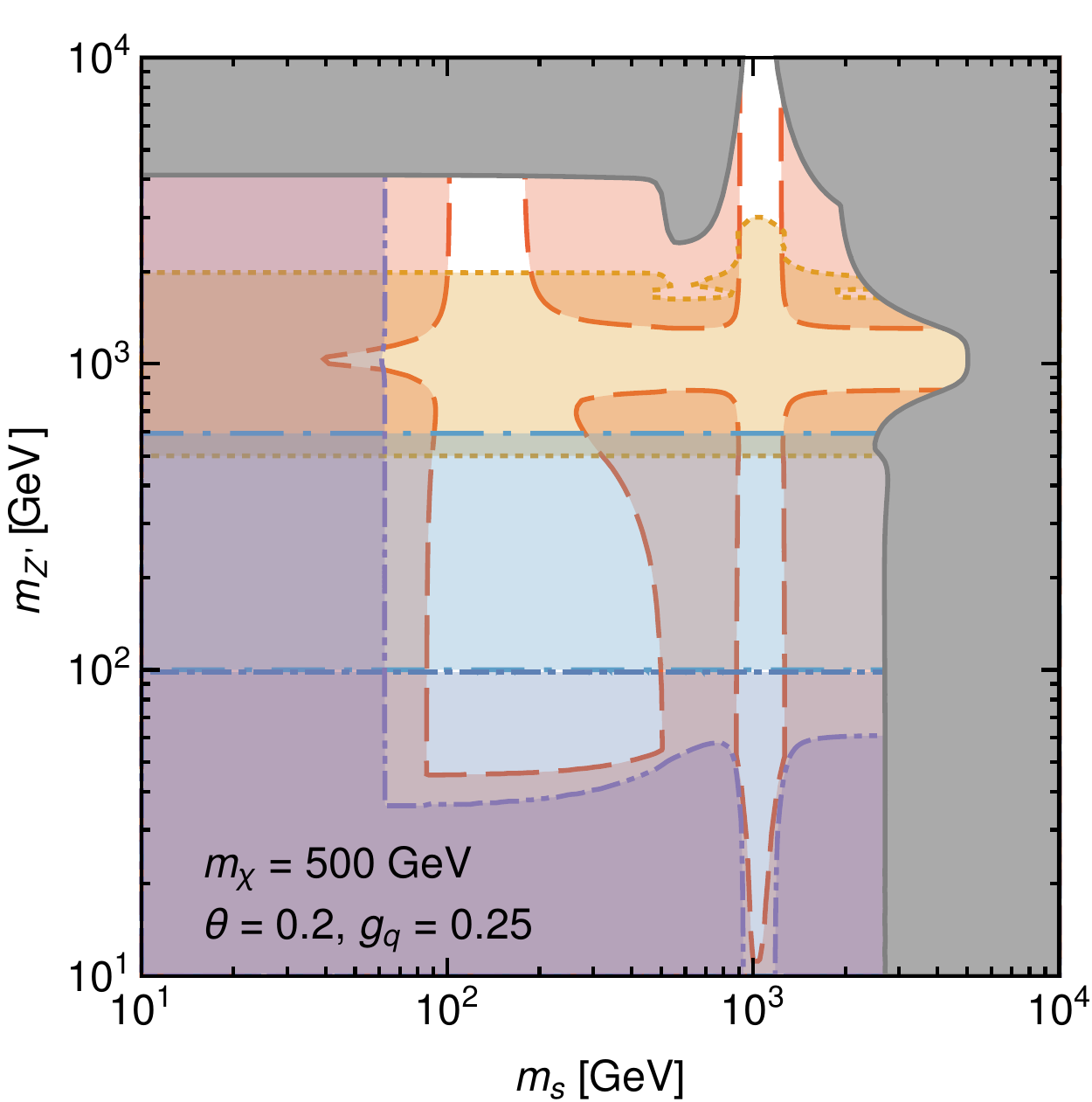}\qquad\includegraphics[height=0.28\textheight]{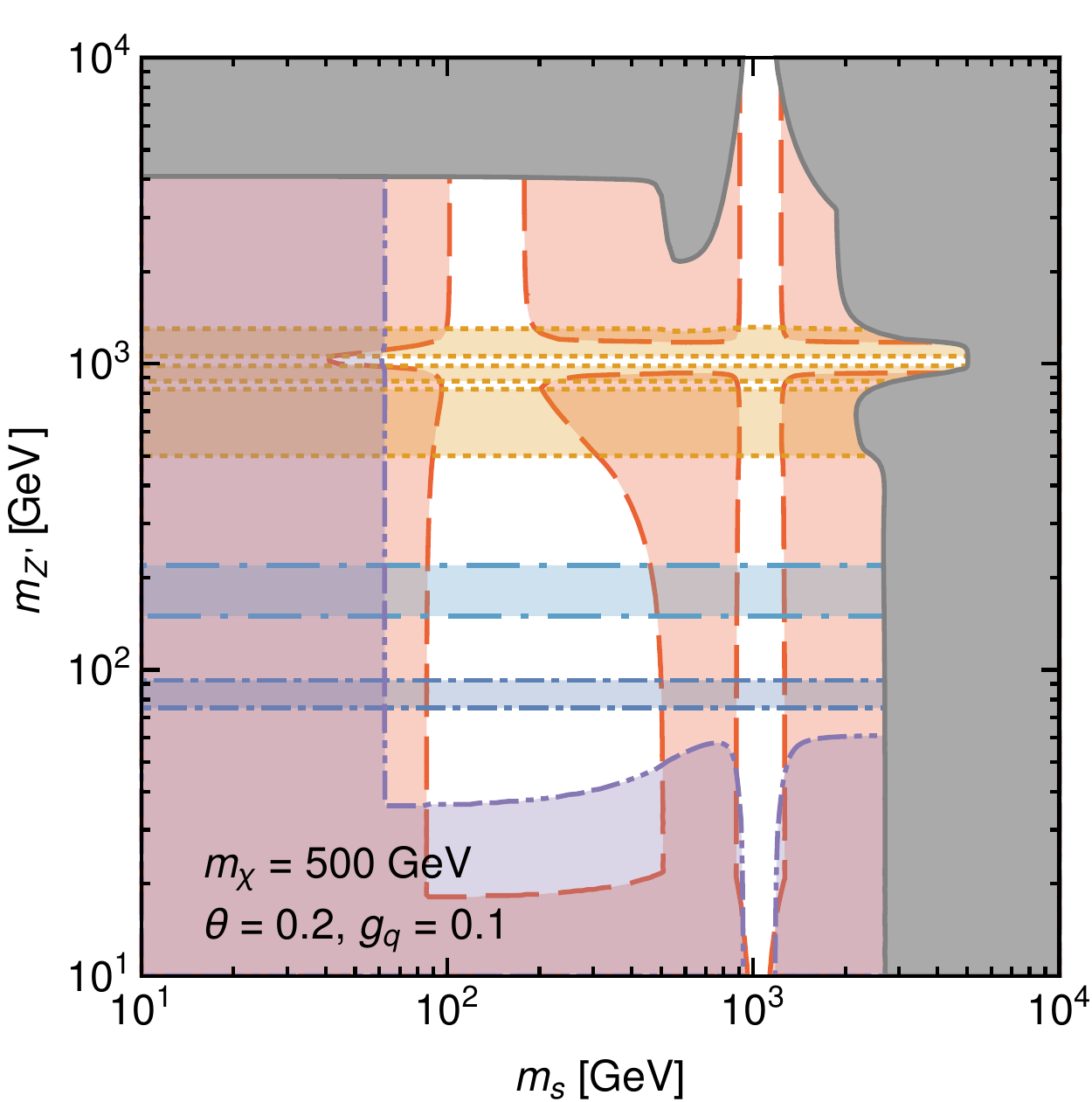}
\includegraphics[height=0.28\textheight]{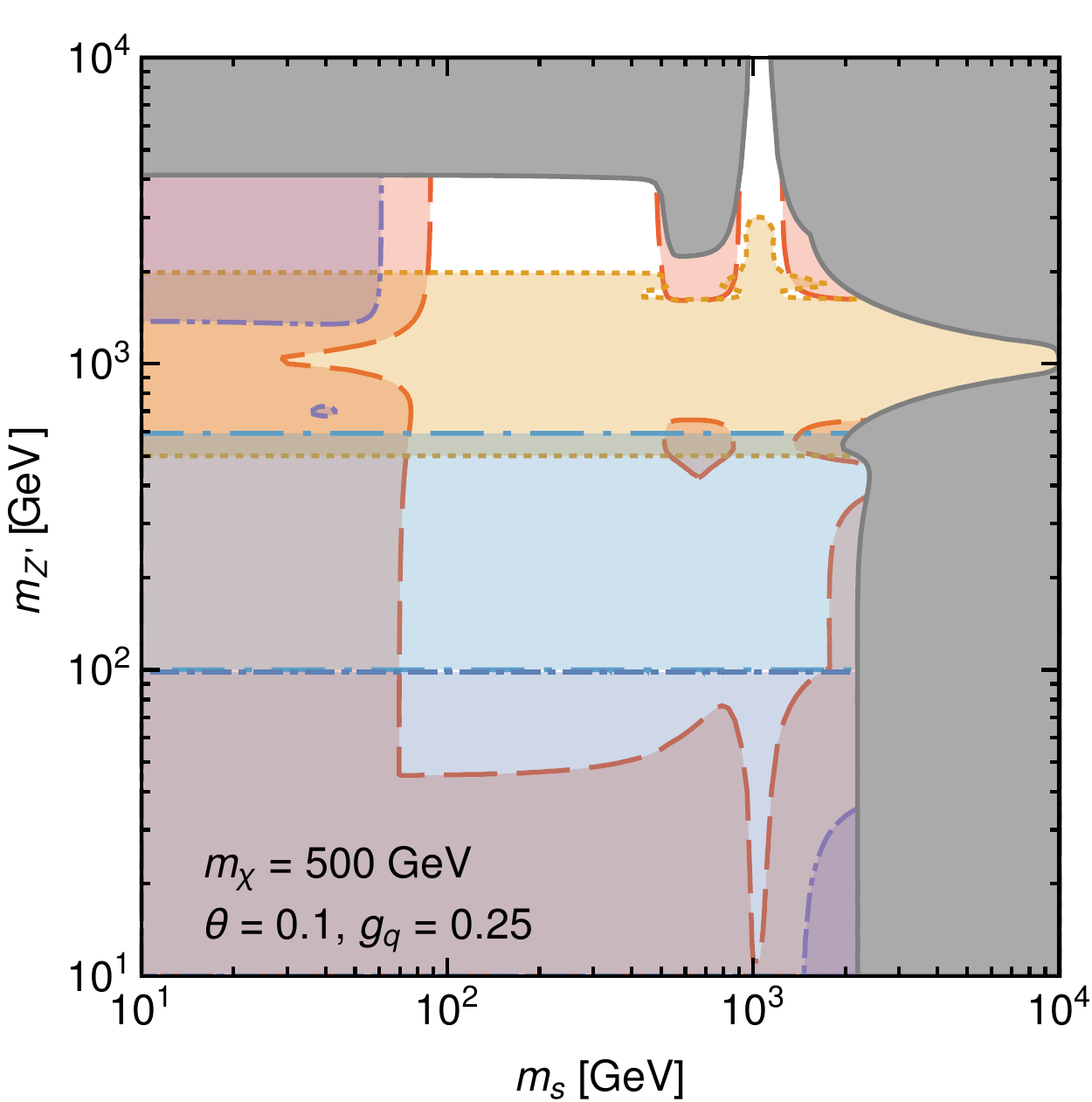}\qquad\includegraphics[height=0.28\textheight]{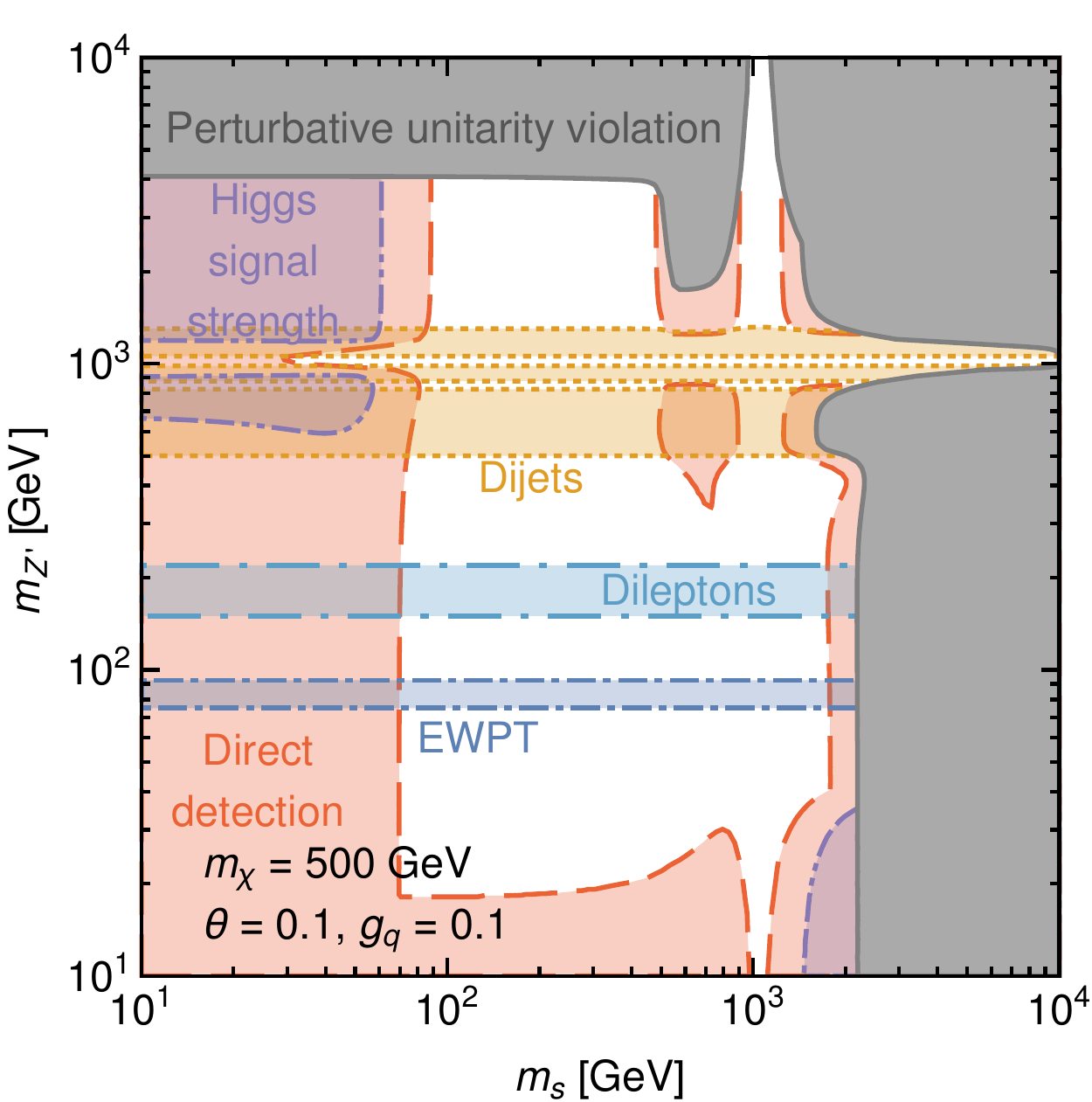}
\caption{Same as figure~\ref{fig:mediators-relic} but for $m_\chi = \unit[500]{GeV}$.}
\label{fig:mediators-relic2}
\end{figure}

Figs.~\ref{fig:mediators-relic} and~\ref{fig:mediators-relic2} show the resulting constraints in the $m_s$--$m_{Z'}$ plane for different combinations of $g_q$ and $\sin \theta$ as well as for $m_\chi = \unit[100]{GeV}$ and $m_\chi = \unit[500]{GeV}$, respectively. We make the following general observations:
\begin{itemize}
 \item As before, the weakest constraints are found if DM annihilations in the early universe receive a resonant enhancement, i.e.\ if either $m_\chi \approx m_{Z'} / 2$ or $m_\chi \approx m_s / 2$.
 \item For $m_\chi < m_s, m_{Z'}$ the relic density is controlled by direct annihilation of DM into SM final states. Depending on the exact values of the couplings and masses, the dominant final state is either $q\bar{q}$ or $W^+W^-$. Both LHC searches and direct detection place significant constraints on this parameter region. 
 \item For either $m_s < m_\chi$ or $m_{Z'} < m_\chi$, the relic density can be easily reproduced via annihilation into two dark terminators. However, if $g_q$ and $\sin \theta$ are fixed to relatively large values, these regions are typically tightly constrained by direct detection and the Higgs signal strength. 
 \item Moreover, for $m_s < m_\chi$ the requirement that $y_\chi$ remains perturbative places an upper bound on $m_{Z'}$, while for $m_{Z'} < m_\chi$ we obtain an upper bound on $m_s$ above which the Higgs potential becomes non-perturbative.
\end{itemize}

For $m_\chi = \unit[100]{GeV}$ we see that~--- for the coupling combinations considered in figure~\ref{fig:mediators-relic}~--- only small regions of parameter space close to the two resonances remain viable, whereas much larger regions are still allowed for $m_\chi = \unit[500]{GeV}$. The reason is that in the latter case, it is possible for $m_s$ or $m_{Z'}$ (or both) to be smaller than $m_\chi$ without immediately being strongly constrained by direct detection or Higgs physics.

The fact that direct detection bounds are so strong in figures~\ref{fig:mediators-relic} and \ref{fig:mediators-relic2} is a direct consequence of the assumption that $g_q$ and $\sin \theta$ are sizeable. For $m_{Z'}, m_s < m_\chi$, however, direct annihilation into two terminators typically gives the dominant contribution to DM annihilation. Consequently, it should be possible to reproduce the required DM relic abundance for much smaller values of $g_q$ and $\sin \theta$, provided $g_\chi$ is sufficiently large.

\begin{figure}[tb]
\centering
\includegraphics[height=0.28\textheight]{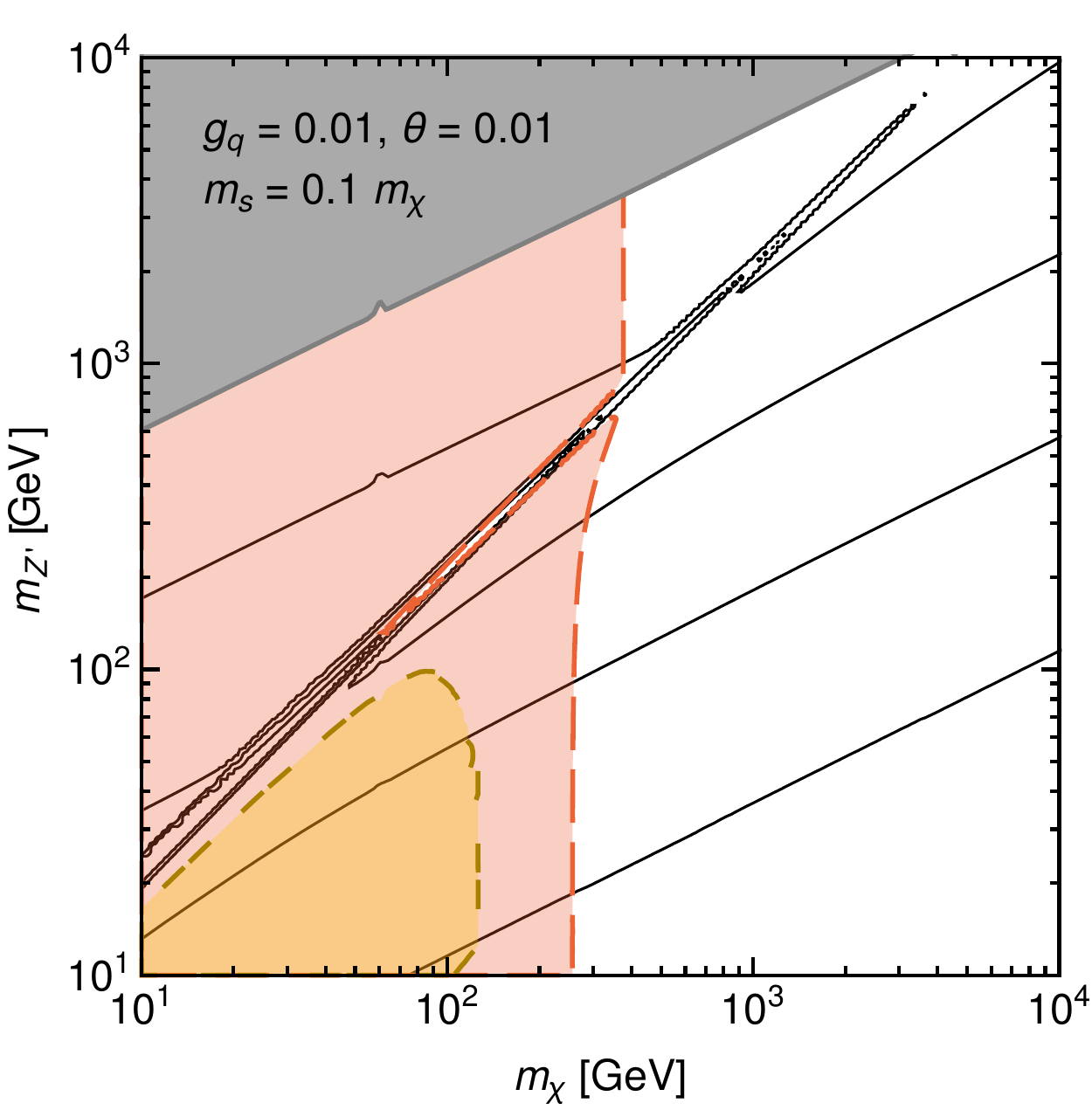}\qquad\includegraphics[height=0.28\textheight]{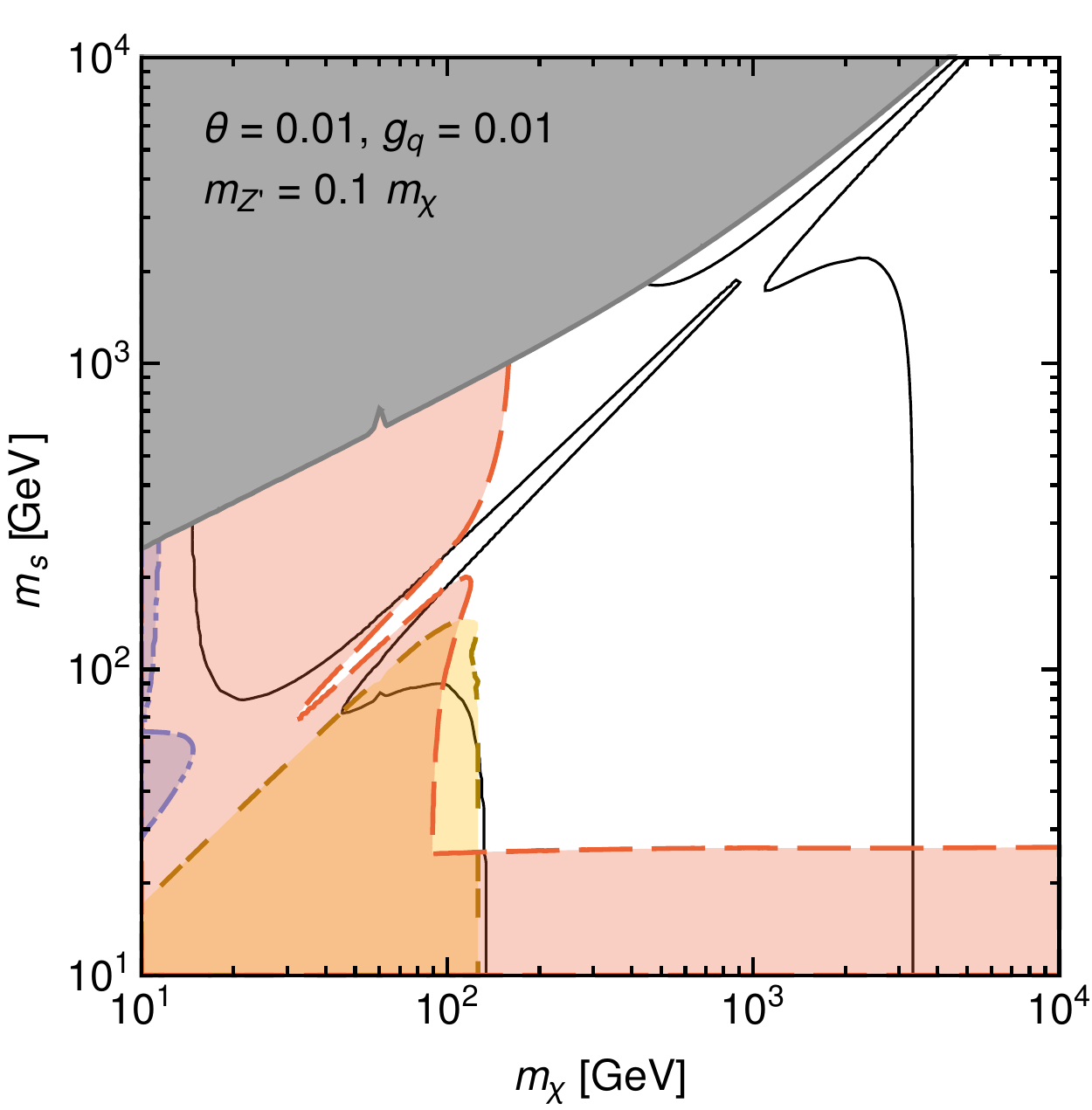}
\includegraphics[height=0.28\textheight]{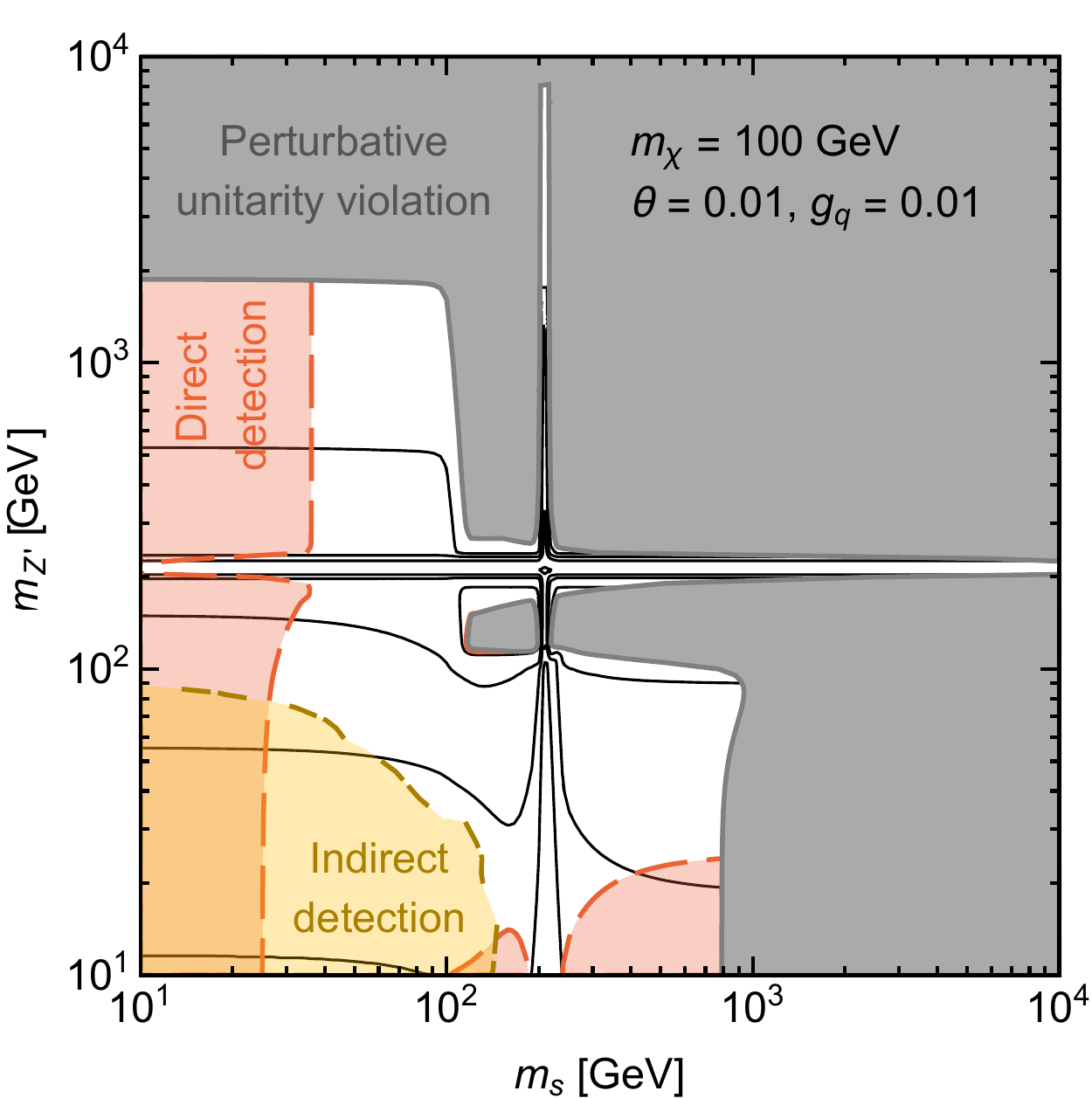}\qquad\includegraphics[height=0.28\textheight]{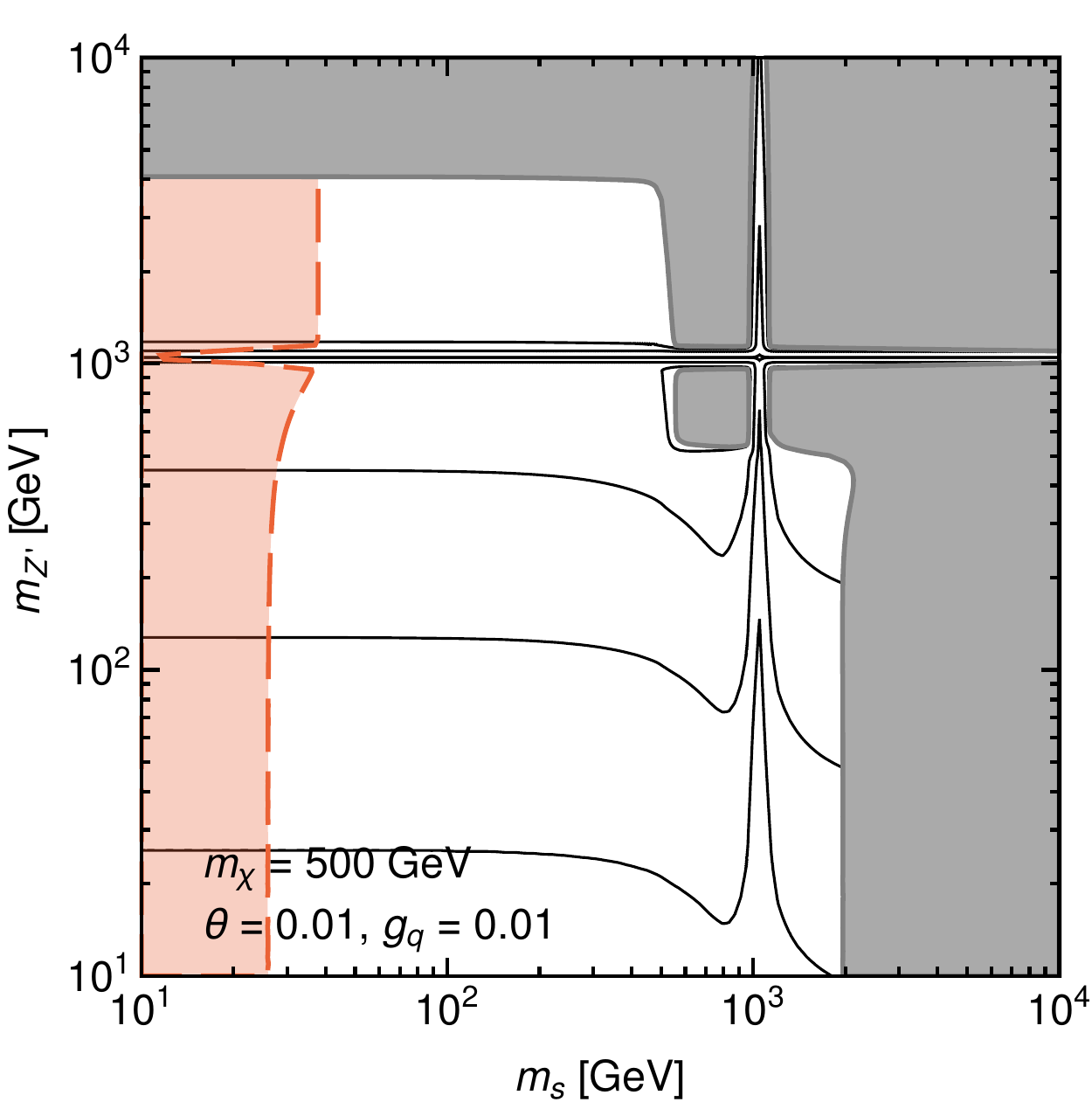}
\caption{The case of very small SM couplings $g_q$ and $\theta$, shown in the $m_\chi$--$m_{Z'}$ parameter plane (top-left), the $m_\chi$--$m_s$ parameter plane (top-right) and the $m_s$--$m_{Z'}$ parameter plane for $m_\chi = \unit[100]{GeV}$ and $m_\chi = \unit[500]{GeV}$ (bottom row). For $m_\chi \lesssim \unit[100]{GeV}$, indirect detection constraints become relevant (yellow region, double-dashed line). The red shaded regions (long dashed) are excluded by direct detection, the purple shaded regions (double dash-dotted) by the observed Higgs signal strength and the bound on invisible Higgs decays. The black contours show constant $g_\chi$.}
\label{fig:small-couplings}
\end{figure}

Figure~\ref{fig:small-couplings} summarises the constraints for $g_q = \theta = 0.01$. The four different panels are constructed in analogy to figures~\ref{fig:spin1-relic}--\ref{fig:mediators-relic2}, i.e.\ they show the $m_\chi$--$m_{Z'}$ plane, the $m_\chi$--$m_s$ plane and the $m_s$--$m_{Z'}$ plane for $m_\chi = \unit[100]{GeV}$ and $m_\chi = \unit[500]{GeV}$, respectively. We find that the parameter region with $m_{Z'}, \, m_s > m_\chi$ is very tightly constrained, essentially because it is impossible to reproduce the relic abundance via annihilation into SM final states using perturbative couplings. For smaller masses of either $m_{Z'}$ or $m_s$, large allowed regions open up, which are very difficult to probe experimentally. If both $m_{Z'}$ and $m_s$ are small, however, the indirect detection constraints discussed above become important, provided the DM mass is sufficiently small so that Fermi LAT is sensitive to the thermal cross section.

To conclude this section, we note that for $m_\chi \sim 30\text{--}\unit[50]{GeV}$ it may be possible within this framework to provide a viable explanation for the Galactic centre excess~\cite{Hooper:2010mq,Hooper:2011ti,Abazajian:2012pn,Gordon:2013vta,Daylan:2014rsa}. For example if $ 0.5 \, m_\chi< m_{Z^\prime} < 0.9 \, m_\chi$ and $m_s + m_{Z'} > 2 \, m_\chi$ the $s$-wave cross section of $\chi \chi \rightarrow Z^\prime Z^\prime$ is naturally close to the thermal value. Alternatively, for $m_s \ll m_{Z'}$ and $m_s + m_{Z'} \approx 2 \, m_\chi$, it is possible that both $\chi \chi \rightarrow Z' s$ and $\chi \chi \rightarrow s s$ contribute at comparable level to thermal freeze-out. The annihilation cross section for the first processes can then be a factor of a few below the thermal cross section, while the second process is velocity suppressed and therefore becomes negligibly small in the present Universe. This way, it may be possible to evade constraints from dwarf spheroidals while still obtaining an acceptable fit to the Galactic centre excess within astrophysical uncertainties.

\section{Global scan of couplings}
\label{sec:2med-scan}

While the various panels shown in figures~\ref{fig:mediators-relic} to \ref{fig:small-couplings} are helpful in order to develop an intuitive understanding of how the different bounds depend on the couplings $g_q$ and $\sin \theta$, it is difficult to derive general conclusions based on these plots. For example, for $m_\chi = \unit[100]{GeV}$ and $m_{Z'} = m_s = \unit[400]{GeV}$, the couplings considered in figure~\ref{fig:mediators-relic} are excluded by LHC searches, while for the couplings considered in figure~\ref{fig:small-couplings}, these masses already violate the perturbativity constraints. This immediately leads to the question whether there might be an intermediate value for $g_q$ and $\theta$ for which LHC constraints can be avoided and the perturbativity requirement satisfied.

To answer this question, we have performed the following global analysis. For fixed masses we scan over $g_q$ and $\theta$ and determine for each choice of those parameters the dark sector coupling by the relic abundance. Each combination of masses then falls into one of three categories:
\begin{enumerate}
 \item All combinations of $g_q$ and $\theta$ are excluded by at least one experimental constraint or perturbative unitarity.
 \item There is at least one combination of $g_q$ and $\theta$ which is consistent with all experimental constraints and perturbative unitarity.
 \item There is at least one combination of $g_q$ and $\theta$, for which current experimental constraints do not apply and which therefore cannot presently be excluded.
\end{enumerate}
The third situation occurs whenever a combination of couplings is most strongly constrained by LHC searches for dijet resonances. These searches typically assume $\Gamma_{Z'} / m_{Z'} < 0.3$, corresponding to $g_q \lesssim 0.8$, so that the bounds do not apply for larger widths/couplings. While it is to be expected that larger quark couplings lead to stronger rather than weaker constraints, we prefer not to perform an extrapolation and instead treat this special case separately.

We have performed such scans for a large number of different values of $m_{Z'}$ and $m_s$. The results of these scans for $m_\chi = \unit[100]{GeV}$ are summarised in the top-left panel of figure~\ref{fig:mediators-couplings}. The three categories discussed above are shown in red (excluded), white (allowed) and orange (inconclusive bound for broad mediator width). To illustrate these different situations, we consider in the remaining panels of figure~\ref{fig:mediators-couplings} five different benchmark cases, each of which corresponds to a set of $m_\chi$, $m_{Z'}$ and $m_s$ (indicated by black dots in the top-left panel). For each benchmark case we show the different constraints as a function of the couplings $g_q$ and $\theta$ (with $g_\chi$ fixed by the relic density requirement) using the same colour coding as in figures~\ref{fig:mediators-relic}--\ref{fig:small-couplings}. We make the following observations:

\begin{figure}[tb]
\centering
\includegraphics[height=0.24\textheight]{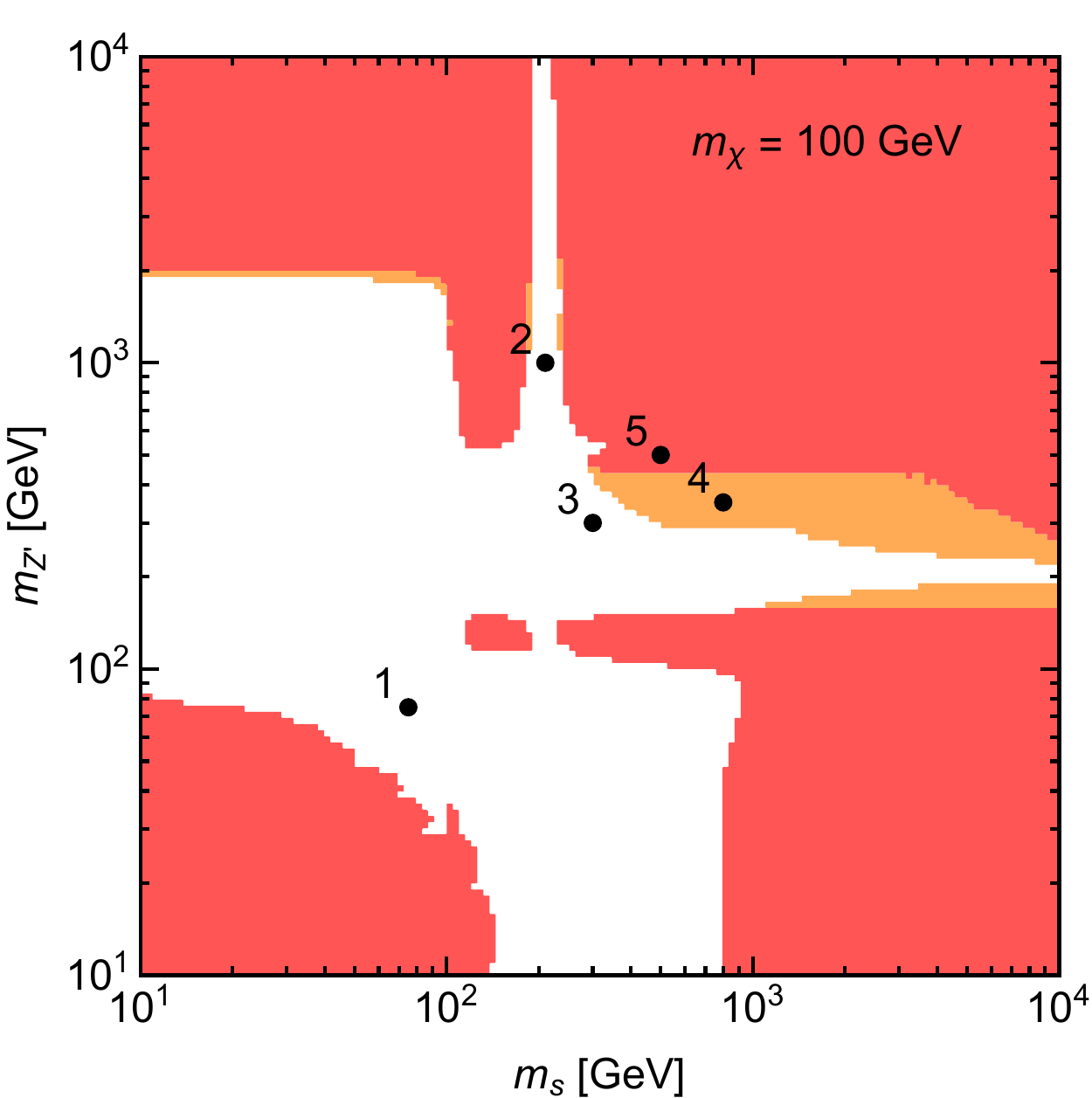}\qquad\includegraphics[height=0.25\textheight]{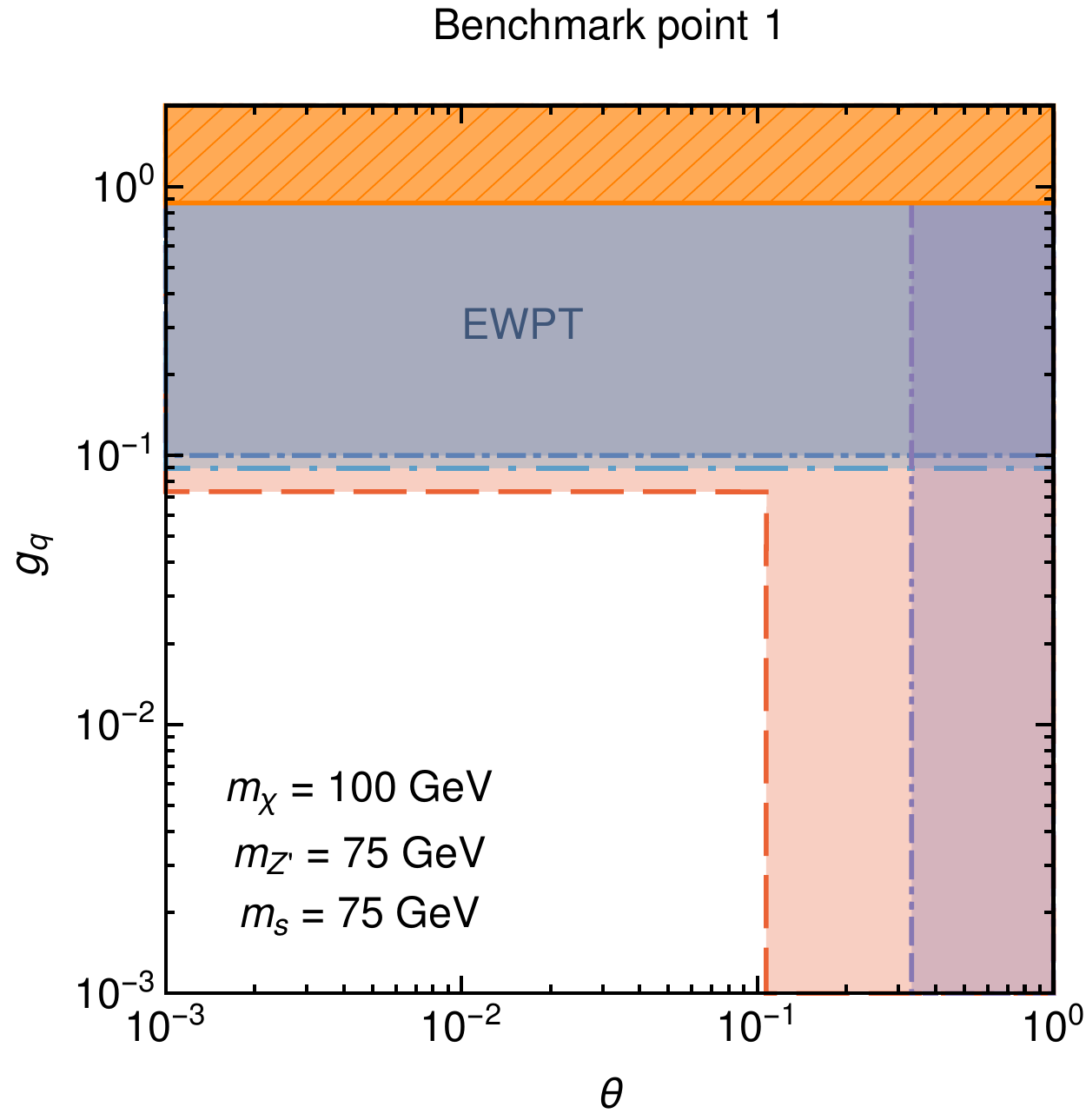}

\vspace{1mm}

\includegraphics[height=0.25\textheight]{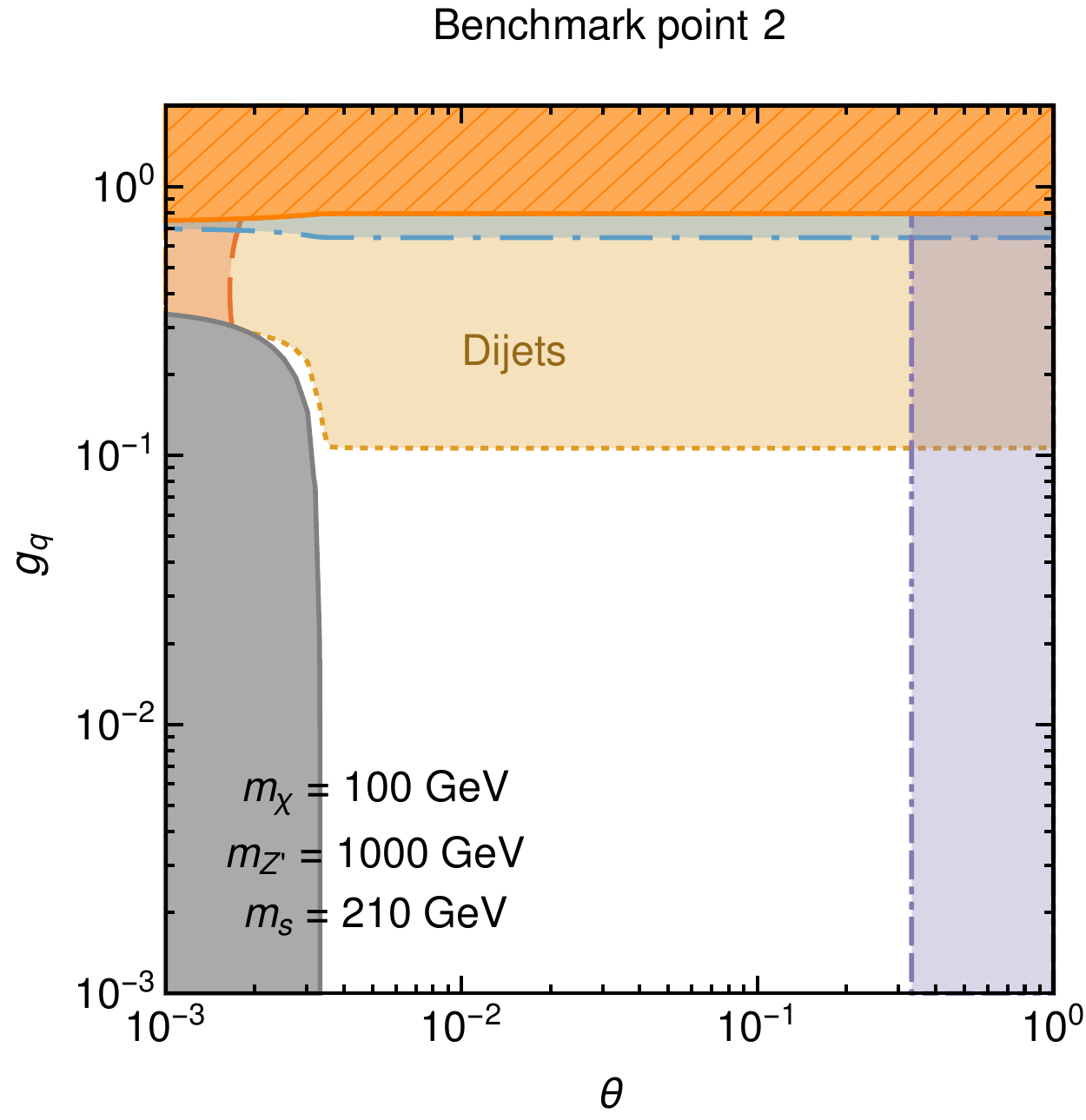}\qquad\includegraphics[height=0.25\textheight]{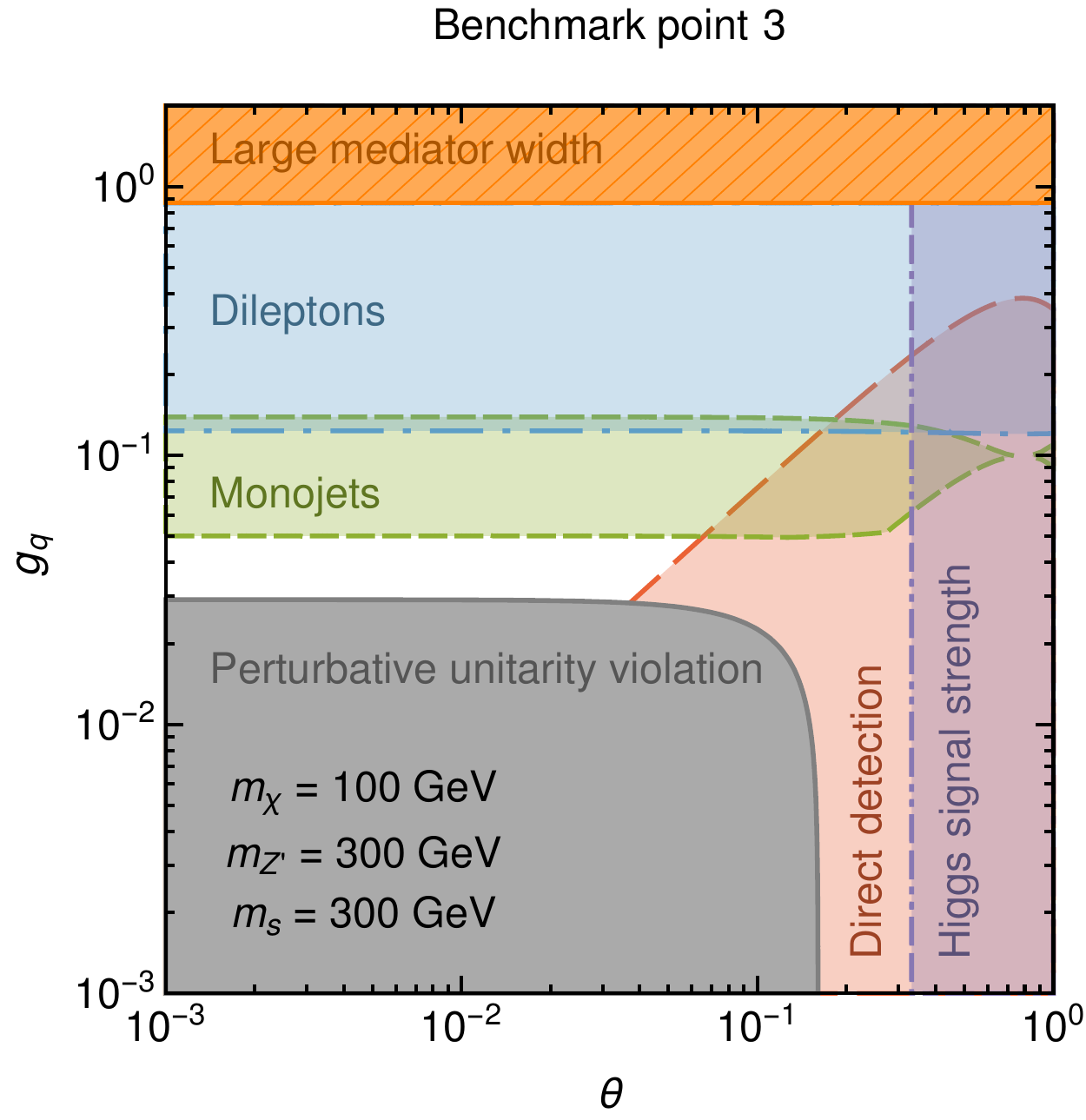}

\vspace{1mm}

\includegraphics[height=0.25\textheight]{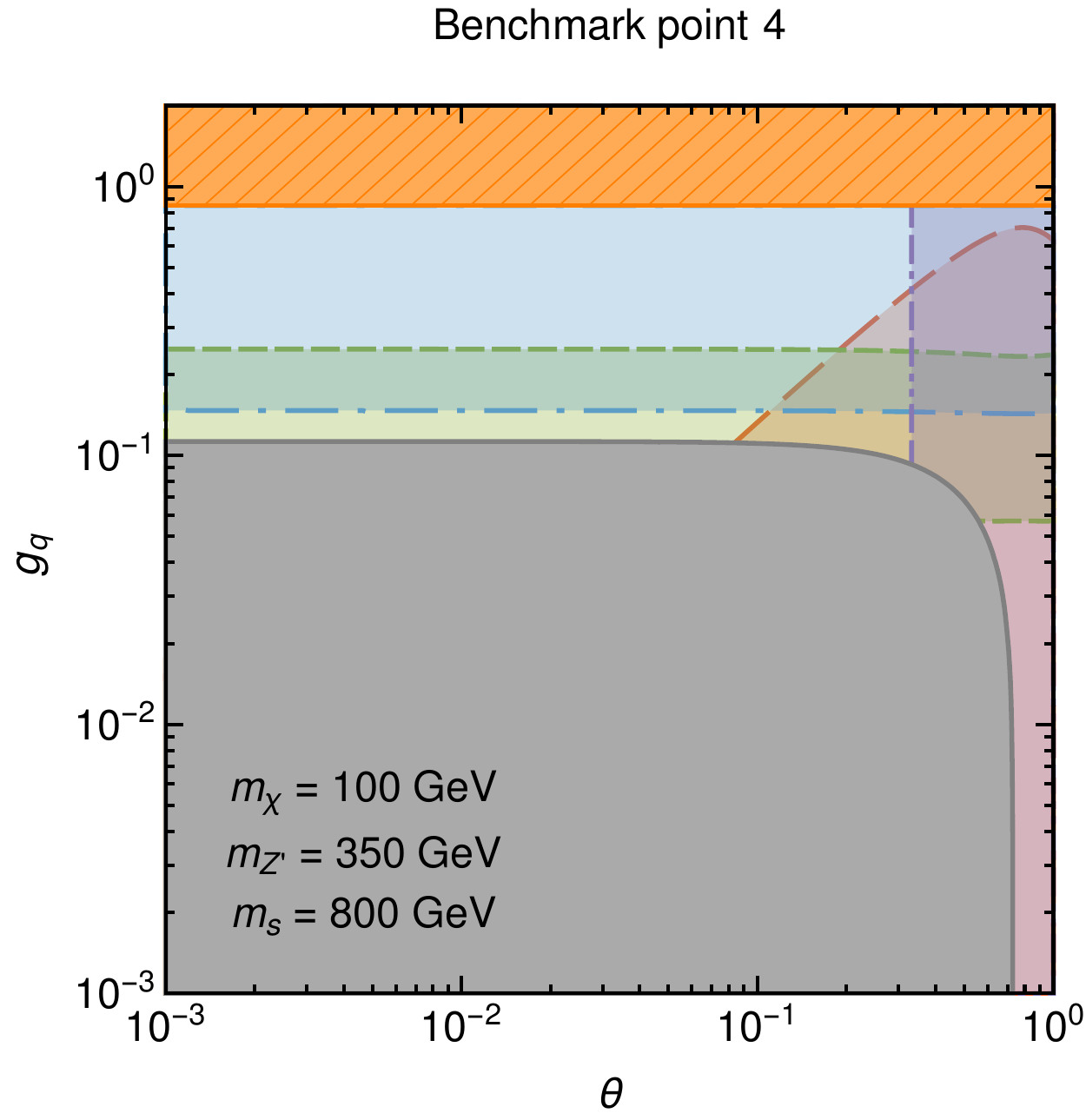}\qquad\includegraphics[height=0.25\textheight]{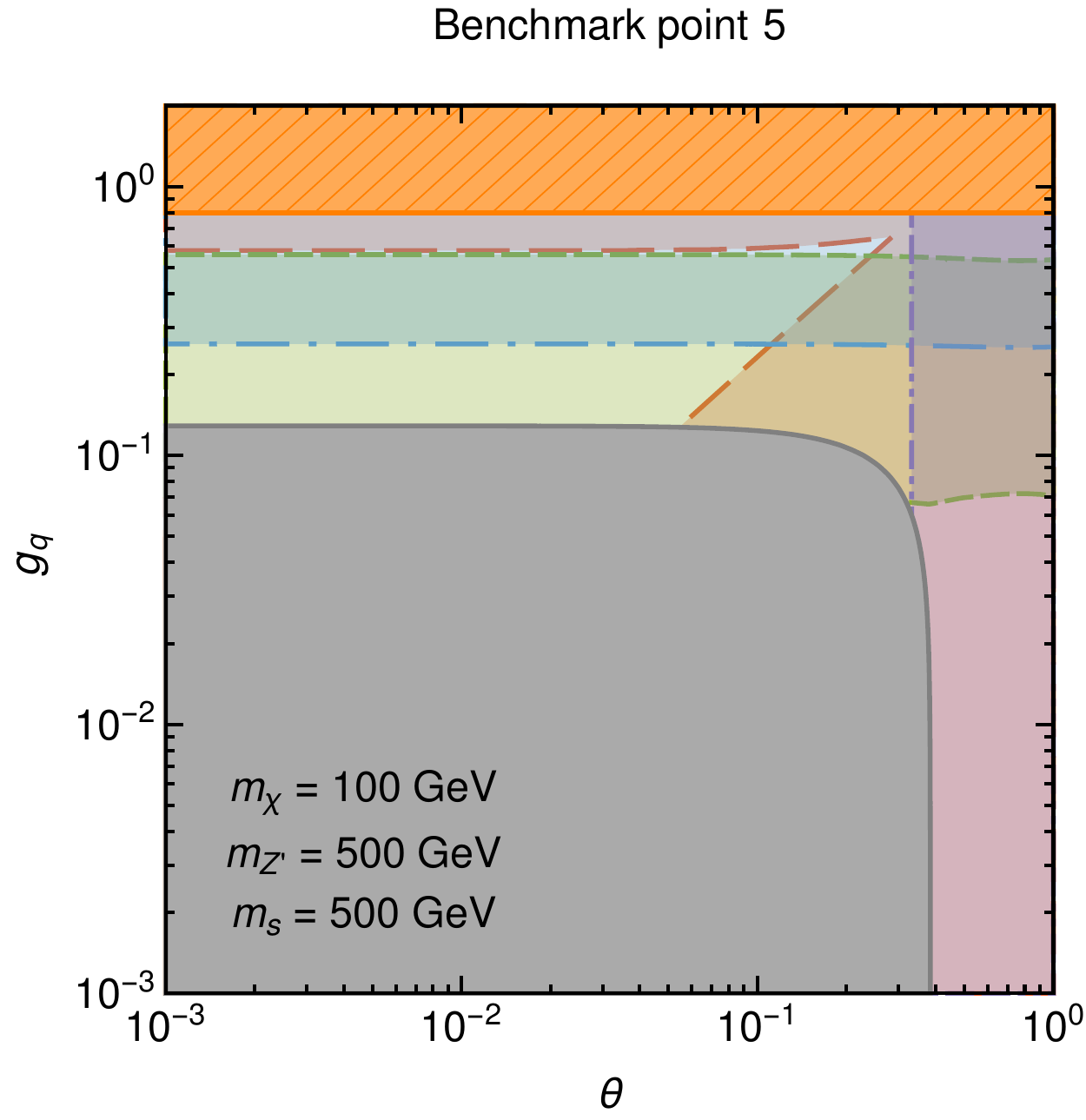}
\caption{Global scan for $m_\chi = \unit[100]{GeV}$. In the first panel (top-left) a global scan has been performed to determine those values of the two SM--mediator couplings that give the weakest constraints (see text for details). The red shaded region is excluded for all possible combinations, while in the white region all constraints can be evaded. In the orange shaded region it is not possible to exclude large values of $g_q$, corresponding to $\Gamma_{Z'} / m_{Z'} > 0.3$. The remaining five panels show the various constraints as a function of the couplings between the two mediators and SM states for different benchmark cases (indicated by black dots in the top-left panel) with the common DM--mediator coupling fixed by the relic density requirement. The colour coding is the same as in figures~\ref{fig:mediators-relic}--\ref{fig:small-couplings}.}
\label{fig:mediators-couplings}
\end{figure}

\begin{itemize}
\item General observations: As expected, the constraints from dijets, dileptons and EWPT exclude large values of $g_q$, while the constraints from the Higgs signal strength exclude large values of $\theta$. Monojet searches are sensitive to an intermediate range of $g_q$, as very large values of $g_q$ suppress the invisible branching ratio of the mediator and hence the monojet signal, while very small values of $g_q$ suppress the total production cross section of the mediator below the observable level. The behavior of the direct detection bound, on the other hand, is less intuitive: if the relic density is dominantly set by annihilation into quarks via the $Z'$, then for small values of $\theta$ both the annihilation and the direct detection rate scale proportionally to $g_q^2 \, g_\chi^2$. In other words, if $g_\chi$ is fixed by the relic density requirement, direct detection will either exclude all values of $g_q$ or no values of $g_q$. For larger values of $\theta$ the direct detection bound on Higgs exchange becomes relevant for small $g_q$ (corresponding to large $g_\chi$ and hence large $y_\chi$). Thus for $m_\chi < m_s, m_{Z'}$, direct detection typically gives the strongest constraint on small $g_q$ and large $\theta$.
\item Benchmark point 1: The combination $m_\chi = \unit[100]{GeV}$, $m_{Z'} = \unit[75]{GeV}$ and $m_s = \unit[75]{GeV}$ is allowed due to the presence of two light terminators. Since DM annihilation proceeds dominantly via $\chi \chi \rightarrow Z' s$, the two couplings $g_q$ and $\theta$ can be made almost arbitrarily small, thus evading all experimental constraints. We note that in principle it is sufficient to have one light terminator, with the second mediator being much heavier than the DM particle. However, perturbativity limits place an upper bound on the mass of the second mediator. For $m_\chi = \unit[100]{GeV}$ we find approximately $m_s \lesssim \unit[800]{GeV}$ and $m_{Z'} \lesssim \unit[2]{TeV}$. Conversely, if both $m_{Z'}$ and $m_s$ are very small, the indirect detection constraints discussed above become relevant.
\item Benchmark point 2: For $m_\chi = \unit[100]{GeV}$, $m_{Z'} = \unit[1000]{GeV}$ and $m_s = \unit[210]{GeV}$, it is impossible to evade all constraints for $\theta \approx 0$. While small values of $g_q$ (and correspondingly large values of $g_\chi$) would be excluded by perturbativity, larger values are constrained by direct detection as well as searches for dijet and dilepton resonances. However, for $\theta > 0$ there is plenty of parameter space to evade all experimental constraints due to the resonant enhancement of the annihilation processes $\chi \chi \rightarrow s \rightarrow \text{SM}$.
\item Benchmark point 3: For $m_\chi = \unit[100]{GeV}$, $m_{Z'} = \unit[300]{GeV}$, $m_s = \unit[300]{GeV}$, one can observe the strong complementarity between different experimental probes. Direct detection experiments and bounds on the Higgs signal strength exclude large values of $\theta$, while searches for monojets and for dilepton resonances exclude wide ranges of $g_q$. Nevertheless, all constraints can be evaded for $g_q \approx 0.04$, so that this parameter point is allowed.
\item Benchmark point 4: For $m_\chi = \unit[100]{GeV}$, $m_{Z'} = \unit[350]{GeV}$, $m_s = \unit[800]{GeV}$ all constraints become somewhat stronger compared to benchmark 3. In particular, the perturbativity constraint now excludes $g_q < 0.1$ (for $\theta \ll 1$), closing the previously unconstrained gap around $g_q \approx 0.04$. Large values of $g_q$ are excluded by searches for dilepton resonances. However, these searches cannot be applied to $g_q \gtrsim 0.85$ due to the broadening of the resonance. We can therefore only reliably exclude smaller couplings of $g_q$.
\item Benchmark point 5: The case $m_\chi = \unit[100]{GeV}$, $m_{Z'} = \unit[500]{GeV}$, $m_s = \unit[500]{GeV}$ is again more strongly constrained than benchmark 4. Crucially, direct detection experiments are now also sensitive to large values of $g_q$, thus covering the parameter region where dilepton searches lose sensitivity. The combination of all these constraints therefore allows to exclude this combination of masses for all coupling values that reproduce the relic abundance.
\end{itemize}

The fact that in many cases a combination of different constraints is necessary to conclusively rule out a given combination of masses (see e.g.\ benchmark point 5) illustrates the necessity for comprehensively scanning over the two couplings. We observe that indeed large parts of the parameter space are excluded for all combinations of couplings that reproduce the relic abundance. For $m_{Z'} > 2\,m_\chi$, LHC constraints are typically very important and push the model to very broad widths. In this parameter region even stronger constraints can be expected in the near future due to improved direct detection and collider constraints. Benchmark point 3 for example may soon be excluded for all combinations of $g_q$ and $\sin \theta$. Indeed, for a traditional dark mediator with mass larger than twice the DM mass, the simple thermal freeze-out picture appears under considerable pressure. Nevertheless, sizeable allowed regions remain for $m_{Z'}, \, m_s < m_\chi$, unless both masses are small enough for indirect detection constraints to become important. 

\begin{figure}[tb]
\centering
\includegraphics[height=0.25\textheight]{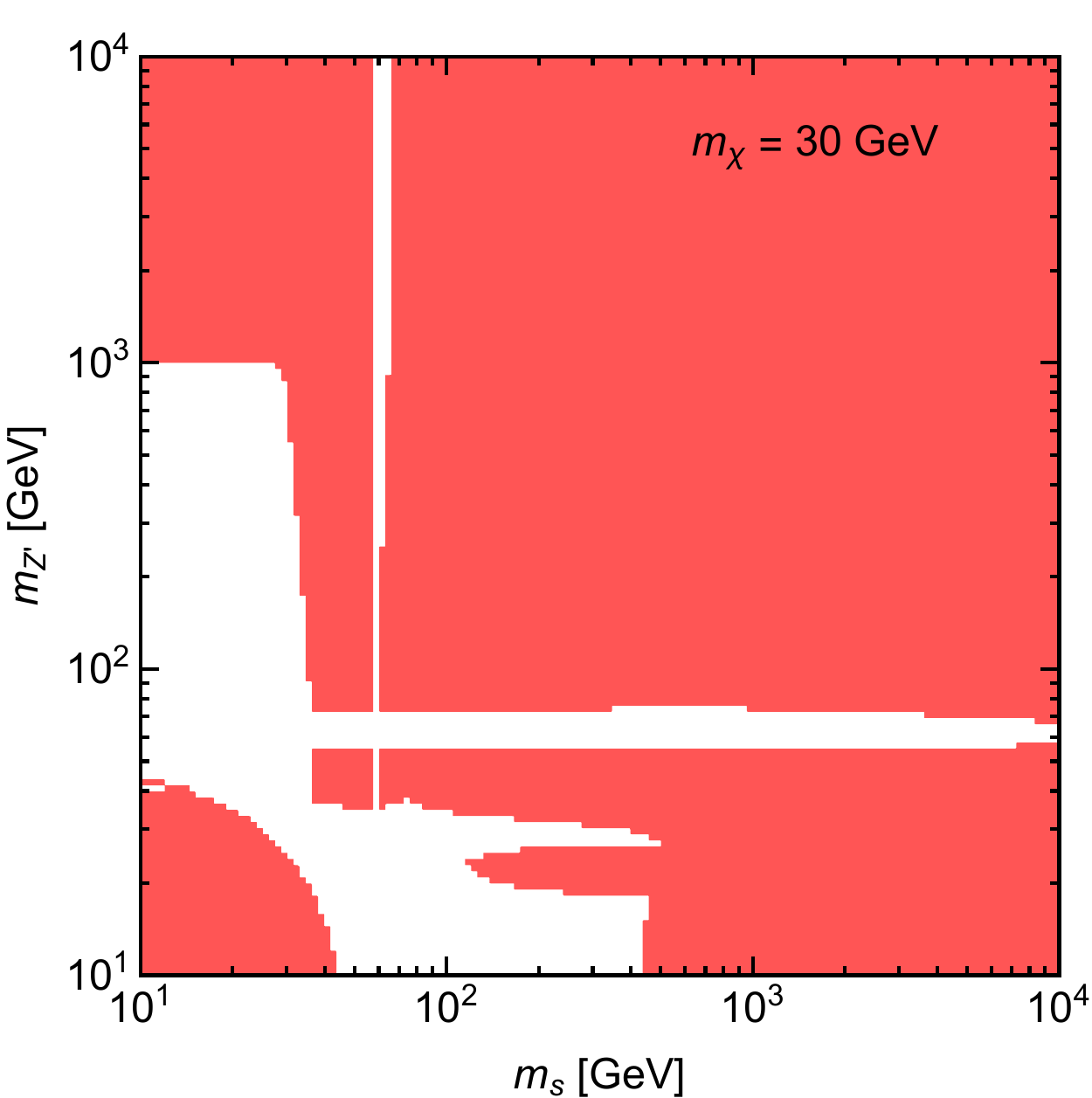}\qquad\includegraphics[height=0.25\textheight]{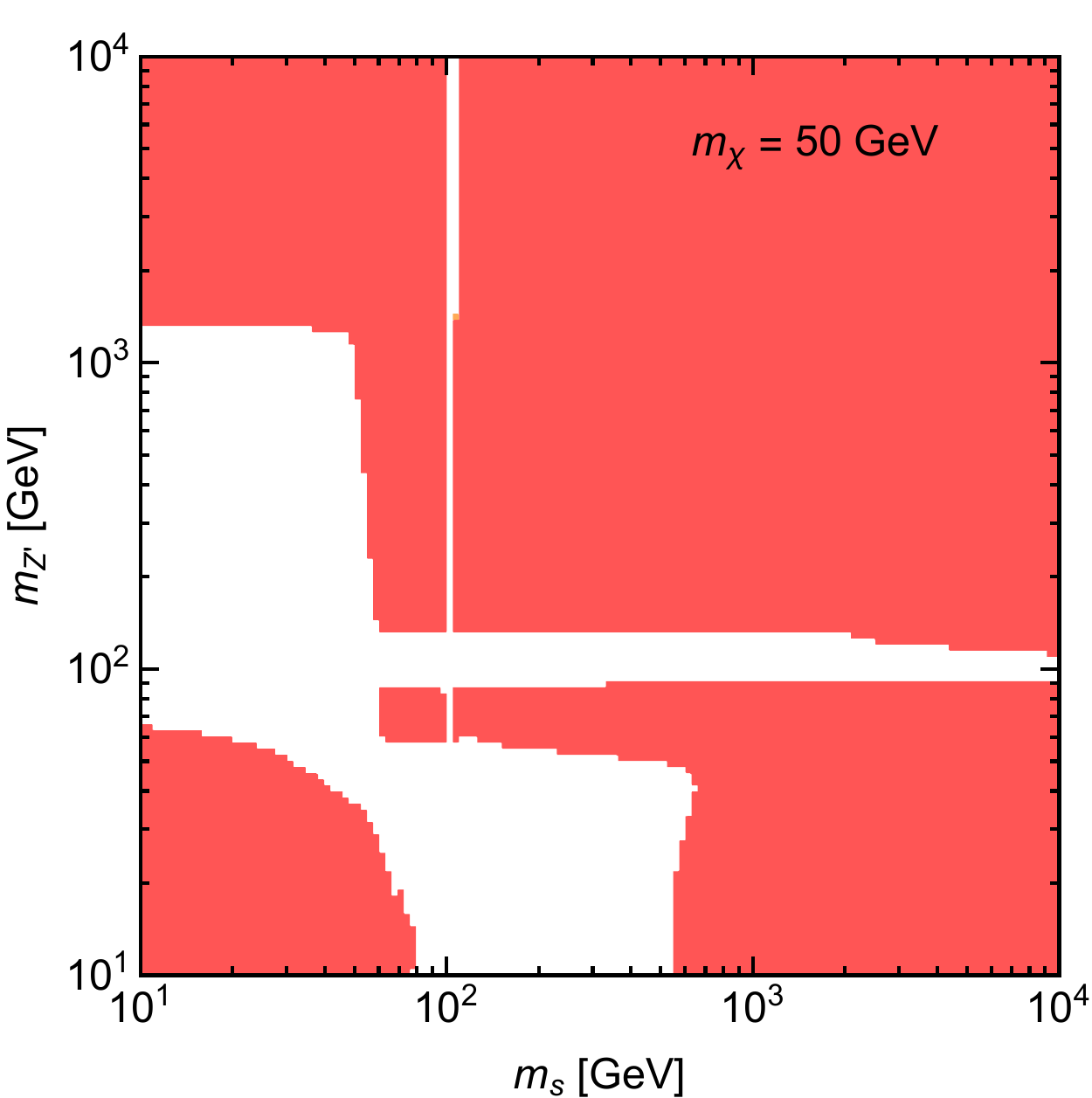}
\includegraphics[height=0.25\textheight]{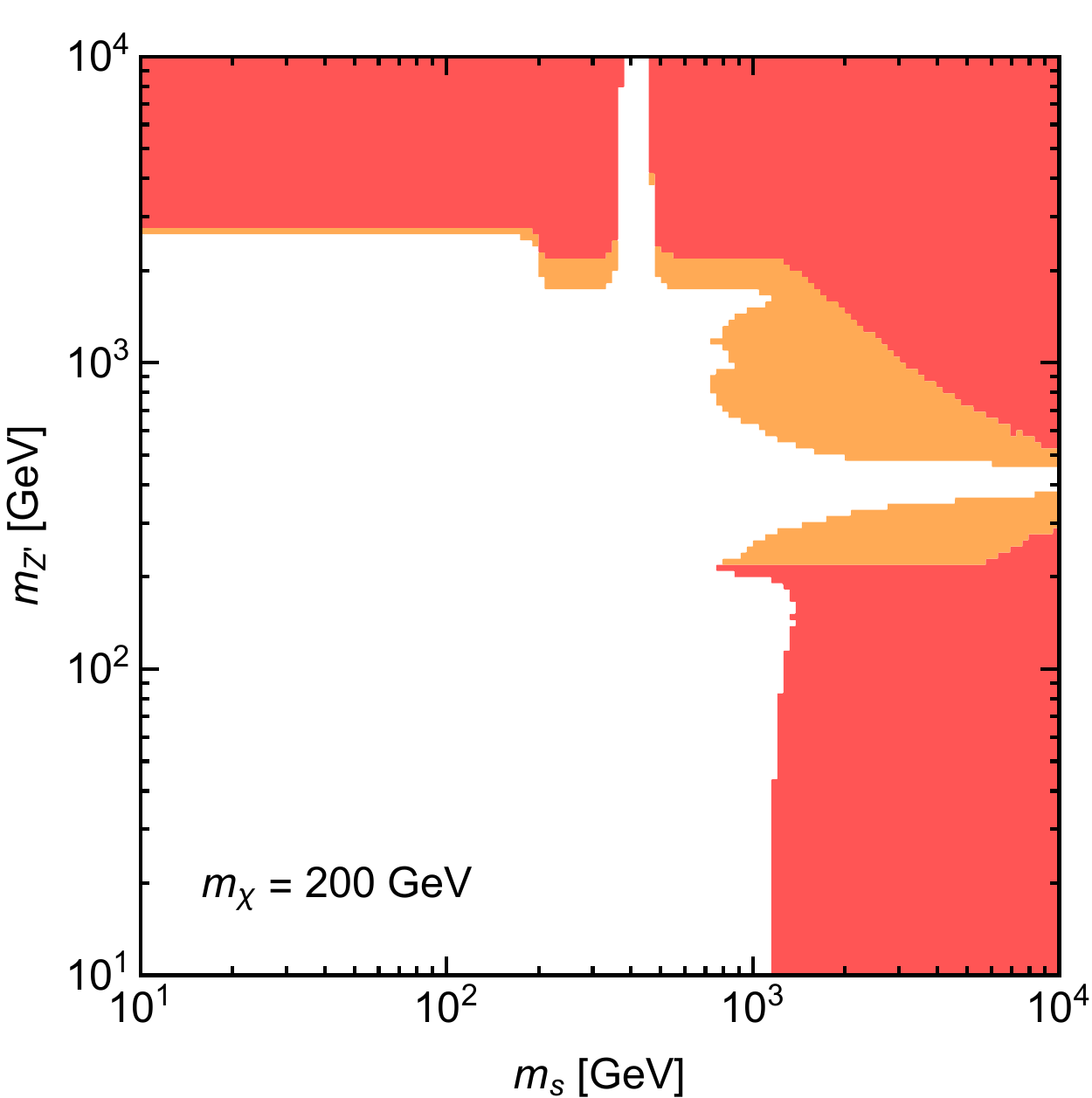}\qquad\includegraphics[height=0.25\textheight]{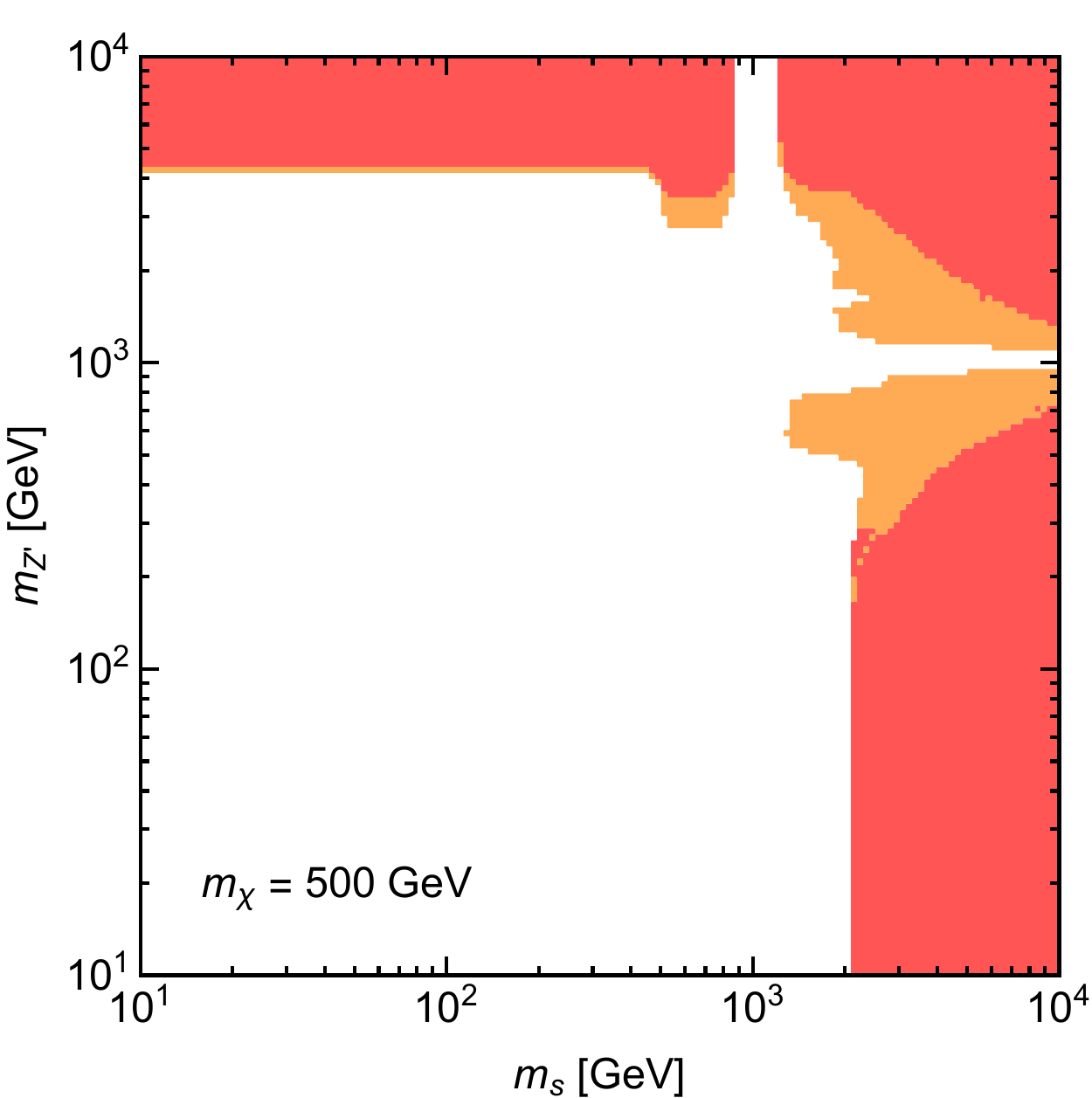}
\caption{Global scans for different values of $m_\chi$. The red shaded region is excluded for all possible combinations of couplings, while in the white region all constraints can be evaded. In the orange shaded region it is not possible to exclude large values of $g_q$, corresponding to $\Gamma_{Z'} / m_{Z'} > 0.3$.
}
\label{fig:globalscans}
\end{figure}

In figure~\ref{fig:globalscans} we present the results of the global scans for different values of $m_\chi$. We make qualitatively the same observations as in the case $m_\chi = \unit[100]{GeV}$, noting that small DM masses are already very tightly constrained by the data. We clearly observe that consistent points can only be obtained at the $s$-channel resonances or with at least one terminator. For $m_\chi < \unit[100]{GeV}$ and $m_s + m_{Z'} < 2 \, m_\chi$ indirect detection constraints from $\chi \chi \rightarrow Z' s$ give a constraint that cannot be avoided by varying $g_q$ and $\sin \theta$. For $m_\chi = \unit[30]{GeV}$ there is an additional constraint from $\chi \chi \rightarrow Z' Z'$, which is relevant for $m_s > m_\chi$ and $0.5\,m_\chi \lesssim m_{Z'} \lesssim 0.9\,m_\chi$.

For larger DM masses the inconclusive large-mediator-width regions become more important, but heavy mediators are still tightly constrained. On the other hand, the size of the parameter region with $m_{Z'}, \ m_s < m_\chi$ grows and indirect detection constraints are absent. The configuration with one or two dark terminators thus remains viable up to $m_\chi \sim \unit[50]{TeV}$, at which point thermal freeze-out becomes incompatible with perturbative unitarity.

\section{Conclusions and outlook}
\label{sec:conclusions}

\subsection{Summary of main assumptions}

Let us summarise the main assumptions of our analysis in the main text
(assumptions \ref{assump:q_coupl} and \ref{assump:kin_mix} are relaxed in appendix~\ref{app:mixing}):
\begin{enumerate}
\item We consider a Majorana DM particle which is a singlet under the SM gauge group but charged under a new $U(1)'$ gauge group.
\item The $U(1)'$ is broken by a dark Higgs which generates a mass for the DM as well as for the $U(1)'$ gauge boson $Z'$. Hence our model predicts two mediators between the SM and the DM. 
\item The dark Higgs mediates DM--SM interactions via mixing with the SM Higgs doublet. To obtain a $Z'$ mediated interaction we assume that quarks are charged under the $U(1)'$ as well.
\item \label{assump:q_coupl} We assume that left-handed and right-handed quarks have the same $U(1)'$ charge (vector-like coupling to $Z'$), that the quark-coupling is flavour universal, and that the charge of leptons is zero. Hence the $U(1)'$ corresponds to gauged baryon number.
\item \label{assump:kin_mix} We take into account only loop-induced kinetic mixing between the SM hypercharge gauge boson and the $Z'$, which implies that the kinetic mixing parameter is not an independent parameter of the model.
\item The DM-related phenomenology is completely described by the model. In particular, we require that the total observed abundance of DM is obtained by thermal freeze-out within this model, where all available annihilation channels are self-consistently determined within the adopted choices of interactions.   
\end{enumerate}

We denote the model by \TMDM. It is characterised by 6
parameters (3 masses and 3 couplings) as given in
table~\ref{tab:parameters}, and it can be considered as a joint
framework for simplified DM models with the simultaneous presence of a
scalar and a vector $s$-channel mediator (see
table~\ref{tab:illustration}).  The \TMDM\ model is gauge
invariant at the Lagrangian level and renormalisable, and we consider
only regions of the parameter space where perturbative unitarity is
guaranteed.  The stability of the DM particle is a consequence
  of the $U(1)'$ gauge symmetry.  In principle the model requires
additional fields in order to cancel the gauge anomalies. However, the
vectorial coupling of the $Z^\prime$ to quarks ensures that the new fermions are
not colour-charged and, therefore, the new particles are not expected
to have a substantial impact on the phenomenology that we consider. Including additional states to cancel anomalies would lead us to recover the models discussed in the context of Baryonic DM~\cite{Duerr:2013dza,Perez:2014qfa}.

\subsection{Main results}

We have analysed the \TMDM\ model under the assumptions specified above, taking
into account a large variety of constraints: direct searches (LUX) and
indirect searches (Fermi-LAT $\gamma$-ray observations of dwarf spheroidals), various observables at colliders (monojets, dijets,
dileptons), electroweak precision tests, and Higgs observables at the LHC.

We find that generally the WIMP hypothesis is under severe
pressure. Within the considered model there are basically only two
possibilities to obtain the correct relic abundance without running
into conflict with experimental constraints and/or perturbative
unitarity:
\begin{itemize}
\item either the DM and mediator masses are tuned close to an $s$-channel resonance, or
\item one or both mediators are lighter than the DM, such that DM annihilations into the dark sector control the relic abundance, while the coupling to the SM can be quite small. In this case, one of the mediators (or both) actually becomes a terminator.
\end{itemize}

The presence of a terminator makes it difficult to test this region of
the parameter space. In our
model, typically the annihilation cross section is $p$-wave and/or
helicity suppressed, which makes indirect detection signals from the
decay of the terminator into SM particles unobservable. However, there
is one exception, namely if the annihilation channels $\chi\chi\to
Z's$ or $\chi\chi\to Z'h$ become kinematically allowed. Those
processes are a generic feature of the two-mediator set-up, they appear
at $s$-wave, and typically dominate the DM annihilation in the present Universe.  Fermi-LAT
observations of dwarf spheroidals constrain this configuration for DM
masses $m_\chi \lesssim \unit[100]{GeV}$.

It is well known that direct detection strongly constrains a thermal
WIMP with an $s$-channel scalar mediator, e.g.,~\cite{LopezHonorez:2012kv}. The vector mediator considered in our
model has axial couplings to DM and vectorial couplings to quarks,
which leads to a velocity suppressed scattering cross section with
nuclei, and therefore bounds from direct detection are usually expected to be
much weaker. However, we have shown that current limits from LUX are
strong enough that despite the velocity suppressed interaction, direct
detection provides a relevant constraint on the parameter space of the
vector-mediator model, under the assumption of the correct thermal
freeze-out abundance.

\subsection{Outlook}

To conclude, let us discuss possible future directions. First of all, significant progress is expected for the near future in our knowledge about the SM-like Higgs boson. More precise measurements of the total signal strength may further constrain or in fact provide evidence for mixing between the observed Higgs boson and additional Higgs bosons. Similarly, relevant constraints are expected from searches for invisibly decaying Higgs bosons and from searches for non-standard Higgs decays such as $h \rightarrow 4b$ or $h \rightarrow 4\tau$. In the presence of a light $Z'$ terminator, it is also conceivable in our model to have a significant branching ratio for $h \rightarrow 4j$.

We furthermore believe that the results presented in the present paper motivate a new class of LHC searches, namely searches for a light dark Higgs. If the DM relic abundance is set via $\chi \chi \rightarrow s s$, the dark Higgs would by definition decay visibly (because $m_s < m_\chi$), with $b\bar{b}$ being the dominant decay mode in large regions of parameter space. The difference to conventional searches for additional light Higgs bosons (as conducted for example in the context of the NMSSM) is that the dark Higgs would generically be produced in association with large amounts of missing transverse energy, because the dominant production modes are dark Higgs Strahlung ($p p \rightarrow Z'^\ast \rightarrow Z' s \rightarrow \chi\chi s$) and final-state radiation of a dark Higgs ($p p \rightarrow Z' \rightarrow \chi \chi \rightarrow \chi \chi s$). In contrast to existing searches for heavy quarks and missing energy, on the other hand, the fact that both $b$-quarks result from the decay of an on-shell resonance means that the distinctive invariant mass distribution can be used to distinguish the signal from potential backgrounds.\footnote{Searches for dijet resonances in association with missing transverse energy have been proposed in various previous works~\cite{Autran:2015mfa,Buschmann:2015awa, Bai:2015nfa, Gupta:2015lfa, Buschmann:2016hkc}. Our proposal differs from these works in that the dark Higgs preferentially decays into heavy quarks.} A detailed study of the LHC sensitivity for this signature will be left for future work.

Another potential constraint not taken into account in the present work arises from DM capture in the Sun and the resulting neutrino signal in IceCube (see e.g.~\cite{Heisig:2015ira,Jacques:2016dqz}). However, since in the set-up we consider DM--nucleus scattering is dominantly spin-independent, we expect solar capture not to be competitive with direct detection experiments. Nevertheless, future progress in indirect detection experiments is crucial in order to further constrain the case of one (or several) dark terminators. In particular, CTA has the potential to substantially extend the sensitivity of indirect searches up to DM masses above the TeV range~\cite{Carr:2015hta}.
Furthermore, it will be interesting to see whether the Galactic centre excess can be attributed to unresolved point sources as recently suggested~\cite{Lee:2015fea,Bartels:2015aea}. If the possibility of a DM interpretation persists, \TMDM\ can easily accommodate such a signal via cascade annihilations. In particular, the observed relic abundance can be obtained with comparable contributions from $s$-wave and $p$-wave annihilation, so that the present-day annihilation cross section would be a factor of a few below the thermal cross section.

To conclude, taking the \TMDM\ model as a benchmark scenario,
  we have seen that the WIMP hypothesis of a thermal relic DM
  abundance from weak-scale physics is under severe pressure. While
  ``classic'' WIMP scenarios with mediators comparable or heavier
  than the DM particle are largely excluded, the \TMDM\ model is
  flexible enough to save the WIMP due to resonances or the presence of a terminator. Although
  our study is performed within the specific \TMDM\ model, we expect our conclusions to apply to a larger class of models where the DM particle is a SM gauge singlet. Examples for alternatives not
  directly covered by our results are DM models where DM is charged
  under the SM gauge group~\cite{ArkaniHamed:2006mb,Banerjee:2016hsk}
  or where co-annihilations are relevant~\cite{Baker:2015qna}.
  Generically we can conclude that if DM is a WIMP then the dark
  sector is most likely more complicated than just a weak-scale DM
  particle with effective interactions with the SM.

\acknowledgments

We thank Torsten Bringmann, Ulrich Haisch, Tilman Plehn and Pedro Schwaller for
discussions. This work is supported by the German Science Foundation
(DFG) under the Collaborative Research Center~(SFB) 676 Particles,
Strings and the Early Universe, the ERC Starting Grant `NewAve'
(638528), and the European Union’s Horizon 2020 research and
innovation programme under the Marie Sklodowska-Curie grant agreement
No~674896.

\begin{appendix}

\section{The model \label{app:Model}}

Assuming that the SM Higgs is uncharged under the $U(1)^\prime$ and that the dark Higgs is a SM singlet, the Lagrangian of the model is given by
\begin{equation}
 \mathcal{L} = \mathcal{L}_\text{SM} + \mathcal{L}_\chi + \mathcal{L}_\text{S} + \mathcal{L}_\text{gauge}^\prime, 
\end{equation}
where $\mathcal{L}_\text{SM}$ is the usual SM Lagrangian, and
\begin{align}
 \mathcal{L}_\chi &= \frac{i}{2} \bar{\chi} \slashed{\partial} \chi - \frac{1}{2} g^\prime q_\chi Z^{\prime \mu} \bar{\chi} \gamma^5 \gamma_\mu \chi - \frac{1}{2} y_\chi \bar{\chi} \left( P_L S + P_R S^\ast \right) \chi, \\
 \mathcal{L}_\text{S} &= \left[ \left( \partial^\mu + i g^\prime q_S Z^{\prime \mu} \right) S\right]^\dagger\left[ \left( \partial_\mu + i g^\prime q_S Z^\prime_\mu \right) S\right] \nonumber \\
      &\quad + \mu_s^2 S^\dagger S - \lambda_s \left( S^\dagger S \right)^2 - \lambda_{hs} \left( H^\dagger H \right) \left( S^\dagger S \right) , \\
  \mathcal{L}_\text{gauge}^\prime &= - g^\prime Z^{\prime \mu} \sum_{f=q} q_f \bar{f} \gamma_\mu f -\frac{1}{4} F^{\prime \mu \nu} F_{\mu\nu}^\prime - \frac{1}{2} \sin \epsilon F^{\prime \mu \nu} B_{\mu\nu} .
\end{align}
Here, $S$ denotes the complex dark Higgs, $H$ is the SM Higgs doublet,
and the gauge kinetic terms are defined as
\begin{align}
 F^{\prime \mu \nu} &= \partial^\mu Z^{\prime \nu} - \partial^\nu Z^{\prime \mu}, \\
 B^{\mu\nu} &= \partial^\mu B^\nu - \partial^\nu B^\mu, 
\end{align}
where $B^\mu$ is the SM $U(1)_Y$ gauge field. 

After spontaneously breaking the $U(1)^\prime$ and the electroweak symmetries we go to the unitary gauge, where
\begin{align}
 S &= \frac{1}{\sqrt{2}} \left( s + w \right) \,, \\
 H &= \frac{1}{\sqrt{2}} \left(0, \ h + v \right)^T \, .
\end{align}
Here $v = \unit[246]{GeV}$ is the SM Higgs vev.
Then, we obtain
\begin{align}
 \mathcal{L}_\chi &= \frac{i}{2} \bar{\chi} \slashed{\partial} \chi - \frac{1}{2} g^\prime q_\chi Z^{\prime \mu} \bar{\chi} \gamma^5 \gamma_\mu \chi -\frac{1}{2} \frac{y_\chi w}{\sqrt{2}} \bar{\chi} \chi - \frac{y_\chi}{2 \sqrt{2}} s \bar{\chi} \chi , \\
\mathcal{L}_\text{S} &=  \frac{1}{2} g^{\prime 2} q_S^2 w^2 Z^{\prime \mu} Z^\prime_\mu + \frac{1}{2} \partial^\mu s \partial_\mu s + \frac{1}{2} g^{\prime 2} q_S^2 Z^{\prime \mu} Z^\prime_\mu \left( s^2 + 2 s w \right) \nonumber \\
&\quad + \frac{\mu_s^2}{2} (s+w)^2 - \frac{\lambda_s}{4} (s+w)^4 - \frac{\lambda_{hs}}{4}(s+w)^2 (h+v)^2 .
 \end{align}
Using the phase freedom for $S$ and $\chi$ we can choose both $w$ and $y_\chi$ real without loss of generality, which ensures a real mass for $\chi$ and a pure scalar coupling of $s$; since $w$ is the only source of $U(1)'$ symmetry breaking we cannot obtain a pseudo-scalar coupling of $s$ to $\chi$~\cite{LopezHonorez:2012kv}.

The model has the following independent new parameters in the Lagrangian:
\begin{equation}
 q_f, \ q_\chi, \  y_\chi, \ w, \ \lambda_s, \ \lambda_{hs}, \ \epsilon, 
\end{equation}
since the dark Higgs can only couple to the DM (and eventually give it a mass) if 
  $q_S = - 2 q_\chi$, 
and the minimisation conditions of the scalar potential enforce the relation
\begin{equation}\label{eq:mus}
 \mu_s^2 = \lambda_s w^2 + \frac{1}{2} \lambda_{hs} v^2 \, .
\end{equation}
Additionally, note that the $g^\prime$ gauge coupling only appears in connection with the charges of the fermion fields, which is expected since the normalisation of an Abelian force is not well defined. Thus, freedom in $g^\prime$ can be hidden in the fermion charges, and $g^\prime$ should not be taken as an independent parameter.

We can define\footnote{Note that $\lambda_h$ is one of the three parameters in the SM Higgs sector, the other two being $\mu_h$ and $v$. Using the minimisation condition of the Higgs potential we can eliminate one of them, and the two physical parameters $m_h = \unit[125]{GeV}$ and $v = \unit[246]{GeV}$ are fixed.}
\begin{align}
 m_\chi &\equiv \frac{1}{\sqrt{2}} y_\chi w , \\
 m_{Z^\prime} &\equiv 2 g^\prime q_\chi w, \\
 g_\chi &\equiv g^\prime q_\chi, \\
 g_f &\equiv g^\prime q_f, \\
 \tan 2 \theta &\equiv \frac{\lambda_{hs} v w}{\lambda_h v^2 - \lambda_s w^2 } ,
\end{align}
and describe the mixing in the Higgs sector as
\begin{align}
 H_1 &= h \cos \theta + s \sin \theta , \nonumber \\
 H_2 &= s \cos \theta - h \sin \theta ,
\end{align}
where $\theta$ is required to lie between $-\pi/4$ and $\pi/4$, such that by definition $H_1$ is the mostly SM-like state. Correspondingly, we denote the mass of $H_1$ as $m_h$ and the mass of $H_2$ as $m_s$. These are given by
\begin{align}
m_h^2 & = v^2 \, \lambda_h + w^2 \, \lambda_s + (v^2 \, \lambda_h - w^2 \, \lambda_s) \sqrt{1 + \frac{v^2 \, w^2 \, \lambda_{hs}^2}{(v^2 \, \lambda_h - w^2 \, \lambda_s)^2}}, \\
m_s^2 & = v^2 \, \lambda_h + w^2 \, \lambda_s - (v^2 \, \lambda_h - w^2 \, \lambda_s) \sqrt{1 + \frac{v^2 \, w^2 \, \lambda_{hs}^2}{(v^2 \, \lambda_h - w^2 \, \lambda_s)^2}} \; .
\end{align}
Note that these expressions allow for both $m_h < m_s$ (if $v^2 \, \lambda_h - w^2 \, \lambda_s < 0$) and $m_h > m_s$ (if $v^2 \, \lambda_h - w^2 \, \lambda_s > 0$).

Then, the free parameters of the model can be taken as
\begin{equation}\label{eq:freeParameters}
 g_\chi, \ g_f, \ m_\chi, \ m_{Z^\prime}, \ m_s , \ \theta, \ \epsilon ,
\end{equation}
and the relevant parts of the Lagrangian are given in terms of these parameters by
\begin{align}
 \mathcal{L}_\chi &= \frac{i}{2} \bar{\chi} \slashed{\partial} \chi - \frac{1}{2} g_\chi Z^{\prime \mu} \bar{\chi} \gamma^5 \gamma_\mu \chi -\frac{1}{2} m_\chi \bar{\chi} \chi - \frac{m_\chi g_\chi}{m_{Z^\prime}} s \bar{\chi} \chi , \\
\mathcal{L}_\text{S} &=  \frac{1}{2} m_{Z^\prime}^2 Z^{\prime \mu} Z^\prime_\mu + \frac{1}{2} \partial^\mu s \partial_\mu s + 2  g_\chi^2 Z^{\prime \mu} Z^\prime_\mu \left( s^2 + \frac{m_{Z^\prime}}{g_\chi} s  \right) \nonumber \\
&\quad + \frac{\mu_s^2}{2} \left(s+\frac{m_{Z^\prime}}{2 g_\chi}\right)^2 - \frac{\lambda_s}{4} \left(s+\frac{m_{Z^\prime}}{2 g_\chi}\right)^4 + \frac{\lambda_{hs}}{4} \left(s+\frac{m_{Z^\prime}}{2 g_\chi}\right)^2 \left( h + v \right)^2, \\
  \mathcal{L}_\text{gauge}^\prime &= -  Z^{\prime \mu} \sum_{f=q} g_f \bar{f} \gamma_\mu f -\frac{1}{4} F^{\prime \mu \nu} F_{\mu\nu}^\prime - \frac{1}{2} \sin \epsilon F^{\prime \mu \nu} B_{\mu\nu}, 
 \end{align}
where the parameters from the Higgs potential are given in terms of the free parameters in eq.~\eqref{eq:freeParameters} and $m_h$ and $v$ as
\begin{align}
 \lambda_h &= \frac{1}{4 v^2} \left[ m_h^2 + m_s^2 + \left( m_h^2 - m_s^2 \right)\cos 2\theta \right],  \label{eq:lambdah} \\
 \lambda_s &= \frac{g_\chi^2}{m_{Z^\prime}^2} \left[ m_h^2 + m_s^2 + \left( m_s^2 - m_h^2 \right)\cos 2\theta \right], \label{eq:lambdas} \\
  \lambda_{hs} &= \frac{g_\chi}{m_{Z^\prime} v} \left( m_h^2 - m_s^2 \right) \sin 2 \theta \; . \label{eq:lambdahs} 
\end{align}
To derive these equations, we have used eq.~\eqref{eq:mus} with $w$ replaced appropriately.

If we allow for a  $U(1)^\prime$ charge for the SM Higgs, the term
\begin{align}
\mathcal{L}_\text{H}^\prime = \left[ (D^\mu H)^\dagger (-i \, g' \, q_H \, Z'_\mu \, H) + \text{h.c.} \right] +
  g'^2 \, q_H^2 \, Z'^\mu Z'_\mu \, H^\dagger H
\end{align}
appears in the Lagrangian of the model. Once the SM Higgs develops a vev, this interaction leads to mass mixing between the SM $Z$ and the $Z^\prime$ gauge bosons: 
\begin{align}
\mathcal{L}_{\text{mixing}} = \delta m^2 Z^\mu Z^\prime_\mu = \frac{1}{2}\frac{e g^\prime q_H}{s_\text{W} c_\text{W}}v^2  Z^\mu Z^\prime_\mu\,,
\end{align}
where $s_\text{W} = \sin \theta_\text{W}$ and $c_\text{W} = \cos \theta_\text{W}$ with the weak mixing angle $\theta_\text{W}$.
Furthermore, the invariance of the SM Yukawa terms under  $SU(2)_\text{L} \times U(1)_Y \times U(1)^\prime$ implies axial interactions of the $Z'$ with the SM fermions,
\begin{align}
\mathcal{L}_{\text{axial}} = - g^A_f Z^{\mu \prime} \bar{f} \gamma^5 \gamma^\mu f \,.  
\end{align}
It should be noted that the interaction strength $g^A_f$ is not a free parameter here and can be related to the Higgs charge by $g^A_f = \frac{1}{2} g^\prime q_\text{H}$. This implies in particular that $g^A_f$ has to be independent of flavour and equal for quarks and leptons. Therefore, searches for dilepton resonances strongly constrain this option (see appendix~\ref{app:mixing}).

\section{Gauge boson mixing}
\label{app:U(1)mixing}

Our treatment of $Z$--$Z^\prime$ mixing follows closely the discussion outlined in~\cite{Kahlhoefer:2015bea} and, therefore, we limit ourselves to a schematic description in the following. In its most general form  the gauge part of the Lagrangian presented in the previous section can be written as
\begin{align}
  {\cal L} =& \; {\cal L}_\text{SM}
  -\frac{1}{4}{\hat{F}}^{\prime \mu\nu}\hat{F}'_{\mu\nu} + {\frac{1}{2}} m_{\hat
    Z'}^2 \hat{X}_\mu \hat{X}^\mu - {\frac{1}{2}} \sin \epsilon\, \hat{B}_{\mu\nu} {\hat{F}}^{\prime\mu\nu} +\delta
  m^2 \hat{Z}_\mu \hat{X}^\mu \;,
\label{eq:Lappendix}
\end{align}
where we use hats to denote non-normalised fields and define $\hat{Z}\equiv \hat{c}_\mathrm{W} \hat{W}^3- \hat{s}_\mathrm{W} \hat{B}$ and ${\hat{F}}^{\prime\mu \nu} \equiv \partial^\mu \hat{X}^\nu - \partial^\nu \hat{X}^\mu$. 
First, the kinetic term can be brought to the canonical form by the transformation 
\begin{align}
\label{eq:Zpmixing}
\left(\begin{array}{c} \hat B_\mu \\ \hat W_\mu^3 \\ 
\hat{X}_\mu \end{array}\right) & = \left(\begin{array}{ccc} 1 & 0 &
    -t_\epsilon \\ 0 & 1 & 0 \\ 0 & 0 &
    1/c_\epsilon \end{array}\right)
\left(\begin{array}{c} B_\mu \\ W_\mu^3 \\ X_\mu \end{array}\right) \ ,
\end{align}
which generates an  additional contribution to the off-diagonal mass term.  
In a second step the mass terms are diagonalised by  the rotation 
\begin{align}
\left(\begin{array}{c} B_\mu \\ W_\mu^3 \\ X_\mu \end{array}\right) &
= \left(\begin{array}{ccc}
    \hat c_\mathrm{W} & -\hat s_\mathrm{W} c_\xi &  \hat s_\mathrm{W} s_\xi \\
    \hat s_\mathrm{W} & \hat c_\mathrm{W} c_\xi & - \hat c_\mathrm{W} s_\xi \\
    0 & s_\xi & c_\xi
\end{array} \right) 
\left(\begin{array}{c} A_\mu \\ Z_\mu \\ Z'_\mu \end{array}\right)
 \; \ ,
\end{align}
with the rotation angle $\xi$ given by
\begin{align}
  t_{2\xi}=\frac{-2c_\epsilon(\delta m^2+m_{\hat Z}^2 \hat
    s_\mathrm{W} s_\epsilon)} {m_{\hat Z'}^2-m_{\hat
      Z}^2 c_\epsilon^2 +m_{\hat Z}^2\hat s_\mathrm{W}^2 s_\epsilon^2
    +2\,\delta m^2\,\hat s_\mathrm{W} s_\epsilon} \; .
\label{eq:xi}
\end{align}
As a consequence of the mixing, the physical mixing angle $s_\mathrm{W}$ differs from the fundamental mixing angle $\hat s_\mathrm{W}$. This difference, however, is only relevant at higher orders in the mixing parameters and will therefore be neglected in the following.

Both kinetic and mass mixing can have a profound impact on the phenomenology  of the model.
The field redefinition induces corrections to the properties of the SM gauge bosons. These deviations from the SM expectation can be parametrized by the electroweak precision observables $S$ and $T$ and, expanding in $\xi$ and $\epsilon$, we find to leading order~\cite{Frandsen:2011cg}
\begin{align}
  \alpha \, S = & 4 \, c_\mathrm{W}^2 \, s_\mathrm{W} \, \xi \, (\epsilon - s_\mathrm{W} \, \xi) \; , \nonumber \\
  \alpha \, T = & \xi^2 \left(\frac{m_{Z'}^2}{m_{Z}^2}-2\right) + 2 \, s_\mathrm{W} \, \xi \, \epsilon \; , \label{eq:ST}
\end{align}
where  $\alpha=e^2/4\pi$. For small values of $m_{Z'}$, it is more appropriate to use the $\rho$ parameter, $\rho \equiv m_W^2 / (m_Z^2 c_\mathrm{W}^2)$, which in the presence of mixing will deviate from unity:
\begin{equation}
 \rho - 1 = \frac{c_\mathrm{W}^2 \, \xi^2}{c_\mathrm{W}^2 - s_\mathrm{W}^2} \left(\frac{m_{Z'}^2}{m_{Z}^2}-1\right) \; .
\end{equation}

Furthermore, due to the mixing with the $Z$, the $Z^\prime$  acquires couplings to all SM particles. From a phenomenological point of view the most interesting of these are the couplings to leptons which can lead to observable signals in searches for dilepton resonances at colliders. If the tree level coupling to leptons is zero, the induced coupling between the new  gauge boson and  leptons is given by 
\begin{align}
  g_{\ell}^\mathrm{V} &= \frac{1}{4}(3 \, g' (s_\mathrm{W} s_\xi-c_\xi t_\epsilon)- g \, c_\mathrm{W} \, s_\xi) \ , & g_{\ell}^\mathrm{A} &= -\frac{1}{4}
  (g' (s_\mathrm{W} \, s_\xi - c_\xi \, t_\epsilon) + g \, c_\mathrm{W} s_\xi) \; ,
\label{eq:dilepton}
\end{align}
where $g$ and $g'$ are the gauge couplings of 
$SU(2)_\text{L}$ and $U(1)_Y$.

\section{Relic density from mixing}
\label{app:mixing}

In this appendix we consider two additional ways in which the $Z'$ can couple to SM states: tree-level kinetic mixing and mass mixing. The latter case is most easily realised if the SM Higgs
carries a charge under the new $U(1)'$. This also implies that the $Z^\prime$ has couplings to SM fermions and, as a result, the axial coupling is directly related to the Higgs charge by $g^A_f = -2 g' q_H$ due to gauge invariance. The $Z^\prime$ vector couplings to fermions, on the other hand, are not uniquely fixed in this model. In the following we
will set them to the minimal value in agreement with gauge invariance, $g^V_f=g^A_f$, see~\cite{Kahlhoefer:2015bea} for a detailed discussion of this choice.    

In principle, either of the two mixing scenarios could conceivably generate large enough couplings of the $Z'$ to the SM to provide the necessary annihilation channels in order to reproduce the observed relic abundance. However, gauge boson mixing is strongly constrained by electroweak precision observables. The contribution of mixing to EWPT is given by eq.~\eqref{eq:ST}. Moreover, both mixing scenarios imply that the $Z'$ couples to leptons either through mixing or directly and, therefore, it can be constrained by searches for dilepton resonances. In the following we use the same experimental constraints as described in section~\ref{sec:vector-collider}.

\begin{figure}[tb]
\centering
\includegraphics[height=0.3\textheight]{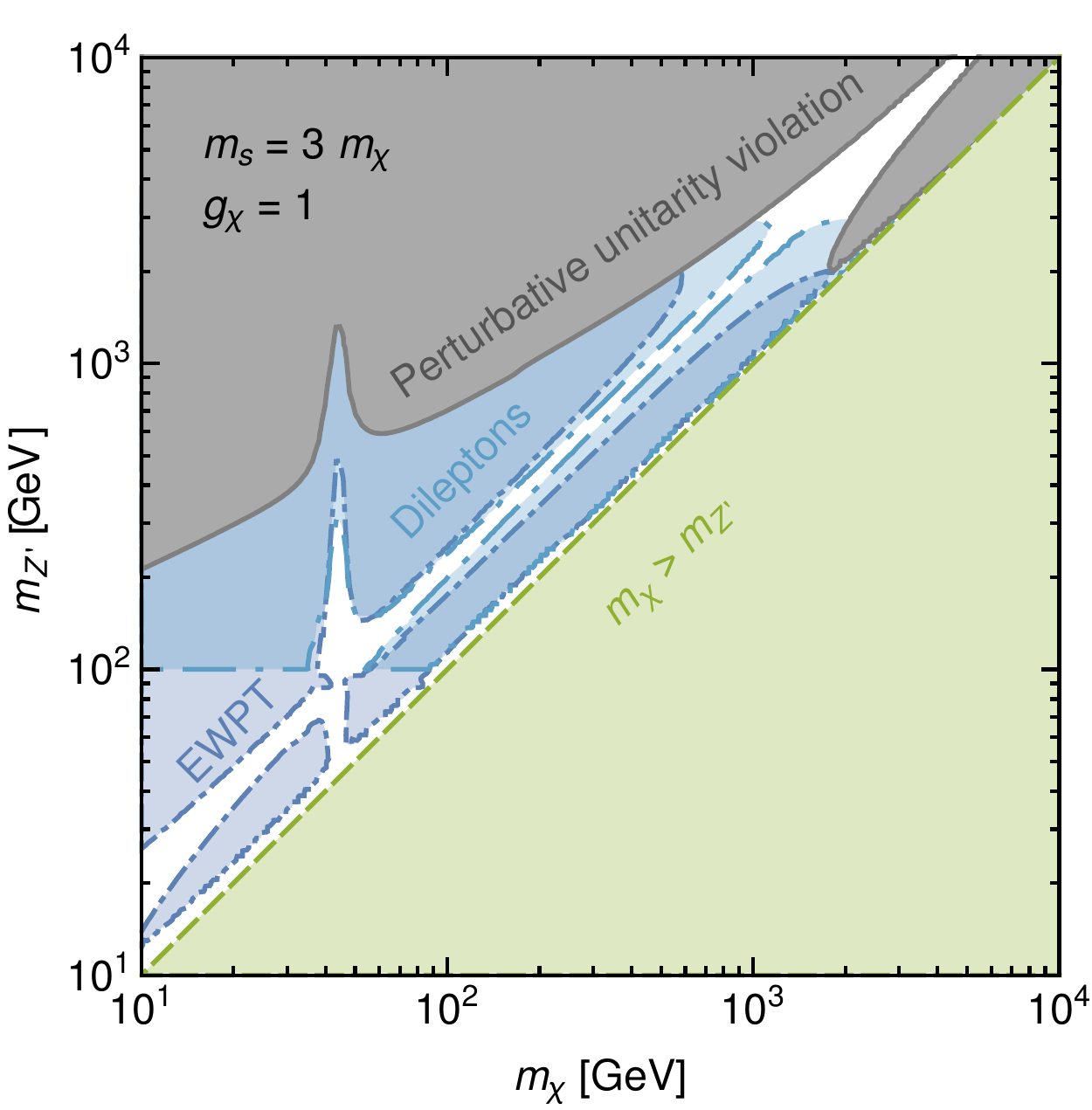}\qquad
\includegraphics[height=0.3\textheight]{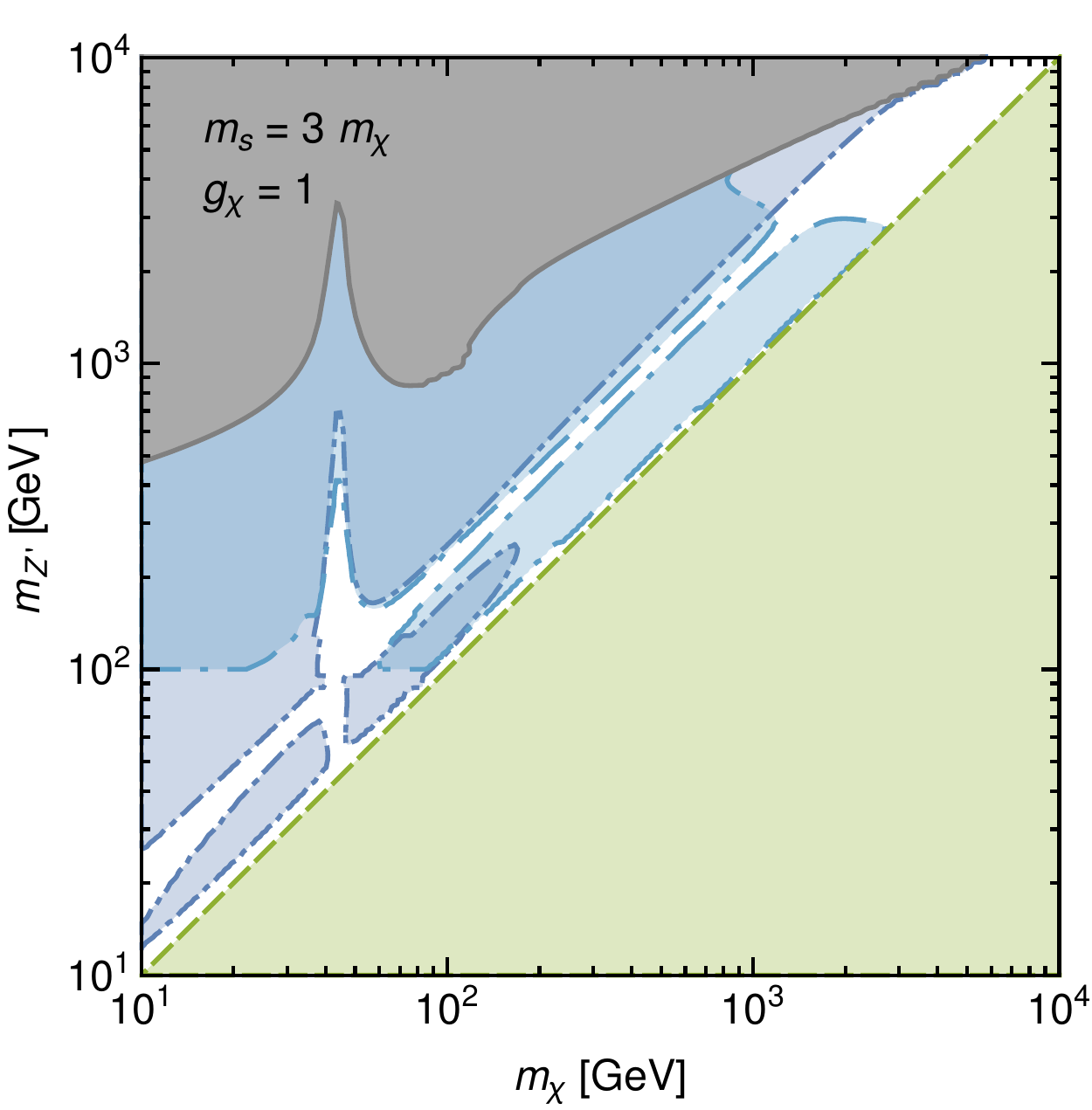}
\caption{Bounds on scenarios with kinetic mixing (left) and axial couplings (right). We fix the dark-sector coupling $g_\chi = 1$ and vary $\epsilon$ (left) or $g_q^A$ (right) in order to obtain the correct relic density. The grey shaded region (solid line) corresponds to $\epsilon > 1$ ($g^A_q > 1$). The light blue (long dash-dotted) and dark blue (short dash-dotted) shaded regions are excluded by searches for dilepton resonances and EWPT, respectively. The green shaded region (dashed) corresponds to $m_\chi > m_{Z'}$. In this parameter region, the relic abundance depends only on $g_\chi$ and not on the mixing parameter, which can therefore be arbitrarily small.}
\label{fig:Zmixing}
\end{figure} 

The bounds discussed above can be compared to the coupling
strength required by the relic density. The results of this analysis
are shown in figure~\ref{fig:Zmixing}. For both panels in the figure we have conservatively  set $g_\chi = 1$ in order to reduce the necessary interaction strength with the SM and fix  $m_s = 3 \, m_\chi$ to remove the contribution of the dark Higgs to the relic density.
  The two panels correspond to varying kinetic mixing and varying
  Higgs charge, respectively, such that the correct relic abundance
  is obtained.  Since we fix the dark sector coupling
to a relatively large value, we generally predict DM underproduction
in the parameter region where $m_\chi > m_{Z'}$ and the process $\chi
\chi \rightarrow Z' Z'$ is allowed. We shade this region in green,
since mixing plays a sub-dominant role here.

As can be seen in the left panel of Fig.~\ref{fig:Zmixing} the values of $\epsilon$ which are necessary to account for $\Omega_\text{DM}$ are excluded by EWPT and dilepton searches in most of the remaining parameter space. Nevertheless, kinetic mixing could  explain thermal DM if the annihilation cross section receives a resonant enhancement from the $Z$ or $Z'$. Similarly, we find that processes induced by mass mixing (i.e.\ axial couplings) cannot account for the observed relic density once EWPT and dilepton searches are taken into account unless there is a resonant enhancement of the annihilation rate.

\section{Annihilation cross sections}
\label{app:annihilation}

\subsection{Standard model final states}

Annihilation into quarks via the $Z'$ is given by
\begin{equation}
\sigma v_\chi (\chi \chi \rightarrow Z' \rightarrow q \bar{q}) = \frac{3\,g_\chi^2\, g_q^2}{12\pi} \frac{(m_q^2 + 2 \, m_\chi^2)(1 - m_q^2 / m_\chi^2)^{1/2}}{(4 \, m_\chi^2 - m_{Z'}^2)^2} v_\chi^2 \; ,
\end{equation}
whereas annihilation into quarks via the dark Higgs is given by
\begin{equation}
\sigma v_\chi (\chi \chi \rightarrow s \rightarrow q \bar{q}) =  \frac{3\,g_\chi^2\,\sin^2 \theta \, \cos^2\theta}{2\pi} \frac{m_q^2 \, m_\chi^4}{m_{Z'}^2 \, v^2}\frac{(1 - m_q^2 / m_\chi^2)^{3/2}}{(4 m_\chi^2 - m_s^2)^2} v_\chi^2 \; .
\end{equation}
The same expression (with $m_s$ replaced by $m_h$) is obtained for annihilation via the SM Higgs. For $m_s$ comparable to $m_h$, the interference between the two diagrams becomes important. Due to the different sign in the mixing matrix, the interference is destructive. The total annihilation cross section is therefore proportional to $(m_s^2 - m_h^2)/[(4 m_\chi^2 - m_h^2)^2 (4 m_\chi^2 - m_s^2)^2]$.

The SM Higgs (as well as the dark Higgs) also mediates additional annihilation processes, most notably annihilation into SM gauge bosons and the SM Higgs:
\begin{align}
\sigma v_\chi (\chi \chi \rightarrow s \rightarrow W^+ W^-) = &  \frac{g_\chi^2\,\sin^2 \theta \, \cos^2\theta}{4\pi} \frac{m_\chi^2 (3 m_W^4 - 4 m_W^2 \, m_\chi^2 + 4 m_\chi^4)}{m_{Z'}^2 \, v^2} \nonumber \\ & \times \frac{\sqrt{1 - m_W^2 / m_\chi^2}}{(4 m_\chi^2 - m_s^2)^2} v_\chi^2 \;.
\end{align}
These processes are enhanced relative to annihilation into quarks by a factor $m_\chi^2 / m_q^2$ and therefore become dominant for large DM mass.

\subsection{Dark terminators}

For the process $\chi \chi \rightarrow s s$, we are only interested in the limit $m_s \ll m_\chi$. We then obtain
\begin{align}
\sigma v_\chi (\chi \chi \rightarrow ss) \simeq  \frac{3 g_\chi^4 m_\chi^2}{8\pi m_{Z'}^4} v_\chi^2= \frac{3 y_{\chi}^4}{ 512 \pi m_{\chi}^2} v_\chi^2\; .
\end{align}

In the limit $\theta \rightarrow 0$ and for small $m_{Z'}$ the annihilation cross section for $\chi \chi \rightarrow Z' Z'$ is given by
\begin{align}
\sigma v_\chi (\chi \chi \rightarrow Z' Z') &\simeq  \frac{g_\chi^4}{16 \pi m_\chi^2} + \frac{(m_s^4 + 8 m_\chi^4) \, g_\chi^4}{12 \pi \, m_\chi^2 (m_s^2 - 4 m_\chi^2)^2} \frac{m_\chi^4}{m_{Z^\prime}^4} v_\chi^2 \nonumber \\ 
&= \frac{m_{Z'}^4 \, y_\chi^4}{1024 \pi \, m_\chi^6} + \frac{(m_s^4 + 8 m_\chi^4) \, y_\chi^4}{768 \pi \, m_\chi^2 (m_s^2 - 4 m_\chi^2)^2} v_\chi^2 \; ,
\end{align}
where for the term proportional to $v_\chi^2$ we only retain the leading $p$-wave contribution; additional terms are relevant if $m_{Z^\prime}$ and $m_\chi$ are comparable. As can be seen the $p$-wave contribution is enhance by a factor $m_\chi^4 / m_{Z^\prime}^4$ relative to the $s$-wave part.  The reason for this behaviour is that the dominant contribution to the $p$-wave channel results from annihilation into pairs of longitudinal $Z^\prime$ bosons. These are the Goldstone modes and couple in the same way as the dark Higgs, so that the annihilation cross section is proportional to $y_\chi^4$ or, equivalently, to $g_\chi^4 \, m_\chi^4 / m_{Z^\prime}^4$.

Finally the annihilation cross section for $\chi \chi \rightarrow Z' s$ is given by
\begin{equation}
\sigma v_\chi (\chi \chi \rightarrow sZ') = \frac{g_\chi^4}{64 \pi} \frac{(m_s^4 + (m_{Z'}^2 - 4 m_\chi^2 )^2 - 2 m_s^2 (4 m_\chi^2 + m_{Z'}^2))^{3/2}}{m_\chi^4 m_{Z'}^4} \; .
\end{equation}
In the limit that both the $Z'$ and the dark Higgs are light compared to the DM particle, this expression reduces to eq.~\eqref{eq:cc-to-sZp}.
When kinematically accessible this processes typically dominates over  
$\chi \chi \rightarrow s s $ and $\chi \chi \rightarrow Z^\prime Z^\prime$ 
due to their velocity suppression and  smaller prefactors.

\end{appendix}

\providecommand{\href}[2]{#2}\begingroup\raggedright\endgroup


\begin{thebibliography}{10}

\bibitem{Frandsen:2012rk}
M.~T. Frandsen, F.~Kahlhoefer, A.~Preston, S.~Sarkar, and K.~Schmidt-Hoberg,
  \href{http://dx.doi.org/10.1007/JHEP07(2012)123}{{\it {LHC and Tevatron
  Bounds on the Dark Matter Direct Detection Cross-Section for Vector
  Mediators}}, } {\em JHEP} {\bf 07} (2012) 123,
  [\href{http://arxiv.org/abs/1204.3839}{{\tt 1204.3839}}].

\bibitem{Arcadi:2013qia}
G.~Arcadi, Y.~Mambrini, M.~H.~G. Tytgat, and B.~Zaldivar,
  \href{http://dx.doi.org/10.1007/JHEP03(2014)134}{{\it {Invisible Z' and Dark
  Matter: LHC vs LUX Constraints}}, } {\em JHEP} {\bf 03} (2014) 134,
  [\href{http://arxiv.org/abs/1401.0221}{{\tt 1401.0221}}].

\bibitem{Garny:2014waa}
M.~Garny, A.~Ibarra, S.~Rydbeck, and S.~Vogl,
  \href{http://dx.doi.org/10.1007/JHEP06(2014)169}{{\it {Majorana Dark Matter
  with a Coloured Mediator: Collider vs Direct and Indirect Searches}}, } {\em
  JHEP} {\bf 06} (2014) 169, [\href{http://arxiv.org/abs/1403.4634}{{\tt
  1403.4634}}].

\bibitem{Chala:2015ama}
M.~Chala, F.~Kahlhoefer, M.~McCullough, G.~Nardini, and K.~Schmidt-Hoberg,
  \href{http://dx.doi.org/10.1007/JHEP07(2015)089}{{\it {Constraining Dark
  Sectors with Monojets and Dijets}}, } {\em JHEP} {\bf 07} (2015) 089,
  [\href{http://arxiv.org/abs/1503.05916}{{\tt 1503.05916}}].

\bibitem{Fairbairn:2016iuf}
M.~Fairbairn, J.~Heal, F.~Kahlhoefer, and P.~Tunney, \href{http://dx.doi.org/10.1007/JHEP09(2016)018}{{\it {Constraints on Z'
  models from LHC dijet searches}}, } {\em JHEP} {\bf 09} (2016) 018, 
  [\href{http://arxiv.org/abs/1605.07940}{{\tt 1605.07940}}].

\bibitem{Jacques:2016dqz}
T.~Jacques, A.~Katz, E.~Morgante, D.~Racco, M.~Rameez, et~al., {\it
  {Complementarity of DM Searches in a Consistent Simplified Model: the Case of
  Z'}},  \href{http://arxiv.org/abs/1605.06513}{{\tt 1605.06513}}.

\bibitem{Bell:2016fqf}
N.~F. Bell, Y.~Cai, and R.~K. Leane, \href{http://dx.doi.org/10.1088/1475-7516/2016/08/001}{{\it {Dark Forces in the Sky: Signals from
  Z' and the Dark Higgs}}, } {\em JCAP} {\bf 08} (2016) 001, [\href{http://arxiv.org/abs/1605.09382}{{\tt
  1605.09382}}].

\bibitem{Alves:2013tqa}
A.~Alves, S.~Profumo, and F.~S. Queiroz,
  \href{http://dx.doi.org/10.1007/JHEP04(2014)063}{{\it {The dark $Z^{'}$
  portal: direct, indirect and collider searches}}, } {\em JHEP} {\bf 04}
  (2014) 063, [\href{http://arxiv.org/abs/1312.5281}{{\tt 1312.5281}}].

\bibitem{Alves:2015mua}
A.~Alves, A.~Berlin, S.~Profumo, and F.~S. Queiroz,
  \href{http://dx.doi.org/10.1007/JHEP10(2015)076}{{\it {Dirac-fermionic dark
  matter in U(1)$_{X}$ models}}, } {\em JHEP} {\bf 10} (2015) 076,
  [\href{http://arxiv.org/abs/1506.06767}{{\tt 1506.06767}}].

\bibitem{Ghorbani:2015baa}
K.~Ghorbani and H.~Ghorbani,
  \href{http://dx.doi.org/10.1103/PhysRevD.91.123541}{{\it {Two-portal Dark
  Matter}}, } {\em Phys. Rev.} {\bf D91} (2015) 123541,
  [\href{http://arxiv.org/abs/1504.03610}{{\tt 1504.03610}}].

\bibitem{Buchmueller:2013dya}
O.~Buchmueller, M.~J. Dolan, and C.~McCabe,
  \href{http://dx.doi.org/10.1007/JHEP01(2014)025}{{\it {Beyond Effective Field
  Theory for Dark Matter Searches at the LHC}}, } {\em JHEP} {\bf 01} (2014)
  025, [\href{http://arxiv.org/abs/1308.6799}{{\tt 1308.6799}}].

\bibitem{Harris:2014hga}
P.~Harris, V.~V. Khoze, M.~Spannowsky, and C.~Williams,
  \href{http://dx.doi.org/10.1103/PhysRevD.91.055009}{{\it {Constraining Dark
  Sectors at Colliders: Beyond the Effective Theory Approach}}, } {\em Phys.
  Rev.} {\bf D91} (2015) 055009, [\href{http://arxiv.org/abs/1411.0535}{{\tt
  1411.0535}}].

\bibitem{Buckley:2014fba}
M.~R. Buckley, D.~Feld, and D.~Goncalves,
  \href{http://dx.doi.org/10.1103/PhysRevD.91.015017}{{\it {Scalar Simplified
  Models for Dark Matter}}, } {\em Phys. Rev.} {\bf D91} (2015) 015017,
  [\href{http://arxiv.org/abs/1410.6497}{{\tt 1410.6497}}].

\bibitem{Abdallah:2015ter}
J.~Abdallah et~al., \href{http://dx.doi.org/10.1016/j.dark.2015.08.001}{{\it {Simplified Models for Dark Matter Searches at the
  LHC}}, } {\em Phys. Dark Univ.} {\bf 9-10} (2015) 8, [\href{http://arxiv.org/abs/1506.03116}{{\tt 1506.03116}}].

\bibitem{Abercrombie:2015wmb}
D.~Abercrombie et~al., {\it {Dark Matter Benchmark Models for Early LHC Run-2
  Searches: Report of the ATLAS/CMS Dark Matter Forum}},
  \href{http://arxiv.org/abs/1507.00966}{{\tt 1507.00966}}.

\bibitem{Kahlhoefer:2015bea}
F.~Kahlhoefer, K.~Schmidt-Hoberg, T.~Schwetz, and S.~Vogl,
  \href{http://dx.doi.org/10.1007/JHEP02(2016)016}{{\it {Implications of
  unitarity and gauge invariance for simplified dark matter models}}, } {\em
  JHEP} {\bf 02} (2016) 016, [\href{http://arxiv.org/abs/1510.02110}{{\tt
  1510.02110}}].

\bibitem{Englert:2016joy}
C.~Englert, M.~McCullough, and M.~Spannowsky, {\it {S-Channel Dark Matter
  Simplified Models and Unitarity}},
  \href{http://arxiv.org/abs/1604.07975}{{\tt 1604.07975}}.

\bibitem{Boveia:2016mrp}
G.~Busoni et~al., {\it {Recommendations on presenting LHC searches for missing
  transverse energy signals using simplified $s$-channel models of dark
  matter}},  \href{http://arxiv.org/abs/1603.04156}{{\tt 1603.04156}}.

\bibitem{Alves:2015dya}
A.~Alves and K.~Sinha, \href{http://dx.doi.org/10.1103/PhysRevD.92.115013}{{\it
  {Searches for Dark Matter at the LHC: A Multivariate Analysis in the Mono-$Z$
  Channel}}, } {\em Phys. Rev.} {\bf D92} (2015) 115013,
  [\href{http://arxiv.org/abs/1507.08294}{{\tt 1507.08294}}].

\bibitem{Jacques:2015zha}
T.~Jacques and K.~Nordstrom,
  \href{http://dx.doi.org/10.1007/JHEP06(2015)142}{{\it {Mapping monojet
  constraints onto Simplified Dark Matter Models}}, } {\em JHEP} {\bf 06}
  (2015) 142, [\href{http://arxiv.org/abs/1502.05721}{{\tt 1502.05721}}].

\bibitem{Harris:2015kda}
P.~Harris, V.~V. Khoze, M.~Spannowsky, and C.~Williams,
  \href{http://dx.doi.org/10.1103/PhysRevD.93.054030}{{\it {Closing up on Dark
  Sectors at Colliders: from 14 to 100 TeV}}, } {\em Phys. Rev.} {\bf D93}
  (2016) 054030, [\href{http://arxiv.org/abs/1509.02904}{{\tt 1509.02904}}].

\bibitem{Bell:2015rdw}
N.~F. Bell, Y.~Cai, and R.~K. Leane,
  \href{http://dx.doi.org/10.1088/1475-7516/2016/01/051}{{\it {Mono-W Dark
  Matter Signals at the LHC: Simplified Model Analysis}}, } {\em JCAP} {\bf 01}
  (2016) 051, [\href{http://arxiv.org/abs/1512.00476}{{\tt 1512.00476}}].

\bibitem{Haisch:2016usn}
U.~Haisch, F.~Kahlhoefer, and T.~M.~P. Tait, \href{http://dx.doi.org/10.1016/j.physletb.2016.06.063}{{\it {On Mono-W Signatures in
  Spin-1 Simplified Models}}, } {\em Phys. Lett.} {\bf B760} (2016) 207, [\href{http://arxiv.org/abs/1603.01267}{{\tt
  1603.01267}}].

\bibitem{Brennan:2016xjh}
A.~J. Brennan, M.~F. McDonald, J.~Gramling, and T.~D. Jacques,
  \href{http://dx.doi.org/10.1007/JHEP05(2016)112}{{\it {Collide and Conquer:
  Constraints on Simplified Dark Matter Models using Mono-X Collider
  Searches}}, } {\em JHEP} {\bf 05} (2016) 112,
  [\href{http://arxiv.org/abs/1603.01366}{{\tt 1603.01366}}].

\bibitem{Buchmueller:2014yoa}
O.~Buchmueller, M.~J. Dolan, S.~A. Malik, and C.~McCabe,
  \href{http://dx.doi.org/10.1007/JHEP01(2015)037}{{\it {Characterising dark
  matter searches at colliders and direct detection experiments: Vector
  mediators}}, } {\em JHEP} {\bf 01} (2015) 037,
  [\href{http://arxiv.org/abs/1407.8257}{{\tt 1407.8257}}].

\bibitem{Fairbairn:2014aqa}
M.~Fairbairn and J.~Heal,
  \href{http://dx.doi.org/10.1103/PhysRevD.90.115019}{{\it {Complementarity of
  dark matter searches at resonance}}, } {\em Phys. Rev.} {\bf D90} (2014)
  115019, [\href{http://arxiv.org/abs/1406.3288}{{\tt 1406.3288}}].

\bibitem{Choudhury:2015lha}
A.~Choudhury, K.~Kowalska, L.~Roszkowski, E.~M. Sessolo, and A.~J. Williams,
  \href{http://dx.doi.org/10.1007/JHEP04(2016)182}{{\it {Less-simplified models
  of dark matter for direct detection and the LHC}}, } {\em JHEP} {\bf 04}
  (2016) 182, [\href{http://arxiv.org/abs/1509.05771}{{\tt 1509.05771}}].

\bibitem{Blennow:2015gta}
M.~Blennow, J.~Herrero-Garcia, T.~Schwetz, and S.~Vogl,
  \href{http://dx.doi.org/10.1088/1475-7516/2015/08/039}{{\it {Halo-independent
  tests of dark matter direct detection signals: local DM density, LHC, and
  thermal freeze-out}}, } {\em JCAP} {\bf 08} (2015) 039,
  [\href{http://arxiv.org/abs/1505.05710}{{\tt 1505.05710}}].

\bibitem{Heisig:2015ira}
J.~Heisig, M.~Kramer, M.~Pellen, and C.~Wiebusch,
  \href{http://dx.doi.org/10.1103/PhysRevD.93.055029}{{\it {Constraints on
  Majorana Dark Matter from the LHC and IceCube}}, } {\em Phys. Rev.} {\bf D93}
  (2016) 055029, [\href{http://arxiv.org/abs/1509.07867}{{\tt 1509.07867}}].

\bibitem{Alves:2015pea}
A.~Alves, A.~Berlin, S.~Profumo, and F.~S. Queiroz,
  \href{http://dx.doi.org/10.1103/PhysRevD.92.083004}{{\it {Dark Matter
  Complementarity and the Z$^\prime$ Portal}}, } {\em Phys. Rev.} {\bf D92}
  (2015) 083004, [\href{http://arxiv.org/abs/1501.03490}{{\tt
  1501.03490}}].

\bibitem{Busoni:2014gta}
G.~Busoni, A.~De~Simone, T.~Jacques, E.~Morgante, and A.~Riotto,
  \href{http://dx.doi.org/10.1088/1475-7516/2015/03/022}{{\it {Making the Most
  of the Relic Density for Dark Matter Searches at the LHC 14 TeV Run}}, } {\em
  JCAP} {\bf 03} (2015) 022, [\href{http://arxiv.org/abs/1410.7409}{{\tt
  1410.7409}}].

\bibitem{Pais:1973mi}
A.~Pais, \href{http://dx.doi.org/10.1103/PhysRevD.8.1844}{{\it {Remark on
  baryon conservation}}, } {\em Phys. Rev.} {\bf D8} (1973) 1844.

\bibitem{Duerr:2013dza}
M.~Duerr, P.~Fileviez~Perez, and M.~B. Wise,
  \href{http://dx.doi.org/10.1103/PhysRevLett.110.231801}{{\it {Gauge Theory
  for Baryon and Lepton Numbers with Leptoquarks}}, } {\em Phys. Rev. Lett.}
  {\bf 110} (2013) 231801, [\href{http://arxiv.org/abs/1304.0576}{{\tt
  1304.0576}}].

\bibitem{Perez:2014qfa}
P.~Fileviez~Perez, S.~Ohmer, and H.~H. Patel,
  \href{http://dx.doi.org/10.1016/j.physletb.2014.06.057}{{\it {Minimal Theory
  for Lepto-Baryons}}, } {\em Phys. Lett.} {\bf B735} (2014) 283,
  [\href{http://arxiv.org/abs/1403.8029}{{\tt 1403.8029}}].

\bibitem{Duerr:2013lka}
M.~Duerr and P.~Fileviez~Perez,
  \href{http://dx.doi.org/10.1016/j.physletb.2014.03.011}{{\it {Baryonic Dark
  Matter}}, } {\em Phys. Lett.} {\bf B732} (2014) 101,
  [\href{http://arxiv.org/abs/1309.3970}{{\tt 1309.3970}}].

\bibitem{Duerr:2014wra}
M.~Duerr and P.~Fileviez~Perez,
  \href{http://dx.doi.org/10.1103/PhysRevD.91.095001}{{\it {Theory for Baryon
  Number and Dark Matter at the LHC}}, } {\em Phys. Rev.} {\bf D91} (2015)
  095001, [\href{http://arxiv.org/abs/1409.8165}{{\tt 1409.8165}}].

\bibitem{Ohmer:2015lxa}
S.~Ohmer and H.~H. Patel,
  \href{http://dx.doi.org/10.1103/PhysRevD.92.055020}{{\it {Leptobaryons as
  Majorana Dark Matter}}, } {\em Phys. Rev.} {\bf D92} (2015) 055020,
  [\href{http://arxiv.org/abs/1506.00954}{{\tt 1506.00954}}].

\bibitem{Duerr:2015vna}
M.~Duerr, P.~Fileviez~Perez, and J.~Smirnov,
  \href{http://dx.doi.org/10.1103/PhysRevD.93.023509}{{\it {Gamma Lines from
  Majorana Dark Matter}}, } {\em Phys. Rev.} {\bf D93} (2016) 023509,
  [\href{http://arxiv.org/abs/1508.01425}{{\tt 1508.01425}}].

\bibitem{Pospelov:2007mp}
M.~Pospelov, A.~Ritz, and M.~B. Voloshin,
  \href{http://dx.doi.org/10.1016/j.physletb.2008.02.052}{{\it {Secluded WIMP
  Dark Matter}}, } {\em Phys. Lett.} {\bf B662} (2008) 53,
  [\href{http://arxiv.org/abs/0711.4866}{{\tt 0711.4866}}].

\bibitem{LopezHonorez:2012kv}
L.~Lopez-Honorez, T.~Schwetz, and J.~Zupan,
  \href{http://dx.doi.org/10.1016/j.physletb.2012.07.017}{{\it {Higgs Portal,
  Fermionic Dark Matter, and a Standard Model Like Higgs at 125 GeV}}, } {\em
  Phys. Lett.} {\bf B716} (2012) 179,
  [\href{http://arxiv.org/abs/1203.2064}{{\tt 1203.2064}}].

\bibitem{Martin:2014sxa}
A.~Martin, J.~Shelton, and J.~Unwin,
  \href{http://dx.doi.org/10.1103/PhysRevD.90.103513}{{\it {Fitting the
  Galactic Center Gamma-Ray Excess with Cascade Annihilations}}, } {\em Phys.
  Rev.} {\bf D90} (2014) 103513, [\href{http://arxiv.org/abs/1405.0272}{{\tt
  1405.0272}}].

\bibitem{Autran:2015mfa}
M.~Autran, K.~Bauer, T.~Lin, and D.~Whiteson,
  \href{http://dx.doi.org/10.1103/PhysRevD.92.035007}{{\it {Searches for dark
  matter in events with a resonance and missing transverse energy}}, } {\em
  Phys. Rev.} {\bf D92} (2015) 035007,
  [\href{http://arxiv.org/abs/1504.01386}{{\tt 1504.01386}}].

\bibitem{Buschmann:2015awa}
M.~Buschmann, J.~Kopp, J.~Liu, and P.~A.~N. Machado,
  \href{http://dx.doi.org/10.1007/JHEP07(2015)045}{{\it {Lepton Jets from
  Radiating Dark Matter}}, } {\em JHEP} {\bf 07} (2015) 045,
  [\href{http://arxiv.org/abs/1505.07459}{{\tt 1505.07459}}].

\bibitem{Bai:2015nfa}
Y.~Bai, J.~Bourbeau, and T.~Lin,
  \href{http://dx.doi.org/10.1007/JHEP06(2015)205}{{\it {Dark matter searches
  with a mono-$Z^{′}$ jet}}, } {\em JHEP} {\bf 06} (2015) 205,
  [\href{http://arxiv.org/abs/1504.01395}{{\tt 1504.01395}}].

\bibitem{Gupta:2015lfa}
A.~Gupta, R.~Primulando, and P.~Saraswat,
  \href{http://dx.doi.org/10.1007/JHEP09(2015)079}{{\it {A New Probe of Dark
  Sector Dynamics at the LHC}}, } {\em JHEP} {\bf 09} (2015) 079,
  [\href{http://arxiv.org/abs/1504.01385}{{\tt 1504.01385}}].

\bibitem{Buschmann:2016hkc}
M.~Buschmann, S.~El Hedri, A.~Kaminska, J.~Liu, M.~de Vries, X.~P.~Wang, F.~Yu and J.~Zurita,
  \href{http://dx.doi.org/10.1007/JHEP09(2016)033}{{\it {Hunting for Dark Matter Coannihilation by Mixing Dijet Resonances and Missing Transverse Energy}}, }
  {\em JHEP} {\bf 09} (2016) 033, 
  [\href{http://arxiv.org/abs/1605.08056}{{\tt 1605.08056}}].
  
\bibitem{Ekstedt:2016wyi}
A.~Ekstedt, R.~Enberg, G.~Ingelman, J.~L\"ofgren, and T.~Mandal, {\it {Ruling
  out minimal anomaly free $\mathrm{U}(1)$ extensions of the Standard Model}},
  \href{http://arxiv.org/abs/1605.04855}{{\tt 1605.04855}}.

\bibitem{Liu:2011dh}
J.-Y. Liu, Y.~Tang, and Y.-L. Wu,
  \href{http://dx.doi.org/10.1088/0954-3899/39/5/055003}{{\it {Searching for
  $Z^{'}$ Gauge Boson in an Anomaly-Free U(1)$'$ Gauge Family Model}}, } {\em
  J. Phys.} {\bf G39} (2012) 055003,
  [\href{http://arxiv.org/abs/1108.5012}{{\tt 1108.5012}}].

\bibitem{Belanger:2014vza}
G.~B\'{e}langer, F.~Boudjema, A.~Pukhov, and A.~Semenov,
  \href{http://dx.doi.org/10.1016/j.cpc.2015.03.003}{{\it {micrOMEGAs4.1: two
  dark matter candidates}}, } {\em Comput. Phys. Commun.} {\bf 192} (2015) 322,
  [\href{http://arxiv.org/abs/1407.6129}{{\tt 1407.6129}}].

\bibitem{Ade:2015xua}
{\bf Planck}, P.~A.~R. Ade et~al., \href{http://dx.doi.org/10.1051/0004-6361/201525830}{{\it {Planck 2015 Results. XIII.
  Cosmological Parameters}}, } {\em Astron.\ \& Astrophys.} {\bf 594} (2016) A13, [\href{http://arxiv.org/abs/1502.01589}{{\tt
  1502.01589}}].

\bibitem{Hedri:2014mua}
S.~El~Hedri, W.~Shepherd, and D.~G.~E. Walker, {\it {Perturbative Unitarity
  Constraints on Gauge Portals}},  \href{http://arxiv.org/abs/1412.5660}{{\tt
  1412.5660}}.

\bibitem{Lee:1977eg}
B.~W. Lee, C.~Quigg, and H.~B. Thacker,
  \href{http://dx.doi.org/10.1103/PhysRevD.16.1519}{{\it {Weak Interactions at
  Very High-Energies: The Role of the Higgs Boson Mass}}, } {\em Phys. Rev.}
  {\bf D16} (1977) 1519.

\bibitem{Kang:2013zba}
S.~K. Kang and J.~Park, \href{http://dx.doi.org/10.1007/JHEP04(2015)009}{{\it
  {Unitarity Constraints in the standard model with a singlet scalar field}}, }
  {\em JHEP} {\bf 04} (2015) 009, [\href{http://arxiv.org/abs/1306.6713}{{\tt
  1306.6713}}].

\bibitem{D'Eramo:2016atc}
F.~D'Eramo, B.~J. Kavanagh, and P.~Panci, \href{http://dx.doi.org/10.1007/JHEP08(2016)111}{{\it {You can hide but you have to
  run: direct detection with vector mediators}}, } {\em JHEP} {\bf 08} (2016) 111,
  [\href{http://arxiv.org/abs/1605.04917}{{\tt 1605.04917}}].

\bibitem{Fitzpatrick:2012ix}
A.~L. Fitzpatrick, W.~Haxton, E.~Katz, N.~Lubbers, and Y.~Xu,
  \href{http://dx.doi.org/10.1088/1475-7516/2013/02/004}{{\it {The Effective
  Field Theory of Dark Matter Direct Detection}}, } {\em JCAP} {\bf 02} (2013)
  004, [\href{http://arxiv.org/abs/1203.3542}{{\tt 1203.3542}}].

\bibitem{Anand:2013yka}
N.~Anand, A.~L. Fitzpatrick, and W.~C. Haxton,
  \href{http://dx.doi.org/10.1103/PhysRevC.89.065501}{{\it {Weakly interacting
  massive particle-nucleus elastic scattering response}}, } {\em Phys. Rev.}
  {\bf C89} (2014) 065501, [\href{http://arxiv.org/abs/1308.6288}{{\tt
  1308.6288}}].

\bibitem{Akerib:2015rjg}
{\bf LUX}, D.~S. Akerib et~al.,
  \href{http://dx.doi.org/10.1103/PhysRevLett.116.161301}{{\it {Improved WIMP
  scattering limits from the LUX experiment}}, } {\em Phys. Rev. Lett.} {\bf
  116} (2016) 161301, [\href{http://arxiv.org/abs/1512.03506}{{\tt
  1512.03506}}].

\bibitem{Akerib:2013tjd}
{\bf LUX}, D.~S. Akerib et~al.,
  \href{http://dx.doi.org/10.1103/PhysRevLett.112.091303}{{\it {First Results
  from the Lux Dark Matter Experiment at the Sanford Underground Research
  Facility}}, } {\em Phys. Rev. Lett.} {\bf 112} (2014) 091303,
  [\href{http://arxiv.org/abs/1310.8214}{{\tt 1310.8214}}].

\bibitem{DelNobile:2013sia}
M.~Cirelli, E.~Del~Nobile, and P.~Panci,
  \href{http://dx.doi.org/10.1088/1475-7516/2013/10/019}{{\it {Tools for
  model-independent bounds in direct dark matter searches}}, } {\em JCAP} {\bf
  10} (2013) 019, [\href{http://arxiv.org/abs/1307.5955}{{\tt 1307.5955}}].

\bibitem{Khachatryan:2014rra}
{\bf CMS}, V.~Khachatryan et~al.,
  \href{http://dx.doi.org/10.1140/epjc/s10052-015-3451-4}{{\it {Search for dark
  matter, extra dimensions, and unparticles in monojet events in proton proton
  collisions at $\sqrt{s} = 8$ TeV}}, } {\em Eur. Phys. J.} {\bf C75} (2015)
  235, [\href{http://arxiv.org/abs/1408.3583}{{\tt 1408.3583}}].

\bibitem{Aad:2015zva}
{\bf ATLAS}, G.~Aad et~al.,
  \href{http://dx.doi.org/10.1140/epjc/s10052-015-3517-3,
  10.1140/epjc/s10052-015-3639-7}{{\it {Search for new phenomena in final
  states with an energetic jet and large missing transverse momentum in pp
  collisions at $\sqrt{s}=$8 TeV with the ATLAS detector}}, } {\em Eur. Phys.
  J.} {\bf C75} (2015) 299, [\href{http://arxiv.org/abs/1502.01518}{{\tt
  1502.01518}}]. [Erratum: \textit{Eur. Phys. J.} \textbf{C75} (2015) 408].

\bibitem{CMS:2016tns}
{\bf CMS}, \href{http://cds.cern.ch/record/2148032}{{\it {Search for dark
  matter production in association with jets, or hadronically decaying W or Z
  boson at $\sqrt{s} = 13$ TeV}}, } Tech. Rep. CMS-PAS-EXO-16-013, 2016.

\bibitem{Aaboud:2016tnv}
{\bf ATLAS}, M.~Aaboud et~al., \href{http://dx.doi.org/10.1103/PhysRevD.94.032005}{{\it {Search for new phenomena in final states
  with an energetic jet and large missing transverse momentum in $pp$
  collisions at $\sqrt{s}=13$ TeV using the ATLAS detector}}, } {\em Phys. Rev.} {\bf D94} (2016) 032005, 
  [\href{http://arxiv.org/abs/1604.07773}{{\tt 1604.07773}}].

\bibitem{Belyaev:2012qa}
A.~Belyaev, N.~D. Christensen, and A.~Pukhov,
  \href{http://dx.doi.org/10.1016/j.cpc.2013.01.014}{{\it {CalcHEP 3.4 for
  collider physics within and beyond the Standard Model}}, } {\em Comput. Phys.
  Commun.} {\bf 184} (2013) 1729, [\href{http://arxiv.org/abs/1207.6082}{{\tt
  1207.6082}}].

\bibitem{Sjostrand:2007gs}
T.~Sjostrand, S.~Mrenna, and P.~Z. Skands,
  \href{http://dx.doi.org/10.1016/j.cpc.2008.01.036}{{\it {A Brief Introduction
  to PYTHIA 8.1}}, } {\em Comput. Phys. Commun.} {\bf 178} (2008) 852,
  [\href{http://arxiv.org/abs/0710.3820}{{\tt 0710.3820}}].

\bibitem{deFavereau:2013fsa}
{\bf DELPHES 3}, J.~de~Favereau, C.~Delaere, P.~Demin, A.~Giammanco,
  V.~Lemaitre, et~al., \href{http://dx.doi.org/10.1007/JHEP02(2014)057}{{\it
  {DELPHES 3, A modular framework for fast simulation of a generic collider
  experiment}}, } {\em JHEP} {\bf 02} (2014) 057,
  [\href{http://arxiv.org/abs/1307.6346}{{\tt 1307.6346}}].

\bibitem{Khachatryan:2015sja}
{\bf CMS}, V.~Khachatryan et~al.,
  \href{http://dx.doi.org/10.1103/PhysRevD.91.052009}{{\it {Search for
  resonances and quantum black holes using dijet mass spectra in proton-proton
  collisions at $\sqrt{s} =$ 8 TeV}}, } {\em Phys. Rev.} {\bf D91} (2015)
  052009, [\href{http://arxiv.org/abs/1501.04198}{{\tt 1501.04198}}].

\bibitem{Aad:2014aqa}
{\bf ATLAS}, G.~Aad et~al.,
  \href{http://dx.doi.org/10.1103/PhysRevD.91.052007}{{\it {Search for new
  phenomena in the dijet mass distribution using $p-p$ collision data at
  $\sqrt{s}=8$ TeV with the ATLAS detector}}, } {\em Phys. Rev.} {\bf D91}
  (2015) 052007, [\href{http://arxiv.org/abs/1407.1376}{{\tt 1407.1376}}].

\bibitem{Khachatryan:2015dcf}
{\bf CMS}, V.~Khachatryan et~al.,
  \href{http://dx.doi.org/10.1103/PhysRevLett.116.071801}{{\it {Search for
  narrow resonances decaying to dijets in proton-proton collisions at $\sqrt(s)
  =$ 13 TeV}}, } {\em Phys. Rev. Lett.} {\bf 116} (2016) 071801,
  [\href{http://arxiv.org/abs/1512.01224}{{\tt 1512.01224}}].

\bibitem{ATLAS:2015nsi}
{\bf ATLAS}, G.~Aad et~al.,
  \href{http://dx.doi.org/10.1016/j.physletb.2016.01.032}{{\it {Search for new
  phenomena in dijet mass and angular distributions from $pp$ collisions at
  $\sqrt{s}=$ 13 TeV with the ATLAS detector}}, } {\em Phys. Lett.} {\bf B754}
  (2016) 302, [\href{http://arxiv.org/abs/1512.01530}{{\tt 1512.01530}}].

\bibitem{Khachatryan:2016ecr}
{\bf CMS}, V.~Khachatryan et~al., \href{http://dx.doi.org/10.1103/PhysRevLett.117.031802}{{\it {Search for narrow resonances in dijet
  final states at sqrt(s)=8 TeV with the novel CMS technique of data
  scouting}}, } {\em Phys. Rev. Lett.} {\bf 117} (2016) 031802, [\href{http://arxiv.org/abs/1604.08907}{{\tt 1604.08907}}].

\bibitem{Holdom:1985ag}
B.~Holdom, \href{http://dx.doi.org/10.1016/0370-2693(86)91377-8}{{\it {Two
  U(1)'s and Epsilon Charge Shifts}}, } {\em Phys. Lett.} {\bf B166} (1986)
  196.

\bibitem{Babu:1996vt}
K.~S. Babu, C.~F. Kolda, and J.~March-Russell,
  \href{http://dx.doi.org/10.1103/PhysRevD.54.4635}{{\it {Leptophobic U(1) $s$
  and the R($b$) - R($c$) crisis}}, } {\em Phys. Rev.} {\bf D54} (1996) 4635,
  [\href{http://arxiv.org/abs/hep-ph/9603212}{{\tt hep-ph/9603212}}].

\bibitem{Carone:1995pu}
C.~D. Carone and H.~Murayama,
  \href{http://dx.doi.org/10.1103/PhysRevD.52.484}{{\it {Realistic models with
  a light U(1) gauge boson coupled to baryon number}}, } {\em Phys. Rev.} {\bf
  D52} (1995) 484, [\href{http://arxiv.org/abs/hep-ph/9501220}{{\tt
  hep-ph/9501220}}].

\bibitem{Aad:2014cka}
{\bf ATLAS}, G.~Aad et~al.,
  \href{http://dx.doi.org/10.1103/PhysRevD.90.052005}{{\it {Search for
  high-mass dilepton resonances in pp collisions at $\sqrt{s}=8$  TeV with
  the ATLAS detector}}, } {\em Phys. Rev.} {\bf D90} (2014) 052005,
  [\href{http://arxiv.org/abs/1405.4123}{{\tt 1405.4123}}].

\bibitem{Aaltonen:2008ah}
{\bf CDF}, T.~Aaltonen et~al.,
  \href{http://dx.doi.org/10.1103/PhysRevLett.102.091805}{{\it {A Search for
  high-mass resonances decaying to dimuons at CDF}}, } {\em Phys. Rev. Lett.}
  {\bf 102} (2009) 091805, [\href{http://arxiv.org/abs/0811.0053}{{\tt
  0811.0053}}].

\bibitem{Agashe:2014kda}
{\bf Particle Data Group}, K.~A. Olive et~al.,
  \href{http://dx.doi.org/10.1088/1674-1137/38/9/090001}{{\it {Review of
  Particle Physics}}, } {\em Chin. Phys.} {\bf C38} (2014) 090001.

\bibitem{Hook:2010tw}
A.~Hook, E.~Izaguirre, and J.~G. Wacker,
  \href{http://dx.doi.org/10.1155/2011/859762}{{\it {Model Independent Bounds
  on Kinetic Mixing}}, } {\em Adv. High Energy Phys.} {\bf 2011} (2011) 859762,
  [\href{http://arxiv.org/abs/1006.0973}{{\tt 1006.0973}}].

\bibitem{Griest:1989wd}
K.~Griest and M.~Kamionkowski,
  \href{http://dx.doi.org/10.1103/PhysRevLett.64.615}{{\it {Unitarity Limits on
  the Mass and Radius of Dark Matter Particles}}, } {\em Phys. Rev. Lett.} {\bf
  64} (1990) 615.

\bibitem{AtlasCMSHiggsSignalStrength}
{\bf ATLAS and CMS}, \href{http://cds.cern.ch/record/2052552}{{\it
  {Measurements of the Higgs boson production and decay rates and constraints
  on its couplings from a combined ATLAS and CMS analysis of the LHC pp
  collision data at $\sqrt{s}$ = 7 and 8 TeV}}, } Tech. Rep.
  ATLAS-CONF-2015-044, 2015.

\bibitem{Pelliccioni:2015hva}
{\bf CMS}, M.~Pelliccioni,
  \href{https://inspirehep.net/record/1370129/files/arXiv:1505.03831.pdf}{{\it
  {CMS High mass WW and ZZ Higgs search with the complete LHC Run1
  statistics}}, } in {\em {Proceedings, 50th Rencontres de Moriond Electroweak
  interactions and unified theories}}, p.~47, 2015.
\newblock \href{http://arxiv.org/abs/1505.03831}{{\tt 1505.03831}}.

\bibitem{Aad:2015kna}
{\bf ATLAS}, G.~Aad et~al.,
  \href{http://dx.doi.org/10.1140/epjc/s10052-015-3820-z}{{\it {Search for an
  additional, heavy Higgs boson in the $H\rightarrow ZZ$ decay channel at
  $\sqrt{s} = 8\;\text{ TeV }$ in $pp$ collision data with the ATLAS
  detector}}, } {\em Eur. Phys. J.} {\bf C76} (2016) 45,
  [\href{http://arxiv.org/abs/1507.05930}{{\tt 1507.05930}}].

\bibitem{Falkowski:2015iwa}
A.~Falkowski, C.~Gross, and O.~Lebedev,
  \href{http://dx.doi.org/10.1007/JHEP05(2015)057}{{\it {A second Higgs from
  the Higgs portal}}, } {\em JHEP} {\bf 05} (2015) 057,
  [\href{http://arxiv.org/abs/1502.01361}{{\tt 1502.01361}}].

\bibitem{Ackermann:2015zua}
{\bf Fermi-LAT}, M.~Ackermann et~al.,
  \href{http://dx.doi.org/10.1103/PhysRevLett.115.231301}{{\it {Searching for
  Dark Matter Annihilation from Milky Way Dwarf Spheroidal Galaxies with Six
  Years of Fermi Large Area Telescope Data}}, } {\em Phys. Rev. Lett.} {\bf
  115} (2015) 231301, [\href{http://arxiv.org/abs/1503.02641}{{\tt
  1503.02641}}].

\bibitem{Hooper:2010mq}
D.~Hooper and L.~Goodenough,
  \href{http://dx.doi.org/10.1016/j.physletb.2011.02.029}{{\it {Dark Matter
  Annihilation in The Galactic Center As Seen by the Fermi Gamma Ray Space
  Telescope}}, } {\em Phys. Lett.} {\bf B697} (2011) 412,
  [\href{http://arxiv.org/abs/1010.2752}{{\tt 1010.2752}}].

\bibitem{Hooper:2011ti}
D.~Hooper and T.~Linden,
  \href{http://dx.doi.org/10.1103/PhysRevD.84.123005}{{\it {On The Origin Of
  The Gamma Rays From The Galactic Center}}, } {\em Phys. Rev.} {\bf D84}
  (2011) 123005, [\href{http://arxiv.org/abs/1110.0006}{{\tt 1110.0006}}].

\bibitem{Abazajian:2012pn}
K.~N. Abazajian and M.~Kaplinghat,
  \href{http://dx.doi.org/10.1103/PhysRevD.86.083511,
  10.1103/PhysRevD.87.129902}{{\it {Detection of a Gamma-Ray Source in the
  Galactic Center Consistent with Extended Emission from Dark Matter
  Annihilation and Concentrated Astrophysical Emission}}, } {\em Phys. Rev.}
  {\bf D86} (2012) 083511, [\href{http://arxiv.org/abs/1207.6047}{{\tt
  1207.6047}}]. [Erratum: \textit{Phys. Rev.} \textbf{D87} (2013) 129902].

\bibitem{Gordon:2013vta}
C.~Gordon and O.~Macias, \href{http://dx.doi.org/10.1103/PhysRevD.88.083521,
  10.1103/PhysRevD.89.049901}{{\it {Dark Matter and Pulsar Model Constraints
  from Galactic Center Fermi-LAT Gamma Ray Observations}}, } {\em Phys. Rev.}
  {\bf D88} (2013) 083521, [\href{http://arxiv.org/abs/1306.5725}{{\tt
  1306.5725}}]. [Erratum: \textit{Phys. Rev.} \textbf{D89} (2014) 049901].

\bibitem{Daylan:2014rsa}
T.~Daylan, D.~P. Finkbeiner, D.~Hooper, T.~Linden, S.~K.~N. Portillo, et~al.,
  \href{http://dx.doi.org/10.1016/j.dark.2015.12.005}{{\it {The
  characterization of the gamma-ray signal from the central Milky Way: A case
  for annihilating dark matter}}, } {\em Phys. Dark Univ.} {\bf 12} (2016) 1,
  [\href{http://arxiv.org/abs/1402.6703}{{\tt 1402.6703}}].

\bibitem{Carr:2015hta}
{\bf CTA Consortium}, J.~Carr et~al.,
  \href{http://inspirehep.net/record/1389681/files/arXiv:1508.06128.pdf}{{\it
  {Prospects for Indirect Dark Matter Searches with the Cherenkov Telescope
  Array (Cta)}}, } in {\em {Proceedings, 34Th International Cosmic Ray
  Conference (Icrc 2015)}}, 2015.
\newblock \href{http://arxiv.org/abs/1508.06128}{{\tt 1508.06128}}.

\bibitem{Lee:2015fea}
S.~K. Lee, M.~Lisanti, B.~R. Safdi, T.~R. Slatyer, and W.~Xue,
  \href{http://dx.doi.org/10.1103/PhysRevLett.116.051103}{{\it {Evidence for
  Unresolved $\gamma$-Ray Point Sources in the Inner Galaxy}}, } {\em Phys.
  Rev. Lett.} {\bf 116} (2016) 051103,
  [\href{http://arxiv.org/abs/1506.05124}{{\tt 1506.05124}}].

\bibitem{Bartels:2015aea}
R.~Bartels, S.~Krishnamurthy, and C.~Weniger,
  \href{http://dx.doi.org/10.1103/PhysRevLett.116.051102}{{\it {Strong support
  for the millisecond pulsar origin of the Galactic center GeV excess}}, } {\em
  Phys. Rev. Lett.} {\bf 116} (2016) 051102,
  [\href{http://arxiv.org/abs/1506.05104}{{\tt 1506.05104}}].

\bibitem{ArkaniHamed:2006mb}
N.~Arkani-Hamed, A.~Delgado, and G.~F. Giudice,
  \href{http://dx.doi.org/10.1016/j.nuclphysb.2006.02.010}{{\it {The
  Well-Tempered Neutralino}}, } {\em Nucl. Phys.} {\bf B741} (2006) 108,
  [\href{http://arxiv.org/abs/hep-ph/0601041}{{\tt hep-ph/0601041}}].

\bibitem{Banerjee:2016hsk}
S.~Banerjee, S.~Matsumoto, K.~Mukaida, and Y.-L.~S. Tsai, {\it {WIMP Dark
  Matter in a Well-Tempered Regime: a Case Study on Singlet-Doublets Fermionic
  WIMP}},  \href{http://arxiv.org/abs/1603.07387}{{\tt 1603.07387}}.

\bibitem{Baker:2015qna}
M.~J. Baker et~al., \href{http://dx.doi.org/10.1007/JHEP12(2015)120}{{\it {The
  Coannihilation Codex}}, } {\em JHEP} {\bf 12} (2015) 120,
  [\href{http://arxiv.org/abs/1510.03434}{{\tt 1510.03434}}].

\bibitem{Frandsen:2011cg}
M.~T. Frandsen, F.~Kahlhoefer, S.~Sarkar, and K.~Schmidt-Hoberg,
  \href{http://dx.doi.org/10.1007/JHEP09(2011)128}{{\it {Direct detection of
  dark matter in models with a light Z'}}, } {\em JHEP} {\bf 09} (2011) 128,
  [\href{http://arxiv.org/abs/1107.2118}{{\tt 1107.2118}}].

\end{thebibliography}
\end{document}